\documentclass[12pt,english]{elsarticle}
\usepackage[T1]{fontenc}
\usepackage[utf8]{inputenc}
\usepackage{color}
\usepackage{mathtools}
\usepackage{bm}
\usepackage{dsfont}
\usepackage{amsmath}
\usepackage{amssymb}
\usepackage{stmaryrd}
\usepackage{graphicx}
\usepackage[export]{adjustbox}
\usepackage[bookmarks=true,bookmarksnumbered=true,bookmarksopen=true,bookmarksopenlevel=4,
 breaklinks=false,pdfborder={0 0 1},backref=false,colorlinks=true]
 {hyperref}
\hypersetup{
 citecolor=blue,linkcolor=blue}

\makeatletter

\@ifundefined{date}{}{\date{}}

\usepackage[T1]{fontenc}
\usepackage{textcomp}
\usepackage[utf8]{inputenc}		
\usepackage{geometry}
\usepackage{color}
\usepackage[english]{babel}
\usepackage{microtype}
\usepackage{float}
\usepackage{amsmath}
\usepackage{amssymb}
\usepackage{stmaryrd}
\usepackage{graphicx}

\usepackage{relsize}
\usepackage{enumitem}

\makeatletter

\floatstyle{ruled}
\newfloat{algorithm}{tbp}{loa}
\providecommand{\algorithmname}{Algorithm}
\floatname{algorithm}{\protect\algorithmname}

\@ifundefined{date}{}{\date{}}
\usepackage{bm}
\usepackage{subfig}
\usepackage{algorithm} 
\usepackage{algpseudocode}
\usepackage{multirow} 
\usepackage{psfrag}
\usepackage{amssymb}
\usepackage[T1]{fontenc}
\usepackage{textcomp}
\usepackage{gensymb}
\usepackage{multicol}
\usepackage{verbatim}
\usepackage{booktabs, tabularx, makecell}
\usepackage{array}
\newcolumntype{Y}{>{\centering\arraybackslash}X}

\usepackage{hyperref}
\usepackage{prettyref}
\newrefformat{fig}{\hyperref[#1]{Figure \ref{#1}}}
\newrefformat{tab}{\hyperref[#1]{Table \ref{#1}}}
\newrefformat{sec}{\hyperref[#1]{Section \ref{#1}}}
\newrefformat{subsec}{\hyperref[#1]{Section \ref{#1}}}

\usepackage{placeins}

\usepackage{nomencl,etoolbox,ragged2e,siunitx,mathtools}

\DeclareRobustCommand{\utilde}[1]{%
  \oalign{%
    $\m@th#1$\crcr
    \hidewidth$\m@th\sim$\hidewidth
  }%
}

\renewcommand\nomgroup[1]{\def\nomtemp{\csname nomhead#1\endcsname}\nomtemp}

\PassOptionsToPackage{authoryear}{natbib}
\newcommand{\pd}[2]{\dfrac{\partial #1}{\partial #2}}

\newcommand{\td}[1]{\oalign{$\bm{#1}$\crcr\hidewidth$\scriptscriptstyle\bm{\sim}$\hidewidth}}

\newcommand{\ten}[1]{\boldsymbol{#1}}

\bibpunct{(}{)}{;}{a}{,}{,}

\newcommand\footnoteref[1]{\protected@xdef\@thefnmark{\ref{#1}}\@footnotemark}
\definecolor{mygreen}{rgb}{0.0, 0.5, 0.0}

\ifdefined\showcaptionsetup
 \PassOptionsToPackage{caption=false}{subfig}
\fi
\usepackage{subfig}
\makeatother

\ifdefined\showcaptionsetup
 \PassOptionsToPackage{caption=false}{subfig}
\fi
\usepackage{subfig}
\makeatother
\usepackage{amsthm}
\theoremstyle{definition}   
\newtheorem{remark}{Remark}

\begin{document}
\begin{frontmatter}
\title{A robust and versatile parallel FFT-based mechanical solver for general non-periodic and periodic boundary conditions}
\journal{Computer Methods in Applied Mechanics and Engineering}
\author[cea]{Yaovi Armand AMOUZOU-ADOUN}
\ead{yaovi.amouzou-adoun@cea.fr}
\author[cea]{Lionel GÉLÉBART\corref{cor1}}
\ead{lionel.gelebart@cea.fr}
\author[UnivPoitiers]{Cédric FLAGEUL}
\ead{cedric.flageul@univ-poitiers.fr}
\author[mdls]{Yushan WANG}
\ead{yushan.wang@cea.fr}
\address[cea]{Université Paris-Saclay, CEA, SRMA, Gif-sur-Yvette, 91191, France} 
\address[UnivPoitiers]{Curiosity Group, Pprime Institute, CNRS - University of Poitiers - ENSMA, France}
\address[mdls]{Université Paris-Saclay, UVSQ, CNRS, CEA, Maison de la Simulation, 91191, Gif-sur-Yvette,France}
\cortext[cor1]{Corresponding author} 
\date{\today}
 
\begin{abstract} 
	General boundary conditions are implemented within a fast Fourier transform framework for linear and non-linear mechanical problems using small or finite transformation formulations. In the context of parallel computing (distributed memory), we present a framework that enables the combination of periodic and non-periodic (Dirichlet or Neumann) boundary conditions. Taking advantage of the link between non-periodic boundary conditions and the symmetries of the relevant components of the fluctuation displacement and stress fields, discrete trigonometric transforms are employed to adapt the classical Moulinec–Suquet fast Fourier transform approach. The present study employs an original displacement-based fixed-point algorithm in combination with a convergence acceleration method  in order to solve boundary value problems. Finite difference approaches are used to build the discrete Green operators associated with a pre-conditioner (reference material), whose choice depends on the loading type and the small or finite transformation frameworks. The newly developed double tetrahedron scheme is employed to investigate non-periodic problems. Outcomes are compared to those of the classical hexahedral scheme. The robustness and computational efficiency of the presented parallel solver is demonstrated through numerical experiments of non-trivial loading scenarios (tension, bending, normal-mixed loading, torsion-bending), complex and densely discretized microstructures and diverse behavior laws (elasticity, isotropic plasticity, crystal plasticity), within small and finite transformation frameworks.    
\end{abstract}
\end{frontmatter}{}

\section{Introduction}
\subsection{State of the art}

Accurate comparisons between experimental results and simulations requires robust and efficient numerical methods capable of handling large 3D grids. Since the seminal work of \cite{Moulinec1994,Moulinec1998}, fast Fourier transform (FFT)-based solvers establish themselves as efficient alternatives to standard finite element (FE) codes, in describing the macroscopic properties of heterogeneous media, as well as the distribution of local fields (\emph{e.g.,} strain, stress). For an extensive overview of classical FFT-based solvers and their applications, we refer the reader to the review articles \citep{schneider_review_2021,lucarini_fft_2022,gierden_review_2022}.

FFT-based solvers exploit the application of a discrete Green operator in Fourier space to solve the governing equations efficiently. The initial proposition of \cite{Moulinec1998} can be regarded as a collocation method, starting from the continuous equation and using a truncation of Fourier series to solve the so-called Lippman-Schwinger equation for an auxiliary
problem defined on a homogeneous material. Another possibility consists of using Galerkin variational formulations \citep{brisard_combining_2012,vondrejc_fft-based_2014,bignonnet_fourier-based_2026} to build the discrete Green operator. Moreover, using a finite difference scheme to obtain the discrete differential operators (\emph{e.g.,} divergence, gradient), a discrete Green operator can directly be built from the local equations defined on discrete grids. In the last three decades, the improvements in the expression of the discrete Green operator coupled with acceleration techniques \citep[\emph{e.g.,} Barzilai-Borwein method, conjugate gradient (linear or non-linear), Anderson acceleration based on the displacement field, domain decomposition strategy see for instance ][]{chen_analysis_2019,schneider_review_2021,to_fourier_2025} of the so-called original basic scheme \citep{Moulinec1994,Moulinec1998}, have allowed to study more complex diffusion and mechanical problems. In this context, the present work proposes a new variant of the Anderson acceleration based on the divergence of the polarization stress field.

Despite their effectiveness and growing popularity, classical FFT-based solvers remain inherently limited to periodic media and periodic boundary conditions (BCs), which restricts their applicability to a broader range of problems. In the recent years, various propositions have been discussed to circumvent this limitation. A buffer zone (with an arbitrary behavior) in which the desired heterogeneous medium is embedded, is used to mimic Neumann and/or Dirichlet BCs \citep{lebensohn_orientation_2008,chen_analysis_2019,gelebart_modified_2020}. This approach has been successfully applied for linear and non-linear material behaviors \citep{nkoumbou_kaptchouang_multiscale_2022,zecevic_extended_2025}, at a minimal cost in classical FFT-based solvers. Nevertheless, the buffer zone is shown to influence the convergence rate \citep{nkoumbou_kaptchouang_multiscale_2022}. Furthermore, studying an effective conductivity problem, \cite{monchiet_fft_2024} employed the mirror transformation method to apply uniform BCs. With this approach, the problem is solved on computation domain 8 times larger than the original desired 3D unit-cell, leading to an increase in the simulation time. \cite{to_fourier_2025} proposed a displacement based boundary domain integral equation that enables the solution of elasticity boundary value problems. The applicability of this approach is demonstrated on 2D elastic composites subjected to homogeneous loading. 

Another strategy to handle non-periodic BCs with FFT-based solvers consists of using  discrete trigonometric transforms \citep[DTTs;][]{wiegmann_fast_1999,grimmstrele_fftbased_2021}. With this method the desired problem (\emph{i.e.,} partial differential equations) is solved without enlarging the domain. It relies on employing sine and cosine real transforms instead of the discrete Fourier transform (DFT) as in the classical FFT-based solvers. In this sense, \cite{grimmstrele_fftbased_2021} develop limited number of mixed uniform BCs for isotropic linear elastic materials. \cite{risthaus_fftbased_2024,risthaus_imposing_2024} also apply this approach to represent only Dirichlet BCs in a deformation-based formulation. Using a Galerkin-based method to build the discrete Green operator, \cite{paux_discrete_2025} were able to discuss heterogeneous elastic structures subjected to non-periodic mixed Dirichlet/Neumann BCs. Meanwhile, \cite{gelebart_fft-based_2024,gelebart_accurate_2025} treated static and transient diffusion problems build on finite difference scheme approach in deriving of the discrete Green operator.  More recently, damage issues in 2D composites are discussed in the work of \cite{cao_modeling_2026} under non-periodic BCs (Dirichlet and free surfaces). The DTTs-based approach for applying non-periodic BCs, is the subject of the present work. We aim to provide a robust and generic FFT-based solver for accounting any possible BCs (non-periodic Neumann-Dirichlet and/or a mix with Periodic), implemented within a parallel framework (distributed memory) and suitable for small and finite transformations. 

As stated earlier, the discrete Green operator can be evaluated with various approaches \citep{schneider_review_2021,bignonnet_fourier-based_2026}. In this work, we build on the finite difference methods. In the context of periodic condition, \cite{willot_fourier-based_2015} proposed a rotated scheme (hereafter referred to as hexahedral scheme in light of FE \citep{schneider_fftbased_2017}) helping to achieve in general case, better convergence in comparison with the basic scheme of Moulinec and Suquet \citep{Moulinec1994,Moulinec1998}. In general, within the fixed-point method, the hexahedral scheme still suffers convergence issue with microstructures that include void phases \citep{risthaus_robust_2025}. Very recently, keeping with periodic BCs, \cite{finel_tetrahedron-based_2025} proposed a double tetrahedron finite difference scheme that remedied these convergence issues. With the latter scheme, stress and strain fields exhibiting fewer numerical artifacts can also be obtained. Staggered grid scheme used in the work of \citep{risthaus_robust_2025} provide ringing-free local fields and good convergence in the context of Dirichlet BCs. The effort is made in this work to compare the double tetrahedron scheme to the hexahedral scheme in the context of non-periodic BCs, and particularly in non-linear conditions.

\subsection{Contributions}

In this context, we propose an efficient, versatile and robust \emph{parallel} FFT-based solver that seamlessly accommodates \emph{arbitrary combinations of periodic and/or non-periodic BCs} in small and \emph{finite transformation} simulations of the linear and \emph{non-linear} mechanical responses of small-scale heterogeneous structures, relying on \emph{a formulation in displacement}.

For the purpose, the mechanical problem for Cauchy continuum theories is formulated in a displacement-based approach. By introducing the fluctuation displacement as an unknown, the general (non-periodic and/or periodic) BCs are linked with the symmetries and/or periodicity of the relevant fluctuation displacement and stress components. We take advantage of the fact that for a given field, symmetry conditions allow to compute the DTTs which themselves are analytically related to the DFT of the \textquotedblleft virtually\textquotedblright{} extended periodic field. The application of the discrete Green operator in the novel generic FFT-based solver can then be interpreted as an appropriate use of the classical periodic discrete Green operator on the later \textquotedblleft virtually\textquotedblright{} extended periodic field. Particularities introduced by non-periodic BCs in the expressions of the discrete Green operators are also discussed based on the hexahedral \citep{willot_fourier-based_2015} and double tetrahedron \citep{finel_tetrahedron-based_2025} finite difference schemes. We show that the classical periodic discrete Green operator based on a homogeneous isotropic elastic medium restricts the total number of applicable BCs. Reliable alternatives \citep{risthaus_imposing_2024,paux_discrete_2025} are discussed. Sec. \ref{sec:theo_gen_approach} details the major steps in building the generic FFT-based solver. 

The implementation is performed in a parallel framework in the open-source \emph{AMITEX$^\star$} solver (new version, in progress, of the \emph{AMITEX\_FTTP} code \citep{amitex_fftp}). The parallel implementation relies on a 2D pencil decomposition of the unit-cell that allows for distributing the simulation over a larger number of processors than a simple 1D slab decomposition, originally available in the FFTW library \citep{frigo_design_2005} (a 1D slab decomposition of a $(n^v)^3$ unit-cell is limited to $n^v$ processors, with $n^v$ the number of voxel in a given direction). This 2D decomposition is managed in \emph{AMITEX$^\star$} by the open-source library 2DECOMP\&FFT \citep{li_2decompfft_nodate,rolfo_2decompfft_2023}. Moreover, DTTs and DFT (collectively referred to as discrete transforms, DTs) are computed using the FFTW library \citep{frigo_design_2005} through the interface 2DECOMP\&FFT. A key contribution of the present work is the extension of 2DECOMP\&FFT to support DTTs and a mix with DFT, which were not available in the original library \citep{li_2decompfft_nodate,rolfo_2decompfft_2023}.

The purpose of Sec. \ref{sec:numerical_simulations} is to demonstrate the robustness and versatility of the solver using non-trivial problems, with various BCs that could not be taken into account by classical FFT-based solvers, linear and non-linear complex behaviors, small and finite transformations (including void growth and coalescence). The advantage of the 2D decomposition is also demonstrated in a crystal plasticity simulation. Finally, concluding remarks and recommendations are presented in Sec. \ref{sec:conclusion}.

\subsection*{Notation}
Unless otherwise indicated, the Einstein index convention is used. Volume mean values are denoted $\left< \star \right> $; discrete Fourier transform: $\widehat{\star} = \mathrm{DFT}(\star)$;  discrete transforms: $\widetilde{\star} = \mathrm{DT}(\star)$ accounting more generally for trigonometric and Fourier transforms; vector $\ten{a}$; second-order tensor $\td{A}$; deviatoric part of a second-order tensor $\mathrm{dev}(\td{A})$; fourth-order tensor $\mathbb{A}$; dyadic product \textquotedblleft $\otimes$\textquotedblright{}  such as $\left(\ten{a} \otimes \ten{b}\right)_{ij} = a_ib_j$; Hadamard product \textquotedblleft $\odot$\textquotedblright{} such as $\left(\td{A}\odot\td{B}\right)_{ij} = A_{ij}B_{ij}$ (no summation on the indices). $(\ten{e}_1,\ten{e}_2,\ten{e}_3)$ represents the classical Cartesian basis of the euclidean space and $(O,\ten{e}_1,\ten{e}_2,\ten{e}_3)$ denote the orthonormal coordinate system.

\section{Generic FFT-based parallel solver for non-periodic and/or periodic BCs }\label{sec:theo_gen_approach}

As aforementioned, the classical FFT-based solver is inherently restricted to the scope of periodic BCs. The objective of this section is to motivate and detail the different modifications associated with non-periodic BCs. To proceed, we hereby present the studied domain and its discretization on a regular grid. This enables us to rewrite the system of equations to be solved for the mechanical problem (based on Cauchy continuum theories). An iterative (accelerated) displacement-based fixed-point method is used. We then describe the trigonometric transforms used with the aim to preserve the structure of the accelerated FFT-based solver. Additionally, the finite difference options employed to build the discrete Green operators are outlined. This section is concluded with implementation specifications related to parallel computing and the treatment of combined periodic and non-periodic BCs.

\subsection{Domain discretization \label{sec:domaindiscretization}}

\begin{figure}[t]
	\centering{}
	\subfloat[3D parallelepiped domain $\Omega$ with face designation.]{
		\label{fig:DomainOmega}
		\includegraphics[width=0.24\textwidth]{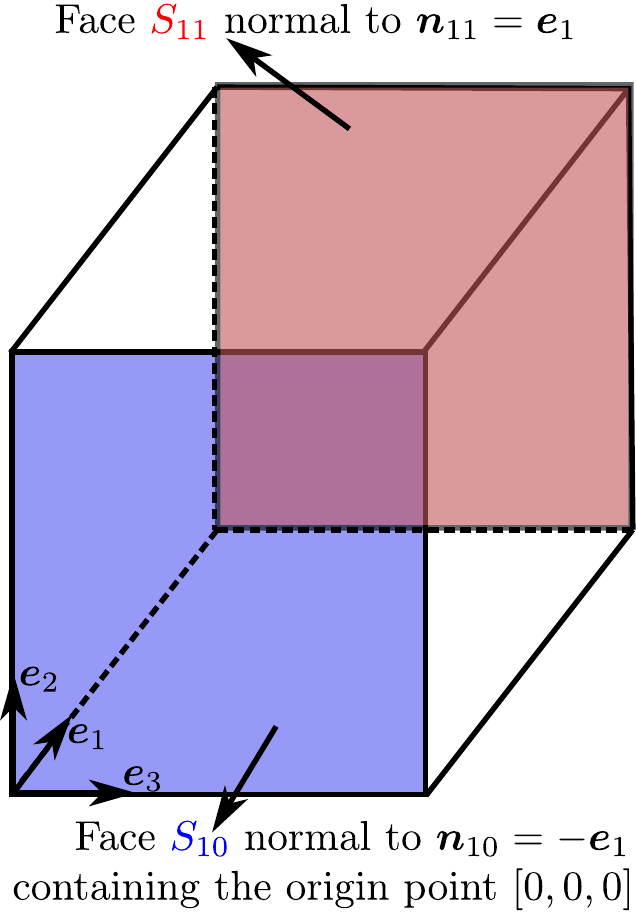}
	}$\hspace{3cm}$ \subfloat[Discretization of 2D projection of the voxels connected to the face $S_{10}$ along $\ten{e}_1$]{
		\label{fig:Domaindiscretization}
		\includegraphics[width=0.39\textwidth]{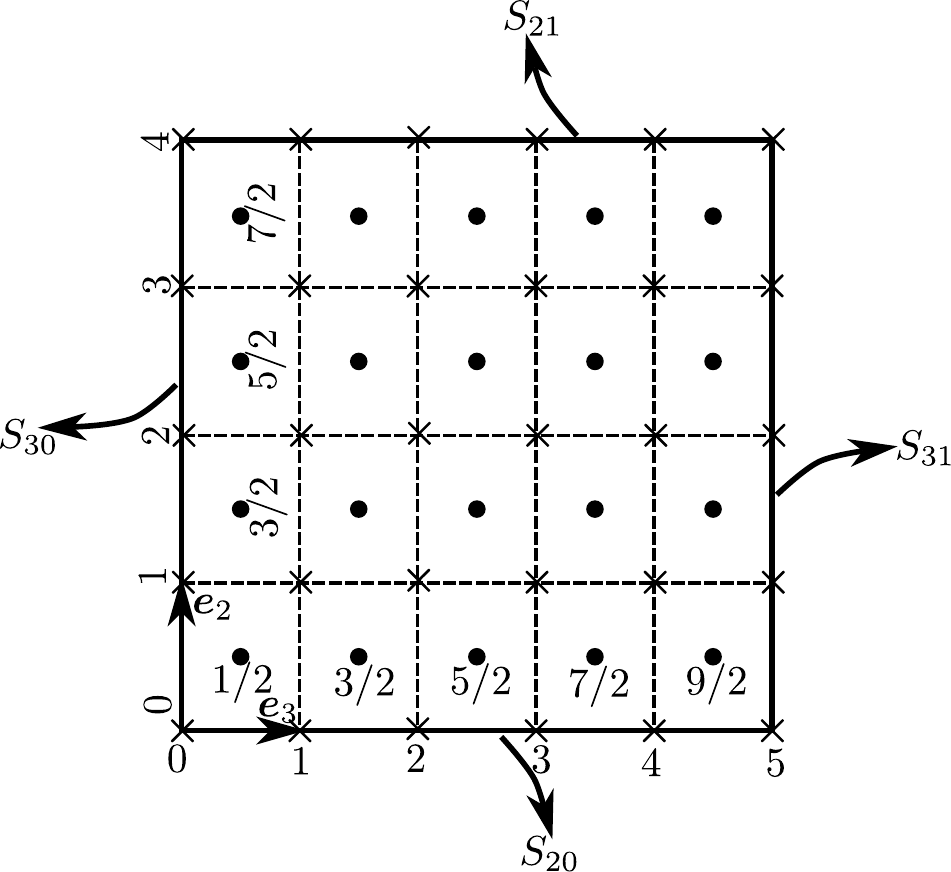}
	}
	\caption{Heterogeneous unit-cell $\Omega$ and its discretization $\Omega^d$: degrees of freedom (typically displacement vector field) are defined at the nodes (cross signs) and behavior law arguments (typically strain or stress second-order tensorial fields) are defined inside the voxel (dot signs).  (For interpretation of the references to color in all the figures, the reader is referred to the web version of this article.)} 
\end{figure}

In the context of FFT-based solver, the domain of interest on which the mechanical problem is to be solved, is considered as a parallelepiped domain $\Omega$ as shown in Fig.~\ref{fig:DomainOmega}. The domain $\Omega$ is delimited by faces denoted by $S_{i\alpha}$ with $i$ being the direction of the normal to the given face and $\alpha \in \{0,1\}$ the position of the face ($\alpha=0$ corresponds to the face that contains the origin point at the coordinates $\left[0,0,0\right]$). $\ten{n}_{i\alpha}$ is the outer normal to the face $S_{i\alpha}$.

The continuum domain $\Omega$ is discretized (\emph{i.e.,} $\Omega^d$) by elementary voxels, of sizes $h_i,\,i \in \{1,2,3\}$ (in each $\ten{e}_i$ direction, in this work, we use $h_1=h_2=h_3$). For a given direction $\ten{e}_i$, the number of considered voxels is labeled $n_i^v$ (leading to a total of $n_1^v \times n_2^v \times n_3^v$ voxels or $N_1^n\times N_2^n \times N_3^n$ nodes with $N_i^n=n_i^v+1$). Each voxel is characterized by its eight nodes with Cartesian coordinates $\left[j_1 h_1, j_2 h_2, j_3 h_3\right]$ with $j_i\in \llbracket 0,n_i^v\rrbracket,\,i \in \{1,2,3\}$. Subsequently, the voxel centers are positioned at the coordinates $\left[(j_1+ 1/2) h_1, (j_2+ 1/2) h_2, (j_3+ 1/2) h_3\right]$ with $j_i\in \llbracket 0,n_i^v-1\rrbracket,\,i \in \{1,2,3\}$.

The degrees of freedom of the mechanical problem (typically the displacement vector field) are classicaly defined on the nodes. According to the finite difference schemes used (which are presented later), the strains and stresses (second-order tensorial fields), which are inputs/outputs of a given behavior law $\mathcal{F}$, are defined at voxel centers. Fig.~\ref{fig:Domaindiscretization} illustrates an example of a $4\times5$ voxels centers (or alternatively $5\times6$ nodes) discretization. This is a 2D projection of the voxels connected to the face $S_{10}$ (\emph{i.e.,} voxels centers and nodes are not located on the same plane).

\subsection{Formulation of the mechanical problem\label{sec:mechanicalproblem}}

To define the mechanical problem, we introduce the following notations and physical fields. $\mathcal{F}$ is the user-defined behavior law, $\ten{u}$ denotes the total displacement vector field, $\ten{u^*}$ is an \emph{arbitrary} displacement vector field which values at the faces $S_{i\alpha}$ correspond to the desired applied Dirichlet BCs. The fluctuation displacement field \citep{risthaus_fftbased_2024,paux_discrete_2025} is then defined as $\ten{u^f} = \ten{u} - \ten{u^*}$. An applied stress traction vector field $\ten{t^*}$ (or $\ten{T^*}$ in finite transformation framework) which values at the faces $S_{i\alpha}$ corresponds to the desired applied Neumann BCs, $\td{\sigma}$ is the Cauchy second-order stress tensor, $\td{P}$ is the first Piola-Kirchoff second-order stress tensor, $\td{\Pi}$ corresponds to the second Piola-Kirchoff second-order stress tensor. $\td{P}$ and $\td{\Pi}$ are related to $\td{\sigma}$ through the gradient of the transformation $\td{F} = \td{1} + \td{\nabla u}$, according to the formulas
\begin{equation}
	\td{P} = \mathrm{det}(\td{F}) \, \td{\sigma}  \td{F}^{-T} \quad;\quad\td{\Pi} = \mathrm{det}(\td{F}) \, \td{F}^{-1} \td{\sigma}  \td{F}^{-T}
\end{equation}  

In the context of general BCs, it is possible to impose a given BC type (periodic, Dirichlet or Neumann) per face $S_{i\alpha},\,i \in \{1,2,3\}, \alpha \in \{0,1\}$ and per component of the displacement or traction vector field. Moreover, small and finite transformation frameworks can be distinguished when formulating the boundary value problem in mechanics. In fact, with small strain hypothesis, the system of equations is given by
\begin{equation}
	\text{Unknown} \,\,\ten{u} /
	\left\{
	\begin{array}{l}
		\ten{u} = \ten{u^f} + \ten{u^*}\\[0.5em]
		\ten{\nabla}\cdot\td{\sigma} =0 \quad \text{in} \quad \Omega \\[0.5em]
		
		\td{\sigma}  = \mathcal{F} ( \td{\nabla^s u}) \quad \text{in} \quad \Omega \\[0.5em]
		
		\text{On each face } S_{i\alpha},\,i \in \{1,2,3\}, \alpha \in \{0,1\} \text{ and for each component } q \in \{1,2,3\} \\[0.5em]
		
		\qquad \bullet \quad \text{ if } \text{ Periodic BC: }  u^f_{q} \text{ periodic and } \sigma_{iq} \text{ periodic}
		 \\[0.5em]
		\qquad \bullet \text{ or if } \text{ Dirichlet BC: } u_{q} = u_{q}^{*} \, \\[0.5em] 
		\qquad \bullet \text{ or if } \text{ Neumann BC: } \sigma_{iq} \left(-1\right)^{\alpha+1}=t^*_{q} 
	\end{array}
	\right.
\end{equation} We recall that, for a given field, periodic condition means that the field has the same value on opposite faces and anti-periodic condition expresses the fact that the considered field has opposite values on opposite faces. For this reason, the stress component $\sigma_{iq}$ being periodic is equivalent to term $\left[\td{\sigma}\,\ten{n}_{i\alpha}\right]_{q}$ being anti-periodic. Within finite transformation framework, we use a Lagrangian formulation so that the system of equation is now
\begin{equation}
	\text{Unknown} \,\,\ten{u} /
	\left\{
	\begin{array}{l}
		\ten{u} = \ten{u^f} + \ten{u^*}\\[0.5em]
		\ten{\nabla}\cdot\td{P} =0  \quad  \text{in} \quad \Omega \\[0.5em]
		
		\td{P}  = \mathcal{F} ( \td{\nabla u})  \quad \text{in} \quad \Omega \\[0.5em]
		
		\text{On each face } S_{i\alpha},\,i \in \{1,2,3\}, \alpha \in \{0,1\} \text{ and for each component } q \in \{1,2,3\} \\[0.5em]
		
		\qquad \bullet \quad \text{ if } \text{ Periodic BC: } u^f_{q} \text{ periodic and } P_{iq} \text{ periodic}   \\[0.5em]
		\qquad \bullet \text{ or if } \text{ Dirichlet BC: } u_{q} = u_{q}^{*} \, \\[0.5em] 
		\qquad \bullet \text{ or if } \text{ Neumann BC: } P_{iq} \left(-1\right)^{\alpha + 1}=T^*_{q}  
	\end{array}
	\right.
	\label{eq:TF_initial}
\end{equation}
As the systems of equations in both small and finite transformation cases are similar, we then focus on the most general not approximated formulation of finite transformation in the following analysis. These continuous systems are solved on the discretized domain $\Omega^d$, replacing continuous derivation operators by finite difference derivation operators. For the purpose, we recall that displacement fields are defined on the voxel nodes and strain/stress fields on voxel centers. 

Observing the well-known particular case of periodic BCs, the periodic extension of displacement and stress fields around the concerned faces, allows to compute the gradient of the displacement and the divergence of the stress in the vicinity of the faces. Actually, to evaluate the displacement gradient at voxel centers connected to the unit-cell faces, using a given finite difference scheme, the displacements at nodes located on the faces are defined by periodicity. Correspondingly, the computation of the stress divergence at nodes located on the faces, using a given finite difference scheme, requires the stress values at mirror voxel centers located outside the computational domain, which are obtained by periodic extension. Following the same idea, non-periodic BCs will rely on the definition of appropriate extensions. To this end, the symmetry (even-symmetry) and anti-symmetry (odd-symmetry) are used. A given field that possesses an anti-symmetric extension \emph{w.r.t.} to a face, has a null value on the concerned face. In contrast, a symmetric extension does not prescribe any specific value on the face, leaving the corresponding quantity unconstrained.

From the definition of the fluctuation displacement field, $\ten{u^{f}} = \ten{u} - \ten{u^*}$, in the case of Dirichlet BC of the component $q \in \{1,2,3\}$, on a given face $S_{i\alpha},\,i \in \{1,2,3\}, \alpha \in \{0,1\}$, the condition $u_{q} = u_{q}^{*}$ then yields to a vanishing fluctuation displacement field of the corresponding component, \emph{i.e.,} $u_q^{f}=0$ on $S_{i\alpha}$. Therefore, Dirichlet BC can be associated with an anti-symmetric extension of the fluctuation field component $u_q^{f}$ \emph{w.r.t.} to the face $S_{i\alpha}$. In addition, in order to always satisfy the corresponding equilibrium condition (\emph{i.e.,} $\left[{}_d\ten{\nabla}\cdot\td{P}\right]_q =0$) at the face $S_{i\alpha}$, symmetric extension of the component $P_{iq}$ and anti-symmetric extension of the other stress components $P_{il}, l \neq q$, are used. Note that for the component $P_{iq}$, the symmetric extension let the quantity free (unconstrained) at the corresponding face.

	\begin{figure}[t]
		\centering{}\includegraphics[width=0.7\textwidth]{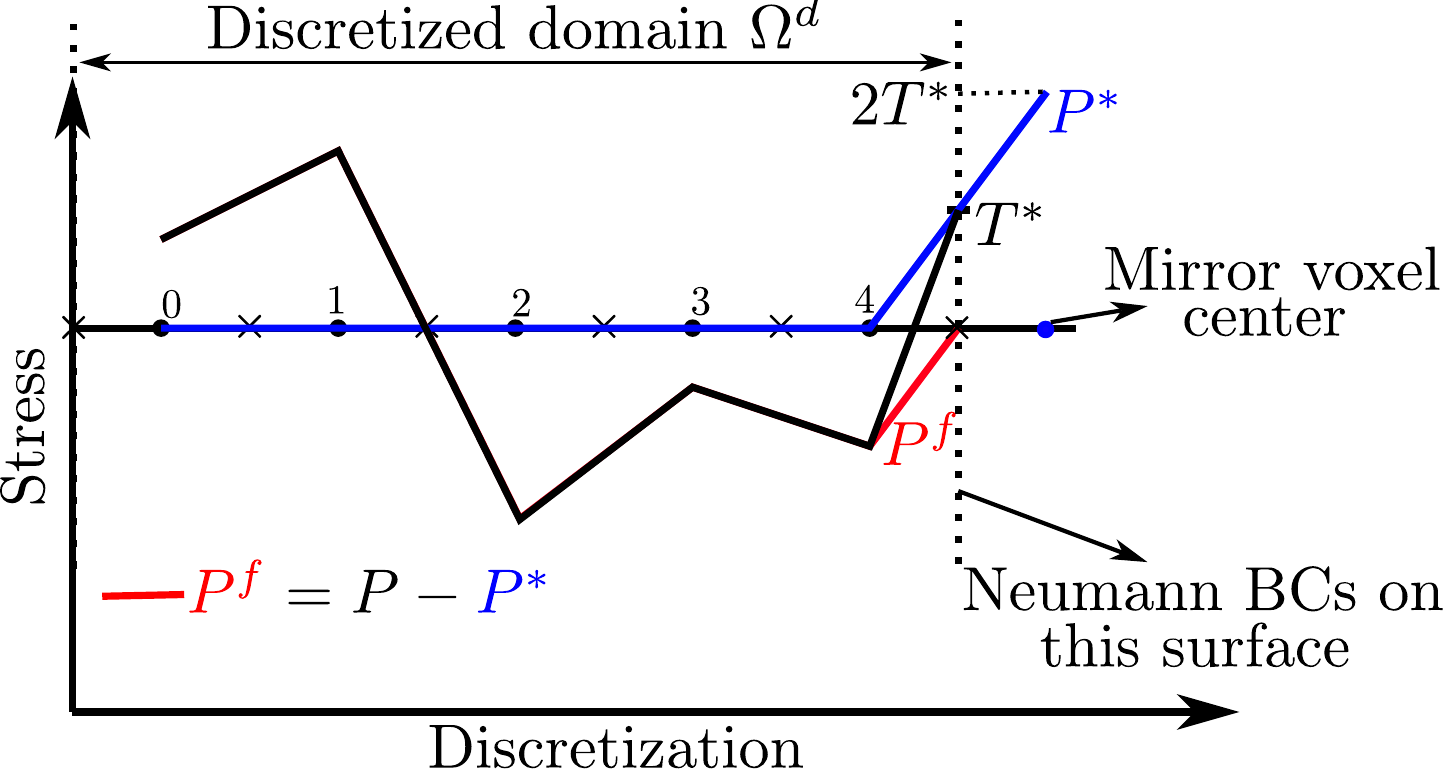}
		\caption{1D illustration of the procedure to apply a general non-vanishing Neumann BC on a given face (right one in the figure): nodes of the discretization are represented by cross signs and the voxel centers by dot signs. Stress fields ($P$ in black, $P^f$ in red and $P^*$ in blue) are defined at voxel centers.}\label{fig:NeumannBC}
	\end{figure}
	
	In case of Neumann BC of the component $q \in \{1,2,3\}$, on a given face $S_{i\alpha},\,i \in \{1,2,3\}, \alpha \in \{0,1\}$, as the stress fields are classically defined on the voxel centers, the Neumann BC, $P_{iq}\left(-1\right)^{\alpha+1}=T^*_{q}$, is not  automatically enforced at the face $ S_{i\alpha}$ (voxels faces). In order to illustrate the analysis on this case and without any loss of generality, we consider an 1D configuration as described in Fig.~\ref{fig:NeumannBC}. The desired stress field $P$ in the domain $\Omega^d$ (represented in solid black line in Fig.~\ref{fig:NeumannBC}) possesses a Neumann BC, \emph{i.e.,} $P=T^*$, only on the right face (for simplicity purpose). Similarly to the Dirichlet BCs, an applied stress field $P^*$ is introduced such as $P^*$ is null at all voxel centers inside the discretized domain $\Omega^d$ and non-null with a value $P^*=2T^*$ at the \textquotedblleft virtual\textquotedblright{} mirror voxel center \emph{w.r.t.} to the faces $ S_{i\alpha}$ (blue line in Fig.~\ref{fig:NeumannBC}). With this choice, the value of $P^*$ interpolated at the face $ S_{i\alpha}$ is equal to $T^*$. A fluctuation stress field $P^f$ can be defined as $P^f = P-P^*$. By construction, the fluctuation stress field strictly coincides with the desired field $P$ inside the unit-cell $\Omega$ and vanishes at the face $S_{i\alpha}$. This is in accordance with the 3D case for the hexahedral and double tetrahedron finite difference schemes that will be discussed in this work. One can conclude that a Neumann BC of a component $q \in \{1,2,3\}$ \emph{w.r.t.} to a face $S_{i\alpha},\,i \in \{1,2,3\}, \alpha \in \{0,1\}$ is characterized by an anti-symmetric extension of the corresponding fluctuation stress component $P^f_{iq}$. The remaining components $P^f_{il}, l\neq q$ are considered to have no particular fixed values (even-symmetry extension \emph{w.r.t.} to the face $S_{i\alpha}$). The displacement component $u_q=u^f_q$ to be determined as consequence of Neumann BC is then also assumed to be symmetric \emph{w.r.t.} to the face $S_{i\alpha}$.

Given the established connection between the non-periodic BCs and the symmetries of the fields, the discrete form (depending on the discretization and a finite difference form) of the boundary value problem in Eq.~\eqref{eq:TF_initial}, is expressed as follows
\begin{equation}
	\text{Unknown} \,\,\ten{u^f} /
	\left\{
	\begin{array}{l}
		\ten{u} = \ten{u^f} + \ten{u^*}\\[0.5em]
		\td{P} = \td{P^f} + \td{P^*} \\[0.5em]
		{}_d\ten{\nabla}\cdot\td{P} =0  \quad  \text{in} \quad \Omega^d \\[0.5em]
		
		\td{P}  = \mathcal{F} ( {}_d\td{\nabla u})  \quad \text{in} \quad \Omega^d \\[0.5em]
		\text{On each face } S_{i\alpha},\,i \in \{1,2,3\}, \alpha \in \{0,1\} \text{ and for each component } q \in \{1,2,3\} \\[0.5em]
		
		\quad \bullet \quad \text{ if } \text{ Periodic BC: } u^f_{q} \text{ periodic and } P_{iq} \text{ periodic} \\[1em]
		\quad \bullet \text{ or if } \text{ Dirichlet BC: } u_{q}^{f} \text{ anti-symmetric, } P_{iq}  \text{ symmetric and } P_{il} \text{ with } l\neq q  \text{ anti-symmetric }
		  \\[2em] 
		\quad \bullet \text{ or if } \text{ Neumann BC: } u_{q}^{f} \text{ symmetric, } P^f_{iq} \text{ anti-symmetric and } P^f_{il} \text{ with } l\neq q \text{ symmetric }
	\end{array}
	\right.
	\label{eq:GeneralBCtobesolved}
\end{equation} In practice, the components of $\td{P}$ (or, equivalently, $\td{P}^f$) exhibit the same symmetries and/or periodicity as the components of $\td{\nabla u^f}$. This prompts an implementation by specifying only the spatial extensions of the fluctuation displacement vector $\ten{u^f}$ for all general BCs (periodic, Dirichlet or Neumann). Subsequently, based on the extensions of the fluctuation displacement fields in the cases of Dirichlet and Neumann BCs, the standard FFT-based iterative solver \citep{Moulinec1998,willot_fourier-based_2015} can not be directly applied. Adaptations are required for a generic FFT-based solver that can be used for all general BCs. 

\subsection{Overview of the displacement-based iterative fixed-point approach \label{sec:iterativeapproach}}

In order to clarify the novel features of the FFT-based solver in the context of the non-periodic BCs, we first review the general displacement-based iterative fixed-point approach used in this work. 

Knowing that FFT-based solvers are classically used in the context of periodic BCs, the mechanical problem to solve is reduced to the following one 
\begin{equation}
	\left\{
	\begin{array}{l}
		\ten{u} = \ten{u^f} + \ten{u^*}\\[0.5em]
		{}_d\ten{\nabla}\cdot\td{P} =0  \quad  \text{in} \quad \Omega^d \\[0.5em]
		
		\td{P}  = \mathcal{F} ( {}_d\td{\nabla u})  \quad \text{in} \quad \Omega^d \\[0.5em]
		
		\text{On all faces } S_{i\alpha},\,i \in \{1,2,3\}, \alpha \in \{0,1\} \text{ and for all components } q \in \{1,2,3\} \\[0.5em]
		
		\qquad \bullet  \text{ Periodic BC: } u^f_{q} \text{ periodic and } P_{iq} \text{ periodic} 
	\end{array}
	\right. 
	\label{eq:fullperiodic}
\end{equation}
\emph{A priori} pre-conditioner $\mathbb{R}_0:\td{\nabla u^f}$ is introduced so that the mechanical problem in Eq.~\eqref{eq:fullperiodic} is rewritten as 
\begin{equation}
	\left\{
	\begin{array}{l}
		\ten{u} = \ten{u^f} + \ten{u^*}\\[0.5em]
		{}_d\ten{\nabla}\cdot \left(\mathbb{R}_0:{}_d\td{\nabla u^f}\right) =  {}_d\ten{\nabla}\cdot(\mathbb{R}_0:{}_d\td{\nabla u^f} - \td{P}) \coloneqq \ten{p} \quad  \text{in} \quad \Omega^d \\[0.5em]
		
		\td{P}  = \mathcal{F} ( {}_d\td{\nabla u})  \quad \text{in} \quad \Omega^d \\[0.5em]
		
		\text{On all faces } S_{i\alpha},\,i \in \{1,2,3\}, \alpha \in \{0,1\} \text{ and for all components } q \in \{1,2,3\} \\[0.5em]
		
		\qquad \bullet  \text{ Periodic BC: } u^f_{q} \text{ periodic and } P_{iq} \text{ periodic} 
	\end{array}
	\right. 
\end{equation}
$\td{\tau} = \mathbb{R}_0:{}_d\td{\nabla u^f} - \td{P}$ is the so-called polarization stress \citep{Moulinec1998}. The principle of the fixed-point algorithm is to predict $\left[\ten{u^f}\right]^{m+1}$ the fluctuation displacement field at an iteration $m+1$ given $\left[\ten{u^f}\right]^{m}$ at the iteration $m$. First, we express, at the iteration $m$, the divergence of the polarization stress $\ten{p}^{m}$
\begin{equation}
	\ten{p}^{m} = {}_d\ten{\nabla}\cdot\left(\mathbb{R}_0:\left[{}_d\td{\nabla u^f}\right]^{m} - \left[\mathcal{F} ( {}_d\td{\nabla u})\right]^{m}\right)
	\label{eq:div_pol}
\end{equation} 
which is used in a second step to deduce $\left[\ten{u^f}\right]^{m+1}$ for the iteration $m+1$, based on the expression
\begin{equation}
	{}_d\ten{\nabla}\cdot \left(\mathbb{R}_0:\left[{}_d\td{\nabla u^f}\right]^{m+1}\right) = \ten{p}^{m}
	\label{eq:div_pol_next_iter}
\end{equation}
The relationship in Eq.~\eqref{eq:div_pol} is evaluated in real space and the last operation in Eq.~\eqref{eq:div_pol_next_iter} can be evaluated in Fourier space using DFT, resulting in the following expression
\begin{equation}
	\widehat{\left[\ten{u^f}\right]^{m+1}} = {}_d\widehat{\mathrm{DGO}}(\widehat{\ten{p}^{m}})
	\label{eq:GreenOp}
\end{equation} defining the Fourier discrete Green operator ${}_d\widehat{\mathrm{DGO}}$ (core of the FFT-based solver), which depends \emph{a priori} on the finite difference scheme and where $\widehat{G}[k] = \displaystyle \sum_{j=0}^{n_{points}-1} g[j] \, \exp\left(-\mathrm{i} \,\frac{2 \pi k}{n_{points}} j\right) $ for a field $G$ defined on $n_{points}$ points (for simplicity, a 1D example is used knowing that a 3D transform is a succession of 1D transform). Finally, $\left[\ten{u^f}\right]^{m+1}$ is obtained by an inverse DFT calculation. $\left[\ten{u^f}\right]^{m+1}$ is then reintroduced in Eq.~\eqref{eq:div_pol} for a new iteration of the fixed-point algorithm up to the equilibrium convergence criterion, expressed as follows (for cuboid voxels, see \cite{nkoumbou_kaptchouang_multiscale_2022} for general non-cubic voxels)
\begin{equation}
\sqrt{\frac{\|{}_d\ten{\nabla}\cdot\td{P}\|_{L^2}}{\|\td{P}\|_{L^2}  } } \,h_1 < \epsilon
\label{eq:equilibrium_condition}
\end{equation} with $\epsilon$ the tolerance. It is well established that this fixed-point approach is efficient in solving the problem in Eq.~\eqref{eq:fullperiodic}. In order to preserve this efficiency in solving the general BCs as described in Eq.~\eqref{eq:GeneralBCtobesolved}, it necessitates to be able to compute DFT and the discrete Green operator in general cases (possibility of mixing Dirichlet, Neumann and/or periodic BCs). The subsequent sections address these points.

In addition, it is important to note that the basic iterative algorithm can be accelerated by different approaches \citep[\emph{e.g.,} Barzilai-Borwein method, conjugate gradient (linear or non-linear),][]{schneider_review_2021,gierden_review_2022}. In this work, an Anderson convergence acceleration technique is used \citep{anderson_iterative_1965,ramiere_iterative_2015}. The idea is directly taken from the acceleration technique used in the FE code CAST3M \citep{castem} to accelerate a modified Newton-Raphson algorithm. Indeed, every 2 or 3 iterations, a displacement field $\left[\ten{u^f}\right]^{m+1}$ for the iteration $m+1$ is proposed based on the obtained fields, $\left[\ten{u^f}\right]^{j}$, and residual fields, $\left[\ten{u^f}\right]^{j} - \left[\ten{u^f}\right]^{j-1}$, at the previous $4$ couples $j$-iterations. This technique described in \cite{chen_analysis_2019}, has been shown to significantly enhance the convergence rate of the basic scheme  in classical FFT-based solver.

\subsection{Generic FFT-based solver features: linking DTTs and DFT \label{sec:spectralcalculations}}

In this section, we build on the link established in the previous Sec. \ref{sec:mechanicalproblem} between the general BCs and the symmetry and/or  periodicity extensions of the desired fields to develop the generic FFT-based solver. This approach aims to facilitate the preservation of the classical FFT-based iterative fixed-point algorithm, with certain modifications, in order to address the problem formulated in Eq.~\eqref{eq:GeneralBCtobesolved} (possibility to combine periodic BCs with non-periodic BCs). Depending on the BCs on each face, symmetry (even-symmetry), anti-symmetry (odd-symmetry) or periodicity extensions are proposed for each component of the fluctuation displacement and stress fields. According to these extensions, the fluctuation displacement and stress fields can be \textquotedblleft virtually\textquotedblright{} extended to obtain periodic fields from which the discrete periodic Green operator can be build (see Eq.~\eqref{eq:GreenOp}). This approach requires to express the relation between the DTs of the original fields and the DFT of the \textquotedblleft virtually\textquotedblright{} extended fields. For the sake of clarity, without any loss of generality, we can restrict to a 1D scalar field defined at the nodes (we recall that to the displacement field $u^f$ is defined at the nodes).

\begin{figure}[t]
	\centering{}
	\subfloat[1D scalar field $G$ (in plain red) with anti-symmetric ($\equiv $ A) on the left and anti-symmetric ($\equiv $ A) on the right.]{
		\includegraphics[width=0.6\textwidth]{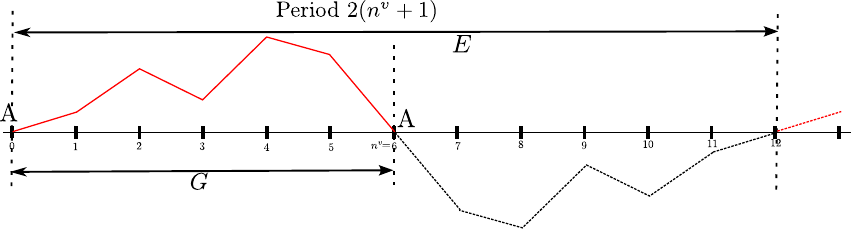}
	}
	\hfill{} \centering{}\subfloat[1D scalar field $G$ (in plain red) with anti-symmetric ($\equiv $ A) on the left and symmetric ($\equiv $ S) on the right.]{
		\includegraphics[width=1.05\textwidth]{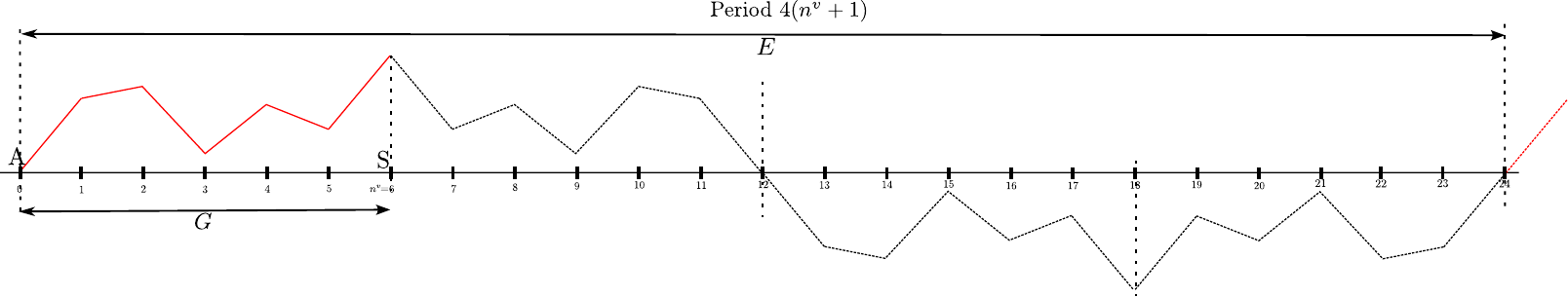}
	}
	\caption{Example of construction of 1D periodic fields $E$ from 1D fields $G$ defined at nodes and using symmetry extensions on the left and right extremities. Original fields $G$ are represented in solid red lines and their extensions are plotted in dot lines. }
	\label{fig:Symmetries}
\end{figure}

\begin{table}[h!]
	\centering{}
	\caption{Relationships between BCs, DTs of the original field $G$ and DFT of the \textquotedblleft virtually\textquotedblright{} extended field $E$: 1D case with $G$ defined on $N^n=n^v+1$ nodes. Notation: Per. $\equiv$ Periodic, Neu. $\equiv$ Neumann, Dir. $\equiv$ Dirichlet, A $\equiv$ anti-symmetric, S $\equiv$ symmetric, P $\equiv$ Periodic. DCT and DST notations correspond to the real DTs in FFTW library \citep{frigo_design_2005}.}
	\begin{tabular}{|c|c|c|}
		\hline
		\vspace{-0.35cm}   &  & \\ 
		BCs and DTs  & $\mathrm{DT}(G)=\widetilde{G}$ & $\mathrm{DFT}(E)=\widehat{E}$ \\
		\hline
		Per./Per. (P/P)  &  $\widetilde{G}\left[k\right] = \displaystyle \sum_{j=0}^{N^n-2}g[j] \exp\left(-\mathrm{i} \,\xi^k j\right)$ & $\widehat{E}\left[k\right] = \alpha^{\mathrm{BC}}\,\widetilde{G}\left[k\right]$ ; $\forall k\in \llbracket 0,N^n-2\rrbracket$ \\[0.5em]
		DFT &  with $\xi^k=\displaystyle \frac{2 \pi k}{N^n-1}~;~ \forall k\in \llbracket 0,N^n-2\rrbracket$ & $\alpha^{\mathrm{BC}}=1$ with $S_f=1$ \\[0.5em]
		\hline
		Neu./Neu. (S/S) &  $\widetilde{G}\left[k\right]= \displaystyle \sum_{j=0}^{N^n-1}\left(2-\delta_{j,0}-\delta_{j,N^n-1}\right)g[j]\cos\left(\displaystyle \xi^kj\right)$ &  $ \widehat{E}\left[k\right] = \alpha^{\mathrm{BC}}\,\widetilde{G}[k] $ ; $\forall k\in \llbracket 0,N^n-1\rrbracket$\\[0.2em]
		DCT-I  &  with $\xi^k=\displaystyle \frac{\pi k}{N^n-1}~;~ \forall k\in \llbracket 0,N^n-1\rrbracket$ & $\alpha^{\mathrm{BC}}=1$ with $S_f=2$\\[0.5em]
		
		\hline
		Neu./Dir. (S/A) &  $ \widetilde{G}\left[k\right]=\displaystyle   \displaystyle \sum_{j=0}^{N^n-2} \left(2-\delta_{j,0}\right)g[j] \cos\left(\xi^kj\right)$ &  $\widehat{E}\left[2k+1\right] =\alpha^{\mathrm{BC}}\,\widetilde{G}[k]$ ; $\forall k\in \llbracket 0,N^n-2\rrbracket$\\[0.2em]
		DCT-III  &  with $\xi^k = \displaystyle \frac{\pi (k+1/2) }{N^n-1}$ ;  $ \forall k\in \llbracket 0,N^n-2\rrbracket$ & $\alpha^{\mathrm{BC}}=2$ with $S_f=4$\\[0.2em]
		
		\hline
		Dir./Dir. (A/A) &  $ \widetilde{G}\left[k\right]= \displaystyle 2\sum_{j=1}^{N^n-2}g[j]\sin\left( \displaystyle \xi^k j\right)$ &  $\widehat{E}\left[k+1\right] =\alpha^{\mathrm{BC}}\,\widetilde{G}[k]$ ; $\forall k\in \llbracket 0,N^n-3\rrbracket$ \\[0.2em]
		DST-I  &  with  $\xi^k = \displaystyle \frac{\pi (k+1)}{N^n-1}$ ; $ \forall k\in \llbracket 0,N^n-3\rrbracket$ & $\alpha^{\mathrm{BC}}=-\mathrm{i}$ with $S_f=2$\\[0.5em]
		
		\hline
		Dir./Neu. (A/S) &  $\widetilde{G}\left[k\right]= \displaystyle \sum_{j=1}^{N^n-1} \left(2-\delta_{j,N^n-1}\right)g[j]\sin\left( \displaystyle \xi^k j\right)  $ &  $\widehat{E}\left[2k+1\right] =\alpha^{\mathrm{BC}}\,\widetilde{G}[k]$ ; $\forall k\in \llbracket 0,N^n-2\rrbracket$\\[0.2em]
		DST-III  &   with  $\xi^k = \displaystyle \frac{\pi (k+1/2) }{N^n-1}$ ; $ \forall k\in \llbracket 0,N^n-2\rrbracket$ & $\alpha^{\mathrm{BC}}=-2\mathrm{i}$ with $S_f=4$\\[0.2em]
		\hline
	\end{tabular}
	\label{Tab:BCsSymmetriesDTTDFT}
\end{table}

We consider a discretized 1D scalar field $G$ with values $g[j],\,j\in\llbracket 0,n^v \rrbracket$ at the nodes ($N^n=n^v+1$ nodes in total) which possesses periodicity or symmetry extensions on respectively the left ($j=0$) and right ($j=n^v$) sides. In the case of periodic condition, the \textquotedblleft virtually\textquotedblright{} extended field $E$ is then considered to be strictly equal to the primary field $G$. In the non-periodic case, four possibilities can be distinguished for the node-defined field. For a field $G$ with symmetry extensions, the original field can be \textquotedblleft virtually\textquotedblright{} extended either twice or four times (symmetry factor $S_f=2$ or $S_f=4$) in order to obtain a \textquotedblleft virtual\textquotedblright{} extended periodic field $E$. Fig.~\ref{fig:Symmetries} illustrates this concept through two scenarios: a field $G$ with anti-symmetric/anti-symmetric extensions (\emph{i.e.,} Dirichlet/Dirichlet BCs) on both sides and another field $G$ with anti-symmetric/symmetric extensions (\emph{i.e.,} Dirichlet/Neumann BCs) on the left and right sides. On the extended periodic field $E$, a computation of  a DFT, $\widehat{E}$, can be performed. It can be demonstrated that there is a connection between the DFT of $E$, $\mathrm{DFT}(E)[k]=\widehat{E}[k] = \displaystyle \sum_{j=0}^{S_f(N^n-1)-1}e[j] \exp\left(-\mathrm{i} \dfrac{2\pi k}{S_f(N^n-1)} j\right) $, and the DTs of $G$, $\mathrm{DT}(G)=\widetilde{G}$. Calculation of the DFT and the DTs is provided in Tab. \ref{Tab:BCsSymmetriesDTTDFT} and specially, their links are summarized in the third column of Tab. \ref{Tab:BCsSymmetriesDTTDFT} depending on the BCs. It is shown that for a field, with a given BC, once the DT of the original field is computed on the desired unit-cell (see index $j$ in the summation sign in Tab. \ref{Tab:BCsSymmetriesDTTDFT} restricted to the domain of interest), the DFT of the extended field is obtained using the constant value (real or purely imaginary), denoted $\alpha^{\mathrm{BC}}$ in Tab. \ref{Tab:BCsSymmetriesDTTDFT}. Furthermore, it is worth noting that the value of fundamental frequencies, denoted by $\xi^k$ with $k$ being the index of selected frequencies and which are used in the calculations of the DTs, $\widetilde{G}$, correspond to appropriately selected frequencies for the DFT, $\widehat{E}$ (see the third column of Tab. \ref{Tab:BCsSymmetriesDTTDFT}). The non-null stored values in Fourier space depend on the BCs. For the generic \emph{AMITEX$^\star$} solver, DTT calculations are performed in the FFTW library \citep{frigo_design_2005} (which also manage the DFT operations), relying on new upgrades of the open-source library 2DECOMP\&FFT \citep{li_2decompfft_nodate,rolfo_2decompfft_2023} as an interface.  

\begin{algorithm}[h!]
	\caption{Overview of the generic FFT-based implementation for solving boundary value problems in \emph{AMITEX$^\star$}}
	\label{alg:algFFT}
	\begin{algorithmic}[1]
		
		\State \textbf{Input:} select BCs (spatial extensions of $\ten{u^f}$), select finite difference scheme, tolerance $\epsilon $, maximum iteration per loading increment $I_{\mathrm{max}}$, total loading increment $N_{\text{step}}$

		\For{loading step $= 1$ to $N_{\text{step}}$}
		\State update the desired loading $\ten{u^*}$,  $\ten{T^*}$
		\State compute ${}_d\td{\nabla u^*}$ and $\ten{D^*} = {}_d\ten{\nabla}\cdot\td{\Sigma^{*}}$
		\State guess $\ten{u^f}$ value based on previous values; iteration $m=0$
		\While{ $m<I_{\mathrm{max}}$ }
		\State compute deformations ${}_d\td{\nabla u^f}$ and ${}_d\td{\nabla u} = {}_d\td{\nabla u^f} + {}_d\td{\nabla u^*}$
		\State compute the voxel-defined stresses based on the behavior law  $\td{\Pi}$, $\td{\sigma}$, $\td{P}$
		\State compute node-defined $\ten{D} = {}_d\ten{\nabla}\cdot\td{P} + \ten{D^*}$

		\State compute $error = \displaystyle \sqrt{\frac{\|\ten{D}\|_{L^2}}{\|\td{P}\|_{L^2}  } } h_1 $  (cuboid voxel)
		\If{$error < \epsilon$}
		\State convergence criterion satisfied at the loading step
		\State break the while loop
		\Else
		\State continue
		\EndIf
		\State compute the divergence of the polarization $\ten{p} = {}_d\ten{\nabla}\cdot(\mathbb{R}_0:{}_d\td{\nabla u^f}) - \ten{D}$
		\State Anderson convergence acceleration technique on the field $\ten{p}$
		
		\State compute the DTs of $\ten{p}$ (requires the spatial extensions of ${}_d\ten{\nabla}\cdot(\mathbb{R}_0:{}_d\td{\nabla u^f})$)
		
		\State apply the discrete Green operator on $4$ times \textquotedblleft virtually\textquotedblright{} extended field $\ten{p}$
		\State update the fluctuation displacement

		\State update increment $m \gets m + 1$
		\EndWhile

		\EndFor
		
	\end{algorithmic}
\end{algorithm}

To sum up, even in the presence of non-periodic BCs, the FFT-based solver (mainly the application of the discrete Green operator, see Eq.~\eqref{eq:GreenOp}) can be used by capitalizing on the relationships between DTs and DFT. In this sense, in order to harmonize the treatment of all possible BCs, we decide to use a $4$ times extended \textquotedblleft virtual\textquotedblright{} domain for fields that possess any type of symmetry or periodicity extensions (\emph{i.e.,} $S_f=4$ for all cases). 

Alg. \ref{alg:algFFT} summarizes the different steps of the resolution with the generic FFT-based solver. Moreover, it can be seen in Alg.~\ref{alg:algFFT} that Anderson acceleration is applied to $\ten{p}$, the divergence of the polarization field. To the best of the authors' knowledge, this constitutes a novel departure from the conventional approach \citep{chen_analysis_2019}, in which Anderson acceleration is applied directly to the fluctuation displacement field, $\ten{u^f}$. Our numerical simulations indicate that the $\ten{p}$-based formulation often yields better convergence performance than the conventional $\ten{u^f}$-based approach.

Going forward, it is still crucial to elucidate the symmetry and/or periodicity extensions of the divergence of the polarization stress, $\ten{p}$, as well as the methodology for deriving the discrete Green operator based on the finite difference scheme and the BCs. To do so, the derivation operations present in the model, here the gradient, the divergence and the Laplacian (combination of the divergence and the gradient), are first expressed in the next section.

\subsection{Finite difference schemes \label{sec:DerivationsOp}}

\begin{figure}[t]
	\begin{center}
		\subfloat[Hexahedral scheme (HEX1): illustration of the derivation of a 3D node-defined field $G$.]{
			\label{fig:DerSchemeHEX1}
			\includegraphics[width=0.45\textwidth]{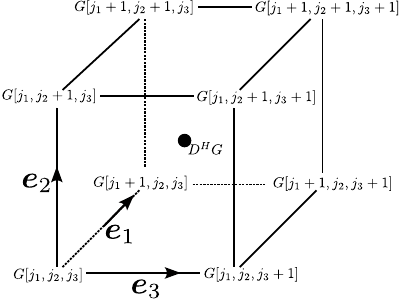}}
	\end{center}
	\centering\subfloat[Double tetrahedron (TETRA2) scheme: illustration of the derivation of a 3D node-defined field $G$. Two derivatives coexist at the same voxel center, first one involving the regular tetrahedron T1 in red and second one the regular tetrahedron T2 in blue.]{
		\label{fig:DerSchemeTETRA2}
		\begin{centering}
			\includegraphics[width=0.45\textwidth]{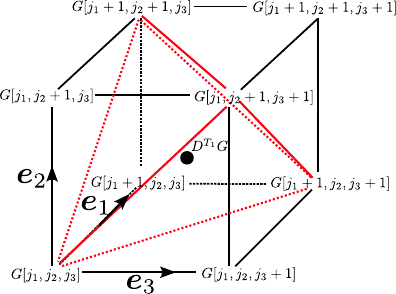}
			$\qquad$
			\includegraphics[width=0.45\textwidth]{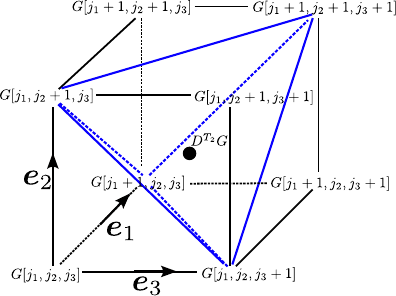}
			\par\end{centering}}

	\caption{Finite difference schemes.}\label{fig:DerScheme}
\end{figure}

In the mechanical problem expressed in Eq.~\eqref{eq:GeneralBCtobesolved}, discrete gradient operations are imperative to quantify strains at the voxel centers and divergence operations to test the equilibrium equation at the nodes. The present work uses finite difference schemes to evaluate the discrete derivation operations. Due to their demonstrated advantages in the context of periodic BCs, the hexahedral scheme \citep[HEX1, ][]{willot_fourier-based_2015,schneider_fftbased_2017} and the recently proposed double tetrahedron scheme \citep[TETRA2, ][]{finel_tetrahedron-based_2025,gelebart_accurate_2025}, are investigated in the present paper. The flexibility of the present implementation in the \emph{AMITEX$^\star$} solver allows for other types of finite difference schemes or finite element types.

To ensure the completeness of the present work, the fundamental concepts of the schemes are reviewed here, prior to the analysis of the modifications introduced by non-periodic BCs. Fig.~\ref{fig:DerScheme} illustrates the two finite difference schemes. The hexahedral scheme (see Fig.~\ref{fig:DerSchemeHEX1}) can be viewed as a linear hexahedral FE with a single \textquotedblleft integration point\textquotedblright{} \citep{schneider_fftbased_2017}, and the double tetrahedron scheme (see Fig.~\ref{fig:DerSchemeTETRA2}) can also be regarded, from an implementation point of view, as a linear hexahedral FE but with two \textquotedblleft integration points\textquotedblright{} \citep{gehrig_elementbased_2025}. This indicates that, for both schemes, the derivation of a node-defined field (\emph{e.g.,} a displacement field) results in a field defined at the voxel center (\emph{e.g.,} a strain field), and vice versa. The hexahedral (HEX1) scheme uses all eight points (voxel nodes or centers) altogether in a single derivation operation. Meanwhile, for the double tetrahedron (TETRA2) case, if the field is initially defined at the voxel nodes, then two derivatives are obtained, each of them being computed with one of the two regular tetrahedron defined in the voxel (four nodes are collectively involved at once). When the field is defined at centers (with two values per voxel, one per \textquotedblleft integration point\textquotedblright{}), the resulting derivative is defined at nodes (with a single value per node). 

The accelerated fixed-point FFT-based solver, necessitates the computation of the gradient, divergence operators in real and Fourier spaces to deduce the discrete Green operator. For the purpose, the following two sections detail the derivation operations in real and Fourier spaces for both finite difference schemes. A cross validation of the derivatives in real and Fourier spaces is performed, for any combination of BCs in the Sec. \ref{sec:discrete_derivation}. Unless otherwise stated and without any loss of generality, we considered a 3D scalar field $G$ defined at the voxel nodes (identified by $j_i \in \llbracket 0, n_i^v \rrbracket,\,i \in \{1,2,3\}$) and with general BCs in a given direction \emph{w.r.t.} to a face. As stated before, the first derivative is obtained at the voxel centers (voxel-defined field). 

\subsubsection{Hexahedral scheme}

In this section, the HEX1 finite difference scheme \citep[see previously described Fig.~\ref{fig:DerSchemeHEX1},][]{willot_fourier-based_2015} is introduced in real and Fourier spaces (depending on the BCs). 
\begin{itemize}
	\item \textbf{Derivation in real space}: 
\end{itemize}

For the 3D node-defined scalar field $G$, we can write down the component of the derivative in each direction at the voxel centers (identified by $j_i \text{ with } j_i + 1/2 \in \llbracket 0, n_i^v-1 \rrbracket,\,i \in \{1,2,3\}$)
	\begin{align}
		\centering
		D_1^HG\left[j_1+\frac{1}{2},j_2+\frac{1}{2},j_3+\frac{1}{2}\right]
		= \displaystyle \frac{1}{4h_1} \Bigl[  
		  & G[j_1+1,j_2,j_3]  -  G[j_1,j_2,j_3] \nonumber \\ 
		+ & G[j_1+1,j_2+1,j_3]  -   G[j_1,j_2+1,j_3] \nonumber \\ 
		+ & G[j_1+1,j_2,j_3+1]  -   G[j_1,j_2,j_3+1] \nonumber \\ 
		+ & G[j_1+1,j_2+1,j_3+1]  -  G[j_1,j_2+1,j_3+1]
		\Bigr] \label{eq:D1HG}
	\end{align}
	\begin{align}
		D_2^HG\left[j_1+\frac{1}{2},j_2+\frac{1}{2},j_3+\frac{1}{2}\right]
		= \displaystyle \frac{1}{4h_2} \Bigl[  
		  & G[j_1,j_2+1,j_3] - G[j_1,j_2,j_3] \nonumber \\ 
		+ & G[j_1+1,j_2+1,j_3] - G[j_1+1,j_2,j_3] \nonumber \\ 
		+ & G[j_1,j_2+1,j_3+1] - G[j_1,j_2,j_3+1] \nonumber \\
		+ & G[j_1+1,j_2+1,j_3+1] - G[j_1+1,j_2,j_3+1]
		\Bigr]	\label{eq:D2HG}
	\end{align}
	\begin{align}
		D_3^HG\left[j_1+\frac{1}{2},j_2+\frac{1}{2},j_3+\frac{1}{2}\right]
		= \displaystyle \frac{1}{4h_3} \Bigl[  
		  & G[j_1,j_2,j_3+1] - G[j_1,j_2,j_3] \nonumber \\ 
		+ & G[j_1+1,j_2,j_3+1] - G[j_1+1,j_2,j_3] \nonumber\\ 
		+ & G[j_1,j_2+1,j_3+1] - G[j_1,j_2+1,j_3] \nonumber \\ 
		+ & G[j_1+1,j_2+1,j_3+1] - G[j_1+1,j_2+1,j_3]
		\Bigr] \label{eq:D3HG}
	\end{align}
	\begin{remark} \label{rem:Derive_from_voxel_center}
		For a field $G$ defined at the voxel centers, the previous Eqs.~\eqref{eq:D1HG}-\eqref{eq:D3HG} are still valid at the condition that the voxel node indicators $j_i$ are changed into the voxel center indicators $j_i+1/2$ with $j_i \in \llbracket 0, n_i^v-1 \rrbracket,\,i \in \{1,2,3\}$ to obtained their derivatives at the nodes. For the nodes positioned at the faces, the BCs of the field $G$ are used.
	\end{remark}
	
	For the mechanical problem in this work, it is interesting to express the gradient and Laplacian (gradient followed by divergence) of a node-defined displacement vector field $\ten{v}$ (with scalar components $v_j,\, j  \in \{1,2,3\}$) and the divergence of a voxel-defined second-order stress/strain tensor $\td{S}$ (with scalar components $S_{ij},\, i,j  \in \{1,2,3\}$). In this sense, we have
	\begin{equation}
		\td{{}_{H}\nabla v} = D_i^Hv_j \,\ten{e}_i \otimes \ten{e}_j 
		\quad; \quad \ten{{}_{H}\nabla}\cdot \td{S} = D_j^HS_{ij} \,\ten{e}_i 
		\quad; \quad \ten{{}_{H}\Delta v} = \ten{{}_{H}\nabla} \cdot \td{{}_{H}\nabla v} =   D_i^H D_i^Hv_{j} \,\ten{e}_j 
	\end{equation}
	
	\begin{itemize}
		\item \textbf{Derivative symmetry and/or periodicity extensions}:
	\end{itemize} 
	
	In general, given the symmetry and/or periodicity extensions of a field \emph{w.r.t.} a face, the extensions of its derivatives can be determined. We consider a field $G$ that possesses  symmetry and/or periodicity extensions, denoted by the triplet $ABC$, where $A$, $B$ and $C$ are respectively the spatial extension in the three directions $\ten{e}_1$, $\ten{e}_2$ and $\ten{e}_3$ (\emph{i.e.,} \emph{w.r.t.} the faces $S_{1\alpha}$, $S_{2\alpha}$ and $S_{3\alpha}$, $\alpha \in \{0,1\}$). $A$, $B$ and $C$ being a couple of conditions XY where X, Y are respectively the BCs at faces with $\alpha=0$ (left side) and $\alpha=1$ (right side). X, Y can be the periodicity (P), even-symmetry (S) or odd-symmetry (A): X and Y $\in \left\{\text{P},\text{A},\text{S}\right\}$. If $\text{X} \equiv \text{A}$, we note $\overline{\text{X}}\equiv \text{S}$ and vice versa. If $\text{X} \equiv \text{P}$ then $\text{Y} \equiv \text{P}$ and $\overline{\text{X}} \equiv \overline{\text{Y}} \equiv \text{P}$. Using the notation $A^{'}B^{'}C^{'}$ for the extensions of the first-order continuous derivatives $D_{i}G$, $i \in \{1,2,3\}$, we have
	\begin{equation}
		\left\{
		\begin{array}{cccccc}
			\text{if } i=1: & A^{'}= \overline{A} &\text{and} & B^{'}= B &\text{and} & C^{'}= C \\[0.2em]
			\text{if } i=2: & A^{'}= A &\text{and} & B^{'}= \overline{B} &\text{and} & C^{'}= C \\[0.2em]
			\text{if } i=3: & A^{'}= A &\text{and} & B^{'}= B &\text{and} & C^{'}= \overline{C}
		\end{array}
		\right.
		\label{eq:Continuous_der_symmetry_per}
	\end{equation} For the HEX1-based discrete derivation, the parallelepiped voxel being itself invariant by symmetry and/or periodicity \emph{w.r.t.} the faces $S_{i\alpha}, \, i  \in \{1,2,3\}$, the extensions of the discrete first-order derivative, $D^H_{i}G$, at the faces, are consistent with the continuous analysis in Eq.~\eqref{eq:Continuous_der_symmetry_per}. This aspect is different for TETRA2 scheme.

	\begin{itemize}
	\item \textbf{Derivation in Fourier space}:
	\end{itemize} 
	
	For any combination of BCs applied to a field $G$, Eqs.~\eqref{eq:D1HG}-\eqref{eq:D3HG} can be written in the Fourier space by applying DFT on the $4$ times \textquotedblleft virtually\textquotedblright{} extended field, $E$. For simplicity of the notation and knowing the established link between the transforms $\widehat{E}$ and $\widetilde{G}$ (see Tab. \ref{Tab:BCsSymmetriesDTTDFT} in Sec. \ref{sec:spectralcalculations}), we use the notation $\widehat{G}$ to signify the DFT of the \textquotedblleft virtually\textquotedblright{} extended field based on the symmetries of $G$. Therefore, combining Eqs.~\eqref{eq:D1HG}-\eqref{eq:D3HG} with the DTT definitions introduced in Tab. \ref{Tab:BCsSymmetriesDTTDFT}, after elementary trigonometric manipulations, the DFT of the derivative components are expressed as
	\begin{equation}
		\widehat{D_i^{H}\,G} =\mathrm{i} \, {}_{H}\xi_i \, \widehat{G} \, H_{tc} \qquad, \qquad i  \in \{1,2,3\}
		\label{eq:derive_Fourier_HEX1}
	\end{equation} with ${}_{H}\xi_i , \, i  \in \{1,2,3\}$ the components of the real value modified frequency vector, defined as 
	\begin{equation}
		{}_{H}\ten{\xi} =  \begin{pmatrix}
			\displaystyle \frac{2}{h_1}\, \mathrm{sin}\left(\frac{\xi_1^{\mathrm{BCs}}}{2}\right)\,\mathrm{cos}\left(\frac{\xi_2^{\mathrm{BCs}}}{2}\right)\,\mathrm{cos}\left(\frac{\xi_3^{\mathrm{BCs}}}{2}\right) \\[1em] 
			\displaystyle \frac{2}{h_2}\, \mathrm{cos}\left(\frac{\xi_1^{\mathrm{BCs}}}{2}\right)\,\mathrm{sin}\left(\frac{\xi_2^{\mathrm{BCs}}}{2}\right)\,\mathrm{cos}\left(\frac{\xi_3^{\mathrm{BCs}}}{2}\right) \\[1em] 
			\displaystyle \frac{2}{h_3}\, \mathrm{cos}\left(\frac{\xi_1^{\mathrm{BCs}}}{2}\right)\,\mathrm{cos}\left(\frac{\xi_2^{\mathrm{BCs}}}{2}\right)\,\mathrm{sin}\left(\frac{\xi_3^{\mathrm{BCs}}}{2}\right) 
		\end{pmatrix}
		\label{eq:modified_freq_hex1}
	\end{equation} and 
	\begin{equation}
		H_{tc} = \exp \left(\mathrm{i} \, \frac{\xi_1^{\mathrm{BCs}} + \xi_2^{\mathrm{BCs}} + \xi_3^{\mathrm{BCs}}}{2}\right)
		\label{eq:Htc_frequencies}
	\end{equation} The frequencies $\xi_i^{\mathrm{BCs}},\, i \in \{1,2,3\}$ are the fundamental ones shown in Tab. \ref{Tab:BCsSymmetriesDTTDFT} (\emph{i.e.,} variables $\xi^k$ in Tab. \ref{Tab:BCsSymmetriesDTTDFT}), in accordance with the symmetry or periodicity extensions of the field $G$. In the Fourier space, a derivative in one direction is then affected by the symmetry and/or periodicity extensions in the other directions. Moreover, when deriving in the Fourier space a field $G$ defined at the voxel centers, the term $H_{tc}$ in the Eq.~\eqref{eq:derive_Fourier_HEX1} has to be replaced by its conjugate $\overline{H_{tc}}$.
	
	Similarly to the real space case, the gradient and Laplacian of a node-defined displacement vector field $\ten{v}$ and, the divergence of a voxel-defined second-order stress/strain tensor $\td{S}$ are expressed in the Fourier space as
	\begin{equation}
		\widehat{\td{{}_{H}\nabla v}} = \mathrm{i} \, \widehat{\ten{v}} \otimes {}_{H}\ten{\xi} \, H_{tc}\quad; \quad \widehat{\ten{{}_{H}\nabla}\cdot \td{S}} = \mathrm{i} \, \widehat{\td{S}}\,{}_{H}\ten{\xi} \, \overline{H_{tc}}\quad; \quad \widehat{\td{{}_{H}\Delta v}} = -\lVert {}_{H}\ten{\xi} \rVert^2 \widehat{\ten{v}}
		\label{eq:derivation_hex1_fourier}
	\end{equation}

\subsubsection{Double tetrahedron scheme}

The finite difference scheme TETRA2 \citep[see Fig.~\ref{fig:DerSchemeTETRA2},][]{finel_tetrahedron-based_2025} is examined in the present section.
\begin{itemize}
	\item \textbf{Derivation in real space}:
\end{itemize}

Defined at the voxel centers (identified by $j_i + 1/2 \text{ with } j_i \in \llbracket 0, n_i^v-1 \rrbracket,\,i \in \{1,2,3\}$), the derivatives in a chosen direction \emph{w.r.t.} the regular tetrahedrons $T_1$ and $T_2$ (derivation supports), can be expressed as 
	\begin{equation}
		\left\{
		\begin{array}{l}
			D_1^{T_1}G
			= \displaystyle \frac{1}{2h_1} \Bigl[ 
			G[j_1+1,j_2+1,j_3] - G[j_1,j_2+1,j_3+1]
			+ G[j_1+1,j_2,j_3+1] - G[j_1,j_2,j_3]\Bigr] \\[0.5cm]
			D_1^{T_2}G
			= \displaystyle \frac{1}{2h_1} \Bigl[ 
			G[j_1+1,j_2,j_3] - G[j_1,j_2,j_3+1] + G[j_1+1,j_2+1,j_3+1] - G[j_1,j_2+1,j_3]\Bigr]
		\end{array}
		\right.  \label{eq:D1T1T2G}
	\end{equation}
	\begin{equation}
		\left\{
		\begin{array}{l}
			D_2^{T_1}G
			= \displaystyle \frac{1}{2h_2} \Bigl[ 
			G[j_1+1,j_2+1,j_3] - G[j_1,j_2,j_3]
			+ G[j_1,j_2+1,j_3+1] - G[j_1+1,j_2,j_3+1]\Bigr] \\[0.5cm]
			D_2^{T_2}G
			= \displaystyle \frac{1}{2h_2} \Bigl[ 
			G[j_1,j_2+1,j_3] - G[j_1+1,j_2,j_3] + G[j_1+1,j_2+1,j_3+1] - G[j_1,j_2,j_3+1]\Bigr]
		\end{array}
		\right.  \label{eq:D2T1T2G}
	\end{equation}
	\begin{equation}
		\left\{
		\begin{array}{l}
			D_3^{T_1}G
			= \displaystyle \frac{1}{2h_3} \Bigl[ 
			G[j_1,j_2+1,j_3+1] - G[j_1,j_2,j_3]
			+ G[j_1+1,j_2,j_3+1] - G[j_1+1,j_2+1,j_3]\Bigr]  \\[0.5cm]
			D_3^{T_2}G
			= \displaystyle \frac{1}{2h_3} \Bigl[ 
			G[j_1,j_2,j_3+1] - G[j_1,j_2+1,j_3] + G[j_1+1,j_2+1,j_3+1] - G[j_1+1,j_2,j_3]\Bigr]
		\end{array}
		\right.  \label{eq:D3T1T2G}
	\end{equation} Remark \ref{rem:Derive_from_voxel_center} is still valid. Additionally,  for the finite difference scheme TETRA2, the second-order derivatives are calculated by combining both tetrahedrons (cross derivations), \emph{i.e.,} $D_p^{T_1}\left(D_q^{T_2}G\right), \, p,q\in \{1,2,3\}$ or $D_p^{T_2}\left(D_q^{T_1}G\right), \, p,q\in \{1,2,3\}$. 
	
	In an analogous manner to the HEX1 scheme, the gradient and the Laplacian of a node-defined displacement vector field $\ten{v}$ (with scalar components $v_j,\, j  \in \{1,2,3\}$) and the divergence of a voxel-defined second-order stress/strain tensor $\td{S}$ (with scalar components $S_{ij},\, i,j  \in \{1,2,3\}$) are given in the form
	\begin{equation}
	\td{{}_{T_r}\nabla v} = D_i^{T_r}v_j \,\ten{e}_i \otimes \ten{e}_j\quad; \quad \ten{{}_{T_r}\nabla}\cdot \td{S} = D_j^{T_r}S_{ij} \,\ten{e}_i \quad \text{with } r \in \{1,2\}
	\end{equation}
	\begin{equation}
	\ten{{}_{T_1T_2}\Delta v} = \displaystyle \frac{1}{2} \left( D_i^{T_1} D_i^{T_2}v_{j} \, + D_i^{T_2} D_i^{T_1} v_{j} \,\right)  \ten{e}_j
	\end{equation}
	 
	\begin{itemize}
		\item \textbf{Derivative symmetry and/or periodicity extensions}:
	\end{itemize} 
	
	In the previous case of the HEX1 scheme, the extensions of a derivative field are shown to be consistent with the continuous analysis. However, for the TETRA2 scheme, the symmetry and/or periodicity analysis are less straightforward. By construction, the tetrahedron $T_2$ (respectively $T_1$) is the symmetry of the tetrahedron $T_1$ (respectively $T_2$) \emph{w.r.t.} the voxel faces (see Fig.~\ref{fig:DerSchemeTETRA2}). This means that the first-order derivation operation on a given tetrahedron is not invariant by symmetry. Considering a single tetrahedron, the first-order derivative of a field extended with appropriate symmetries do not recover any specific symmetries. In order to clarify this assertion, we consider the face $S_{10}$ normal to the direction $\ten{e}_1$ and containing here the node $\left[0,0,0\right]$ (see Fig.~\ref{fig:DerSchemeTETRA2} with $\left[j_1,j_2,j_3\right] = \left[0,0,0\right] $). For the purpose of this demonstration, $G$ is a field that possesses an even-symmetry \emph{w.r.t.} the face $S_{10}$. Combining the definition of the derivative in Eq.~\eqref{eq:D1T1T2G}$_1$ and the symmetry condition of $G$ \emph{w.r.t.} the face $S_{10}$, it follows that the derivative $D_1^{T_1}G$ at the mirror voxel center is given by the relationship
	\begin{equation}
		D_1^{T_1}G\left[j_1-\frac{1}{2},j_2+\frac{1}{2},j_3+\frac{1}{2}\right] = - D_1^{T_2}G\left[j_1+\frac{1}{2},j_2+\frac{1}{2},j_3+\frac{1}{2}\right]
	\end{equation} This expression shows no obvious symmetry property  of the first-order derivative field $D_1^{T_1}G$ \emph{w.r.t.} the face $S_{10}$ (the same holds true for the derivative field $D_1^{T_2}G$). The symmetries appear when considering a change of derivation support from $T_1$ to $T_2$. Meanwhile, for periodic BCs, both first-order derivative fields $D_i^{T_1}G$ and $D_i^{T_2}G$, $i\in \{1,2,3\}$, are periodic. It is important to note that for the second-order derivatives, \emph{i.e.,} $D_p^{T_1}\left(D_q^{T_2}G\right), \, p,q\in \{1,2,3\}$ or $D_p^{T_2}\left(D_q^{T_1}G\right), \, p,q\in \{1,2,3\}$, the derived discrete fields have the same spatial support and also the same symmetry and/or periodicity as the field $G$. 
	
	This discussion has a major consequence in the computation of the DFT of the first-order derivatives, in Fourier space. Actually, for the non-periodic case, the fact that the first-order derivative fields $D_i^{T_r}G$, $i \in \{1,2,3\}, r \in \{1,2\}$ have no specific symmetries, makes it impossible to use DTTs (on the original field to be derived) to evaluate these first-order derivatives in Fourier space. Nevertheless, the DTTs can be employed in the evaluation of the second-order derivatives, which are essential for the discrete Green operators. These issues do not emerge for full periodic BCs.

	\begin{itemize}
	\item \textbf{Derivation in Fourier space}:
	\end{itemize} 
	
	Given the absence of discernible symmetry extensions of the first-order derivatives, it is not possible to evaluate these derivatives using DTTs. However, in the periodic case, since differentiation preserves periodicity, the DFT remains applicable. For the purpose, considering a periodic field, we have the DFT of the first-order derivatives as
	\begin{align}
		&\widehat{D_m^{T_1}\,G} = \mathrm{i} \, {}_{T_1}\xi_m \, \widehat{G} \, H_{tc} \quad; \quad m  \in \{1,2,3\} \label{eq:derive_Fourier_T1}\\
		&\widehat{D_m^{T_2}\,G} = \mathrm{i} \, {}_{T_2}\xi_m \,  \widehat{G} \, H_{tc} \quad; \quad m  \in \{1,2,3\} \label{eq:derive_Fourier_T2}
	\end{align} with $H_{tc}$ given in Eq.~\eqref{eq:modified_freq_hex1}, ${}_{T_1}\xi_m$ and ${}_{T_2}\xi_m , \, m  \in \{1,2,3\}$ the components of the complex values modified frequency vectors, given by
	\begin{equation}
		{}_{T_1}\ten{\xi} = \begin{pmatrix}
			\displaystyle \frac{2}{h_1}\, \mathrm{sin}\left(\frac{\xi_1^{\mathrm{BCs}}}{2}\right)\,\mathrm{cos}\left(\frac{\xi_2^{\mathrm{BCs}}}{2}\right)\,\mathrm{cos}\left(\frac{\xi_3^{\mathrm{BCs}}}{2}\right) \\[1em] 
			\displaystyle \frac{2}{h_2}\, \mathrm{cos}\left(\frac{\xi_1^{\mathrm{BCs}}}{2}\right)\,\mathrm{sin}\left(\frac{\xi_2^{\mathrm{BCs}}}{2}\right)\,\mathrm{cos}\left(\frac{\xi_3^{\mathrm{BCs}}}{2}\right) \\[1em] 
			\displaystyle \frac{2}{h_3}\, \mathrm{cos}\left(\frac{\xi_1^{\mathrm{BCs}}}{2}\right)\,\mathrm{cos}\left(\frac{\xi_2^{\mathrm{BCs}}}{2}\right)\,\mathrm{sin}\left(\frac{\xi_3^{\mathrm{BCs}}}{2}\right) 
		\end{pmatrix} - \mathrm{i} \begin{pmatrix}
			\displaystyle \frac{2}{h_1}\, \cos\left(\frac{\xi_1^{\mathrm{BCs}}}{2}\right)\,\sin\left(\frac{\xi_2^{\mathrm{BCs}}}{2}\right)\,\sin\left(\frac{\xi_3^{\mathrm{BCs}}}{2}\right) \\[1em] 
			
			\displaystyle \frac{2}{h_2}\, \sin\left(\frac{\xi_1^{\mathrm{BCs}}}{2}\right)\,\cos\left(\frac{\xi_2^{\mathrm{BCs}}}{2}\right)\,\sin\left(\frac{\xi_3^{\mathrm{BCs}}}{2}\right) \\[1em] 
			
			\displaystyle \frac{2}{h_3}\, \sin\left(\frac{\xi_1^{\mathrm{BCs}}}{2}\right)\,\sin\left(\frac{\xi_2^{\mathrm{BCs}}}{2}\right)\,\cos\left(\frac{\xi_3^{\mathrm{BCs}}}{2}\right) 
		\end{pmatrix} 
		\label{eq:freq_TETRA2}
	\end{equation}
	\begin{equation}
		{}_{T_2}\ten{\xi} = \overline{{}_{T_1}\ten{\xi}}
	\end{equation} Here, the frequencies $\xi_i^{\mathrm{BCs}}$, $i  \in \{1,2,3\}$ corresponds to the periodic case values. For a full periodic field, its gradient, divergence and Laplacian can be expressed. In this regard, for a node-defined displacement vector field $\ten{v}$ and a voxel-defined second-order stress/strain tensor $\td{S}$, we can write
	\begin{equation}
		\widehat{\td{{}_{T_r}\nabla v}} = \mathrm{i} \, \widehat{\ten{v}} \otimes {}_{T_r}\ten{\xi} \,H_{tc} \quad; \quad \widehat{\ten{{}_{T_r}\nabla}\cdot \td{S}} = \mathrm{i} \, \widehat{\td{S}}\,{}_{T_r}\ten{\xi}  \,\overline{H_{tc}}
	\end{equation}
	\begin{equation}
		\widehat{\ten{{}_{T_1T_2}\Delta v}} =  \frac{1}{2}\left(\widehat{D_m^{T_1}D_m^{T_2}\,\ten{v}} + \widehat{D_m^{T_2}D_m^{T_1}\,\ten{v}}\right) = - \lVert {}_{T_1}\ten{\xi} \rVert^2 \, \widehat{\ten{v}} = - \lVert {}_{T_2}\ten{\xi} \rVert^2 \, \widehat{\ten{v}}
		\label{eq:laplacian_tetra2}
 	\end{equation}
	
	It is important to recall that for the second-order derivatives (\emph{e.g.,} Laplacian operator in Eq.~\eqref{eq:laplacian_tetra2}), the issue associated with the first derivative do not hold (\emph{i.e.,} a field and its Laplacian possess the same symmetries). For fields with symmetry extensions, the second-order derivatives can then be evaluated using appropriate connection between DTTs and DFT on extended field (see Tab. \ref{Tab:BCsSymmetriesDTTDFT} in Sec. \ref{sec:spectralcalculations}). In this case, the frequencies $\xi_i^{\mathrm{BCs}}$, $i  \in \{1,2,3\}$ used in Eq.~\eqref{eq:freq_TETRA2} corresponds to the non-periodic values in Tab. \ref{Tab:BCsSymmetriesDTTDFT}.

\subsection{Discrete Green operators \label{sec:DisGreenOp}}

The previous sections elaborated on the accelerated fixed-point approach, as well as the evaluation of discrete derivatives (in real and Fourier space). The current section is dedicated to the expression of the discrete Green operator ${}_d\widehat{\mathrm{DGO}}$, \emph{i.e.,} in general words, the deduction of the fluctuation displacement field, $\ten{u^f}$, knowing the divergence of polarization field, $\ten{p}$ (see Eq.~\eqref{eq:GreenOp}). In this work, we distinguish the discrete Green operator as function of the pre-conditioner $\mathbb{R}_0$, the derivation scheme and the loading conditions. Classically, in the periodic case, the pre-conditioner $\mathbb{R}_0$ (generally called \textquotedblleft reference material\textquotedblright) is often chosen as an isotropic elastic stiffness tensor. In the non-periodic case, its usage is limited to specific BCs, and two alternatives \citep{risthaus_imposing_2024,paux_discrete_2025} are discussed.

\subsubsection{ $\mathbb{C}_0$-discrete Green operator \label{sec:DG01}}

Within the finite transformation framework, we first consider the usual choice of pre-conditioner such as
 \begin{equation}
 	\mathbb{R}_0:{}_d\td{\nabla u^f} = \mathbb{C}_0:{}_d\td{\nabla u^f} = 2\mu_0 \, {}_d\td{\nabla u^f} + \lambda_0  \, \mathrm{tr}\left({}_d\td{\nabla u^f}\right) \ten{1}
 	\label{eq:TF_C0_behavior}
 \end{equation} with $\mathbb{C}_0$ the stiffness-like matrix, $\lambda_0$ and $\mu_0$ respectively the first and second Lamé parameters. Choosing these parameters appropriately is crucial for ensuring convergence. Within the small-strain framework, Eq.~\eqref{eq:TF_C0_behavior} is updated by replacing the fluctuation displacement gradient ${}_d\td{\nabla u^f}$ with its symmetric part ${}_d\td{\nabla^s u^f}$. In that case, the subsequent analysis remain valid. Focusing on the finite transformation framework, the discrete Green operator, ${}_d\widehat{\mathrm{DGO}}$, is then deduced by detailing the following expression in Fourier space
 \begin{align}
 	\widehat{\ten{p}} &= \widehat{\ten{{}_d\nabla}\cdot\left(\mathbb{C}_0: \td{{}_d\nabla u^f}\right)} \nonumber\\
 	& = 2\mu_0 \, \widehat{\ten{{}_d\nabla}\cdot\td{{}_d\nabla u^f}} +  \lambda_0 \, \widehat{\ten{{}_d\nabla}\cdot\td{{}_d\nabla^T u^f}} \label{eq:FourierSGO}
 \end{align}

 In the context of general (periodic and non-periodic) BCs, the calculation of  $\widehat{\ten{p}}$ can be done using DTs (depending on the corresponding symmetries and/or periodicity). Based on Eq.~\eqref{eq:FourierSGO}, the periodicity and or symmetry extensions of $\ten{p}$ are required to be the same as the ones of $\ten{{}_d\nabla}\cdot\td{{}_d\nabla u^f}$ and $\ten{{}_d\nabla}\cdot\td{{}_d\nabla^T u^f}$. For the sake of this investigation, we consider that a given component $u^f_i,\,i \in \left\{1,2,3\right\}$ has the symmetries and/or periodicity $A_iB_iC_i$, where $A_i$, $B_i$ and $C_i$ are respectively the extensions in the three directions $\ten{e}_1$, $\ten{e}_2$ and $\ten{e}_3$ (\emph{i.e.,} \emph{w.r.t.} the faces $S_{1\alpha}$, $S_{2\alpha}$ and $S_{3\alpha}$, $\alpha \in \{0,1\}$). Based on the previous analysis of derivative properties in Eq.~\eqref{eq:Continuous_der_symmetry_per} and examining the components $\left[\ten{\nabla}\cdot\td{\nabla u^f}\right]_{j}=u^f_{j,ii}$ and $\left[\ten{\nabla}\cdot\td{\nabla^T u^f}\right]_{j}=u^f_{i,ji}$, we can write the extensions of each term involved in these operations as follows
 \begin{equation}
 	\ten{\nabla}\cdot\td{\nabla u^f} = 
 	\begin{pmatrix}
 		\displaystyle \underbrace{u^f_{1,11}}_{A_1B_1C_1} + \underbrace{u^f_{1,22}}_{A_1B_1C_1} + \underbrace{u^f_{1,33}}_{A_1B_1C_1}\\[2em] 
 		
 		\displaystyle \underbrace{u^f_{2,11}}_{A_2B_2C_2} + \underbrace{u^f_{2,22}}_{A_2B_2C_2} + \underbrace{u^f_{2,33}}_{A_2B_2C_2} \\[2em] 
 		
 		\displaystyle \underbrace{u^f_{3,11}}_{A_3B_3C_3} + \underbrace{u^f_{3,22}}_{A_3B_3C_3} + \underbrace{u^f_{3,33}}_{A_3B_3C_3} 
 	\end{pmatrix} \quad; \quad 
 	\ten{\nabla}\cdot\td{\nabla^T u^f} = 
 	\begin{pmatrix}
 		\displaystyle \underbrace{u^f_{1,11}}_{A_1B_1C_1} + \underbrace{u^f_{2,12}}_{\overline{A_2B_2}C_2} + \underbrace{u^f_{3,13}}_{\overline{A_3}B_3\overline{C_3}}\\[2em] 
 		
 		\displaystyle \underbrace{u^f_{1,21}}_{\overline{A_1B_1}C_1} + \underbrace{u^f_{2,22}}_{A_2B_2C_2} + \underbrace{u^f_{3,23}}_{A_3\overline{B_3C_3}} \\[2em] 
 		
 		\displaystyle \underbrace{u^f_{1,31}}_{\overline{A_1}B_1\overline{C_1}} + \underbrace{u^f_{2,32}}_{A_2\overline{B_2C_2}} + \underbrace{u^f_{3,33}}_{A_3B_3C_3} 
 	\end{pmatrix}
 	\label{eq:lapl_laplT_SGO_BCs}
 \end{equation} In Eq.~\eqref{eq:lapl_laplT_SGO_BCs}, the underbrace expression corresponds to the symmetries and/or periodicity of a given field. Based on the BC analysis in Eq.~\eqref{eq:lapl_laplT_SGO_BCs}, in the general case, no obvious symmetries and/or periodicity can be identified for the field $\ten{\nabla}\cdot\td{\nabla^T u^f}$ and consequently on the field $\ten{p}$. Nevertheless, choosing appropriately the BCs, it is still possible to constrain both terms $\ten{\nabla}\cdot\td{\nabla u^f}$ and $\ten{\nabla}\cdot\td{\nabla^T u^f}$ to have the same symmetries and/or periodicity. The appropriate choice to fulfill this condition is such as
 \begin{equation}
		\left\{
		\begin{array}{c}
			A_1 =  \overline{A_2} =  \overline{A_3} \\[0.2em]
			B_1 =  \overline{B_2} =  B_3 \\[0.2em]
			C_1 =  C_2 =  \overline{C_3}
		\end{array}
		\right.
		\label{eq:SGO_BCs}
	\end{equation} so that $\ten{p}$ and $\ten{u^f}$ exhibit identical symmetry and/or periodicity extensions. In practice, Eq.~\eqref{eq:SGO_BCs} imposes to the fluctuation displacement field $\ten{u^f}$ to have the following periodicity and/or symmetry extensions:
\begin{enumerate}[label=\bfseries\Roman*-]
	\item per face, the full anti-symmetry of the normal components and full symmetry of the tangential ones,
	\item or, conversely, per face, the full anti-symmetry of the tangential components and full the symmetry
	of the normal ones,
	\item per opposite faces, the periodicity of the three components
\end{enumerate} These categories of conditions can be referred to the so-called \textquotedblleft normal-mixed\textquotedblright{} BCs. In total the conditions in Eq.~\eqref{eq:SGO_BCs} lead to a maximum of $5^3$ loading possibilities (5 possibilities per couple of opposite faces and per component). Loading conditions used in the work of \cite{grimmstrele_fftbased_2021} in the context of homogenization are particular cases of the \textquotedblleft normal-mixed\textquotedblright{} BCs as introduced here. For \textquotedblleft normal-mixed\textquotedblright{} BCs, depending on the finite difference scheme, the relation in Eq.~\eqref{eq:FourierSGO} can be detailed. 

\begin{itemize}
	\item \textbf{HEX1 scheme}:
\end{itemize}
Considering the HEX1 scheme and combining the relations in Eq.~\eqref{eq:derivation_hex1_fourier} with Eq.~\eqref{eq:FourierSGO}, we have 
\begin{equation}
	-\widehat{\ten{p}} = 2\mu_0 \lVert {}_{H}\ten{\xi} \rVert^2 \widehat{\ten{u^f}} + \lambda_0 \left({}_{H}\ten{\xi} \otimes {}_{H}\ten{\xi}\right) \widehat{\ten{u^f}} = 2\mu_0 \lVert {}_{H}\ten{\xi} \rVert^2 \widehat{\ten{u^f}} + \lambda_0 \left({}_{H}\ten{\xi} \cdot \widehat{\ten{u^f}}\right) {}_{H}\ten{\xi} 
\end{equation} Multiplying this relation by ${}_{H}\ten{\xi}$ helps to get the term $\widehat{\ten{u^f}}\cdot{}_{H}\ten{\xi}$ as a function of $\widehat{\ten{p}}\cdot{}_{H}\ten{\xi}$. It follows that 
	\begin{equation}
		\widehat{\ten{u^{f}}} = - \displaystyle \frac{1}{2\mu_0 \lVert {}_{H}\ten{\xi} \rVert^2} \left(\widehat{\ten{p}} - \frac{ \lambda_0}{2\mu_0 + \lambda_0} \frac{\widehat{\ten{p}} \cdot {}_{H}\ten{\xi}}{ \lVert  {}_{H}\ten{\xi} \rVert^2} {}_{H}\ten{\xi} \right) = {}_H\widehat{\mathrm{DGO}}(\widehat{\ten{p}})
		\label{eq:FourierSGO_HEX1}
	\end{equation} This expression corresponds to the full periodic case \citep[][and see also \ref{sec:SS_DGO1}]{nkoumbou_kaptchouang_multiscale_2022,to_fourier_2025}, with the notable distinction that the modified frequency vector ${}_{H}\ten{\xi}$ must account for the BCs as described in Eq.~\eqref{eq:modified_freq_hex1}.
\begin{itemize}
	\item \textbf{TETRA2 scheme}:
\end{itemize}	
With the TETRA2 scheme, due to the second-order derivatives in Eq.~\eqref{eq:FourierSGO}, the DFT of the divergence of the polarization stress $\widehat{\ten{p}}$ is written as
\begin{equation}
	\widehat{\ten{p}} = \displaystyle \frac{1}{2} \left(\widehat{{}_{T_2T_1}\ten{p}} + \widehat{{}_{T_1T_2}\ten{p}}\right)
\end{equation}
with 
\begin{equation}
	\left\{
	\begin{array}{l}
		\widehat{{}_{T_2T_1}\ten{p}} = 2\mu_0 \, \widehat{\ten{{}_{T_2}\nabla}\cdot\td{{}_{T_1}\nabla u^f}} +  \lambda_0 \, \widehat{\ten{{}_{T_2}\nabla}\cdot\td{{}_{T_1}\nabla^T u^f}} \\[0.5em]
		\widehat{{}_{T_1T_2}\ten{p}} = 2\mu_0 \, \widehat{\ten{{}_{T_1}\nabla}\cdot\td{{}_{T_2}\nabla u^f}} +  \lambda_0 \, \widehat{\ten{{}_{T_1}\nabla}\cdot\td{{}_{T_2}\nabla^T u^f}}
	\end{array}
	\right.
\end{equation}
Consequently, 
\begin{equation}
	-\widehat{\ten{p}} = \displaystyle 2 \mu_0 \, \lVert {}_{T_2}\ten{\xi} \rVert^2 \, \widehat{\ten{u^f}} +  \frac{\lambda_0}{2} \, \left[ \left(\widehat{\ten{u^f}}\cdot \overline{{}_{T_2}\ten{\xi}} \right) {}_{T_2}\ten{\xi} + \left(\widehat{\ten{u^f}}\cdot {}_{T_2}\ten{\xi} \right) \overline{{}_{T_2}\ten{\xi}} \right]
	\label{eq:p_u_tetra2}
\end{equation} Multiplying Eq.~\eqref{eq:p_u_tetra2} by ${}_{T_2}\ten{\xi}$ and $\overline{{}_{T_2}\ten{\xi}}$ leads to the system of equations
\begin{equation}
	\left\{
	\begin{array}{l}
		-\widehat{\ten{p}} \cdot {}_{T_2}\ten{\xi} = \displaystyle \frac{2\mu_0+\lambda_0}{2} \, \lVert {}_{T_2}\ten{\xi} \rVert^2 \, \left(\widehat{\ten{u^f}}\cdot {}_{T_2}\ten{\xi} \right) + \frac{\lambda_0}{2} \, \left({}_{T_2}\ten{\xi} \cdot {}_{T_2}\ten{\xi}\right) \, \left(\widehat{\ten{u^f}}\cdot \overline{{}_{T_2}\ten{\xi}} \right)\\[1em]
		-\widehat{\ten{p}} \cdot \overline{{}_{T_2}\ten{\xi}} = \displaystyle \frac{\lambda_0}{2} \, \left(\overline{{}_{T_2}\ten{\xi}} \cdot \overline{{}_{T_2}\ten{\xi}}\right)  \, \left(\widehat{\ten{u^f}}\cdot {}_{T_2}\ten{\xi} \right) + \frac{2\mu_0+\lambda_0}{2} \, \lVert {}_{T_2}\ten{\xi} \rVert^2 \, \left(\widehat{\ten{u^f}}\cdot \overline{{}_{T_2}\ten{\xi}} \right)
	\end{array}
	\right.
\end{equation} It is possible to invert this system by considering the terms $\left(\widehat{\ten{u^f}}\cdot {}_{T_2}\ten{\xi} \right)$ and $\left(\widehat{\ten{u^f}}\cdot \overline{{}_{T_2}\ten{\xi}} \right)$ as the unknowns. Introducing these quantities back in Eq.~\eqref{eq:p_u_tetra2}, the Fourier discrete Green operator for TETRA2 is given by
\begin{equation}
	\widehat{\ten{u^{f}}} = \displaystyle - \frac{1}{2\mu_0\,\lVert {}_{T_2}\ten{\xi} \rVert^2} \left[\widehat{\ten{p}} + \displaystyle \frac{ \lambda_0}{2}\left(p_{\mathrm{int}1}\,\overline{{}_{T_2}\ten{\xi}} + p_{\mathrm{int}2}\,{}_{T_2}\ten{\xi} \right)\right] = {}_{T_1T_2}\widehat{\mathrm{DGO}}(\widehat{\ten{p}})
	\label{eq:FourierSGO_TETRA2_1}
\end{equation} with $p_{\mathrm{int}1}$ and $p_{\mathrm{int}2}$ intermediate fields defined as
\begin{equation}
	 \left\{
		\begin{array}{l}
			
			p_{\mathrm{int}1}  = \displaystyle \frac{1}{D_s} \left[\displaystyle -\frac{2\mu_0+\lambda_0}{2}\,\lVert {}_{T_2}\ten{\xi} \rVert^2\,\left( \widehat{\ten{p}} \cdot {}_{T_2}\ten{\xi}\right)  + \displaystyle \frac{\lambda_0}{2}\, \left({}_{T_2}\ten{\xi} \cdot {}_{T_2}\ten{\xi}\right) \left( \widehat{\ten{p}} \cdot \overline{{}_{T_2}\ten{\xi}}\right)\right]  \\[1em]
			
			p_{\mathrm{int}2} = \displaystyle \frac{1}{D_s} \left[ \displaystyle \frac{\lambda_0}{2}\,\left(\overline{{}_{T_2}\ten{\xi}} \cdot \overline{{}_{T_2}\ten{\xi}}\right)\,\left( \widehat{\ten{p}} \cdot {}_{T_2}\ten{\xi}\right) -  \displaystyle\frac{2\mu_0+\lambda_0}{2} \, \lVert {}_{T_2}\ten{\xi} \rVert^2 \left( \widehat{\ten{p}} \cdot \overline{{}_{T_2}\ten{\xi}}\right)\right] \\[1em]
			\qquad \text{where } D_s = \displaystyle \left(\frac{2\mu_0+\lambda_0}{2}\,\lVert {}_{T_2}\ten{\xi} \rVert^2\right)^2 - \left(\frac{\lambda_0}{2}\,\lvert {}_{T_2}\ten{\xi} \cdot {}_{T_2}\ten{\xi} \rvert\right)^2
		\end{array}
		\right.
		\label{eq:FourierSGO_TETRA2_2}
	\end{equation}

\ref{sec:SS_DGO1} shows the small strain version of the discrete Green operator in Eqs.~\eqref{eq:FourierSGO_HEX1} and~\eqref{eq:FourierSGO_TETRA2_1}. It is important to recall that the discrete Green operators in Eqs.~\eqref{eq:FourierSGO_HEX1} and~\eqref{eq:FourierSGO_TETRA2_1} are valid for \textquotedblleft normal-mixed\textquotedblright{} BCs. However, it should be noted that, \emph{e.g.,} a bending loading BCs or a full Dirichlet BCs, are not included in this list of $5^3$ eligible possibilities. To overcome this limitation, the reference material behavior must be defined alternatively. Two options are examined hereafter.

\subsubsection{ $2\mu_0$-discrete Green operator}

As discussed in the previous section, the restriction to specific BCs (see Eq.~\eqref{eq:SGO_BCs}) stems from the symmetries and/or periodicity of certain derivative terms appearing in the computation of $ \widehat{\ten{\nabla}\cdot\td{\nabla^T u^f}}$ in Eq.~\eqref{eq:FourierSGO}. To overcome this limitation, a simple choice is to set the first Lamé parameter to zero, \emph{i.e.,} $\lambda_0=0$ \citep{risthaus_imposing_2024,paux_discrete_2025}. This means that the pre-conditioner is described by 
\begin{equation}
	\mathbb{R}_0:\td{{}_d\nabla u^f} = 2\mu_0\,\td{{}_d\nabla u^f}
\end{equation}
This choice eliminates the need for any constraints on the periodicity or symmetry extensions of the fluctuation displacement field $\ten{u^f}$. The result of this process is the possible combination of any Neumann, Dirichlet and periodic BCs for all faces and components, which yields a total of $5^9$ possibilities.

Exploiting the previous analysis and fixing $\lambda_0=0$, we deduce the Fourier discrete Green operator as 
\begin{itemize}
	\item \textbf{HEX1 scheme case}
	\begin{equation}
		\widehat{\ten{u^{f}}} = - \displaystyle \frac{1}{2\mu_0 \lVert \ten{\xi}_{H} \rVert^2} \widehat{\ten{p}} = {}_{H}\widehat{\mathrm{DGO}}(\widehat{\ten{p}})
		\label{eq:NSGO1_HEX1}
	\end{equation}
	\item \textbf{TETRA2 scheme case}
	\begin{equation}
		\widehat{\ten{u^{f}}} = - \displaystyle \frac{1}{2\mu_0 \lVert {}_{T_2}\ten{\xi} \rVert^2} \widehat{\ten{p}} = {}_{T_1T_2}\widehat{\mathrm{DGO}}(\widehat{\ten{p}})
		\label{eq:NSGO1_TETRA2}
	\end{equation}
\end{itemize}
The constructed discrete Green operators are analogous for both finite difference schemes. This formulation offers the advantage of simplicity, requiring only the choice of the second Lamé parameter, $\mu_0$, and is suitable for all types of BCs.

	\subsubsection{$\mathbb{B}_0$-discrete Green operator}
	
	In contrast with the previous approach of setting the first Lamé parameter of the reference material to zero, another potential solution is investigated in this section. Following the work of \cite{paux_discrete_2025}, an alternative pre-conditioner is 
	\begin{equation}
		\mathbb{R}_0: \td{\nabla u^f} = 2\mu_0 \td{\nabla u^f} + \lambda_0 \td{\nabla u^f} \odot \td{I} =\mathbb{B}_0: \td{\nabla u^f}
	\end{equation}
	Similarly to the previous cases, the discrete Green operator ${}_d\widehat{\mathrm{DGO}}$, for a given finite difference scheme, is determined based on the following relationship
	\begin{equation}
		\widehat{\ten{p}} = \widehat{\ten{{}_d\nabla}\cdot\left(\mathbb{B}_0: \td{{}_d\nabla u^f}\right)}  = 2\mu_0 \, \widehat{\ten{{}_d\nabla}\cdot\td{{}_d\nabla u^f}} +  \lambda_0 \, \widehat{\ten{{}_d\nabla}\cdot\left(\td{{}_d\nabla u^f} \odot \td{I}\right)} \label{eq:FourierNSGO2}
	\end{equation}
	In order to detail this discrete Green operator, the index formulation of Eq.~\eqref{eq:FourierNSGO2} is used 
	\begin{align}
		\widehat{p_i} &= \displaystyle 2 \mu_0 \, \sum_{j=1}^{3} \widehat{D_j^d D_j^d u^f_i} + \lambda_0 \,  \sum_{j=1}^{3} \widehat{D_j^d\left(D_j^d u^f_i \, \delta_{ij}\right)} \qquad \text{ (no summation on the index } i \text{)}\\
		& = \displaystyle 2 \mu_0 \, \sum_{j=1}^{3} \widehat{D_j^d D_j^d u^f_i} + \lambda_0 \, \widehat{D_i^d D_i^d u^f_i} \qquad \text{ (no summation on the index } i \text{)}
	\end{align} This expression indicates that the term $\ten{{}_d\nabla}\cdot\left(\td{{}_d\nabla u^f} \odot \td{I}\right)$ yields equivalent periodicity and/or symmetry extensions to those of $\ten{{}_d\nabla}\cdot\td{{}_d\nabla u^f}$, which are also identical to the ones of $\ten{u^f}$. Consequently, the constructed discrete Green operator can be used for any choice of BCs.
	\begin{itemize}
		\item Accordingly to the \textbf{HEX1 scheme} and building on Eq.~\eqref{eq:derive_Fourier_HEX1}, we have
	\end{itemize}
	\begin{equation}
		\widehat{D_j^H D_j^H u^f_i} = - \lVert {}_{H}\ten{\xi} \rVert^2 \, \widehat{u^f_i}
	\end{equation}
	\begin{equation}
		\widehat{D_i^H D_i^H u^f_i} = - \left({}_H \xi_i\right)^2 \, \widehat{u^f_i} \qquad \text{ (no summation on the index } i \text{)}
	\end{equation}
	It follows that
	\begin{equation}
		\widehat{p_i} = - 2 \mu_0 \, \lVert {}_{H}\ten{\xi} \rVert^2 \, \widehat{u^f_i} - \lambda_0 \, \left({}_H \xi_i\right)^2 \, \widehat{u^f_i} \qquad \text{ (no summation on the index } i \text{)}
	\end{equation}
	Finally,
	\begin{equation}
		\widehat{u^f_i} = - \displaystyle \frac{1}{2\mu_0 \lVert {}_{H}\ten{\xi} \rVert^2 + \lambda_0 \, \left({{}_{H}\xi}_i\right)^2}\, \widehat{p_i} = {}_{H}\widehat{\mathrm{DGO}}(\widehat{\ten{p}}) \qquad \text{ (no summation on the index } i \text{)}
		\label{eq:NSGO2_HEX1}
	\end{equation}
	\begin{itemize}
		\item Using the same previous calculation steps for the \textbf{TETRA2 scheme}, we obtain the discrete Green operator as 
	\end{itemize}
	\begin{equation}
		\widehat{u^f_i} = - \displaystyle \frac{1}{2\mu_0 \lVert {}_{T_2}\ten{\xi} \rVert^2 + \lambda_0 \, \lvert {{}_{T_2}\xi}_i\rvert^2} \,\widehat{p_i} = {}_{T_1T_2}\widehat{\mathrm{DGO}}(\widehat{\ten{p}})  \qquad \text{ (no summation on the index } i \text{)}
		\label{eq:NSGO2_TETRA2}
	\end{equation}
	
	Unless otherwise specified, the discrete Green operator based on the pre-conditioner, $\mathbb{B}_0: \td{\nabla u^f}$ (reference material) is the one used in the simulations (small and finite transformations). It is worth noting that for all the three versions of the discrete Green operators (see Eqs.~\eqref{eq:FourierSGO_HEX1}, \eqref{eq:FourierSGO_TETRA2_1}-\eqref{eq:FourierSGO_TETRA2_2}, \eqref{eq:NSGO1_HEX1}, \eqref{eq:NSGO1_TETRA2}, \eqref{eq:NSGO2_HEX1} and \eqref{eq:NSGO2_TETRA2}), when the corresponding denominators are found to be null, we enforce $\widehat{\ten{u^{f}}} = 0$.

\subsection{Implementation specificities \label{sec:Impl_Spec}}

The purpose of this section is to emphasize the most important evolutions (\emph{i.e., w.r.t.} a classical periodic implementation) required to introduce non-periodic BCs together with various types of finite difference schemes in a distributed memory parallel solver.

\subsubsection{Field implementation}

\begin{itemize}
	\item \textbf{Fields in real space}
\end{itemize}

In most classical implementations of periodic FFT-based solvers, fields in real space are defined as arrays of size ($n_1^v$,$n_2^v$,$n_3^v$,$n_{comp}$), with $n_{comp}$ the number of components (3 for a vector, 6 for a symmetric second-order tensor and 9 for a non-symmetric tensor). The displacement and strain/stress fields are defined on grids $\llbracket 0,n_1^v-1\rrbracket \times \llbracket 0,n_2^v-1\rrbracket \times \llbracket 0,n_3^v-1\rrbracket$.

In our implementation for generic BCs, the strain field is still defined on a grid $\llbracket 0,n_1^v-1\rrbracket \times \llbracket 0,n_2^v-1\rrbracket \times \llbracket 0,n_3^v-1\rrbracket$ while the displacement is now defined on a larger grid $\llbracket 0,n_1^v\rrbracket \times \llbracket 0,n_2^v\rrbracket \times \llbracket 0,n_3^v\rrbracket$. In addition, for an extension to various types of finite difference schemes \citep[TETRA2 among others, see][]{gehrig_elementbased_2025} 
requiring more than one point per voxel, fields are defined as arrays of size ($n_1^v$,$n_2^v$,$n_3^v$,$n_{comp}$,$n_{ppv}$), with $n_{ppv}$ the number of points per voxel (\emph{e.g.,} using TETRA2 scheme, $n_{ppv}=2$ for a strain/stress field and 1 for a displacement field). As a consequence, new field objects are defined to distinguish the two kinds of fields and introduce multi-points per voxel. 

\begin{itemize}
	\item \textbf{Fields in Fourier space}
\end{itemize}

In classical (periodic) implementations, all fields in Fourier space are complex numbers with a size ($n_1^v/2+1$,$n_2^v$,$n_3^v$,$n_{comp}$). Actually, to reduce memory footprint and building on the fact that the input fields are real values, the size of the DFT in the first direction is classically divided by (almost) two without any loss of information (only non-redundant values are stored).

To account for possibly non-periodic BCs, a field component in Fourier space may be obtained either through a DFT, when periodicity is prescribed in at least one direction, or through DTTs when non-periodic BCs are considered. Consequently, depending on the selected BCs, a component of a field, in Fourier space, may be either complex-valued (if a DFT is involved in at least one direction) or purely real-valued (if DTTs are employed in all three directions). In addition, if a periodicity condition appears firstly in a given direction $l\in \{1,2,3\}$, the array dimension divided by $2$ to save memory corresponds to the precise direction $l$. As a consequence, to propose a versatile code accounting for general BCs (\emph{i.e.,} arbitrary combinations of periodic and/or non-periodic BCs on each face and each component), the novel implementation in this work includes Fourier field object that are able to handle components of different types (complex or real) and sizes.

\subsubsection{Domain decomposition and discrete transformations}

\begin{figure}[t]
	\centering{}\includegraphics[width=0.8\textwidth]{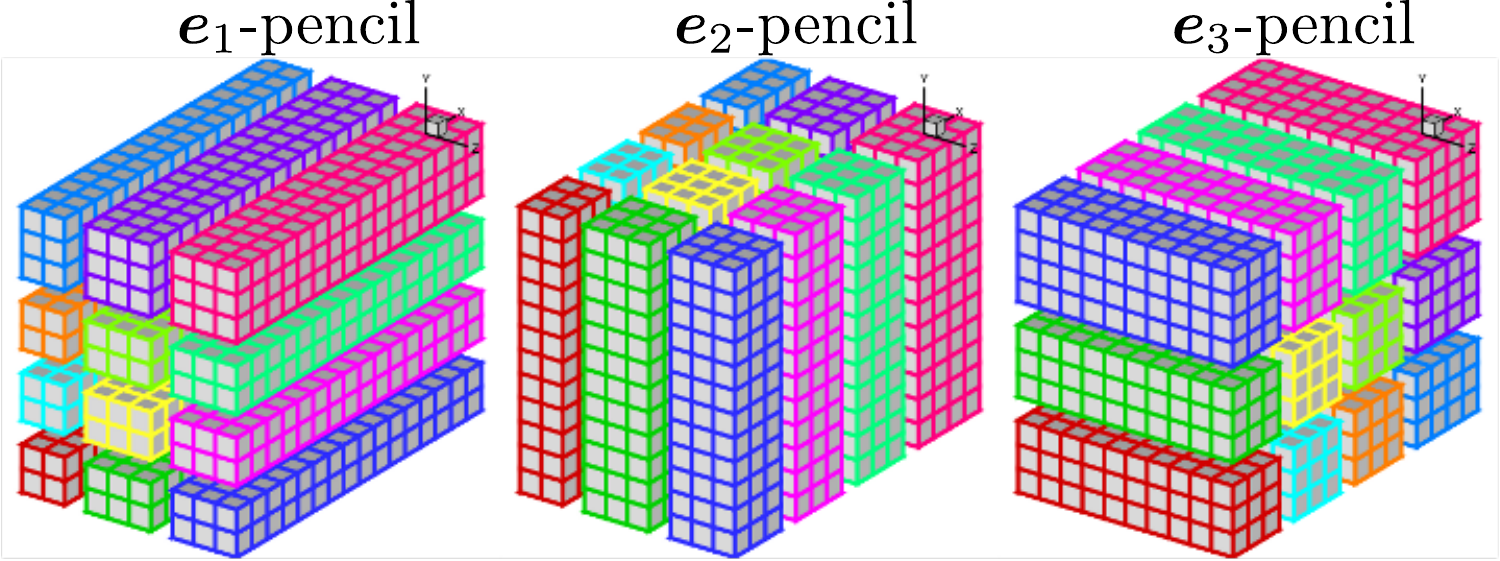}
	\caption{Illustration of parallel 2D domain decomposition using $12$ processors (adapted from \citep{li_2decompfft_nodate}).}\label{fig:Domain2Decomp}
\end{figure}

The distributed memory implementation of the FFTW library \citep{frigo_design_2005} is based on a 1D slab decomposition of the unit-cell. Such an implementation exhibits an important limit regarding the maximum number of processors. Actually, a $(n^v)^3$ unit-cell can be distributed on a maximum of $n^v$ processors. To overcome this limitation, the 2DECOMP\&FFT library \citep{li_2decompfft_nodate,rolfo_2decompfft_2023} provides the possibility of using 2D pencils to decompose the unit-cell. Fig.~\ref{fig:Domain2Decomp} shows an example of decomposition with $12$ processors, each processor managing a part of the unit-cell. This MPI-based library is a foundation of the classical periodic solver \cite{amitex_fftp}, and the present work necessitated an evolution of 2DECOMP\&FFT to account for DTTs in a massively parallel codes. In \emph{AMITEX$^\star$}, for field components in real space, the pencil direction is $\ten{e}_1$. The 3D DTs consist of a succession of 1D transforms in each direction. The first transform is in the direction $\ten{e}_1$. The second transform, in direction $\ten{e}_2$, can not be made directly because $e_1$-pencils do not have the full length of data in direction $\ten{e}_2$. Hence, a transposition of the array is done (via a MPI\_ALLTOALL communication), before applying the DTs, to get it distributed in $e_2$-pencils (see Fig.~\ref{fig:Domain2Decomp}). After the application of the DT in the third direction, requiring a second transposition step, the array is distributed in $e_3$-pencils. Consequently, field components in Fourier space are distributed in $e_3$-pencils.

Recalling that the displacement field is defined on the grid $\llbracket 0,n_1^v\rrbracket \times \llbracket 0,n_2^v\rrbracket \times \llbracket 0,n_3^v\rrbracket$, it is important to notice that, within the novel \emph{AMITEX$^\star$} solver, the application of the DTs in each direction must be done taking into account \textquotedblleft dismissed\textquotedblright{} points (see index $j$ Tab. \ref{Tab:BCsSymmetriesDTTDFT} in Sec. \ref{sec:spectralcalculations}). For example, consider the sine transform DST1 of a 1D signal, the value of the signal at the two sides are set to zero and the DST1 is applied to a reduced signal (skipping the two extreme points). Depending on the DT, the first and last point could be dismissed or not. The \textquotedblleft dismissed\textquotedblright{} points are identified during \emph{AMITEX$^\star$} initialization and provided to 2DECOMP\&FFT when initializing a 3D transform in the new library.

Actually, when a 2DECOMP\&FFT 3D transform is initialized, an array of integers is provided to describe the DTs in the three directions. The first three components of the array are mandatory and describe the forward DTs in the three directions $\ten{e}_1$, $\ten{e}_2$ and $\ten{e}_3$. Additionally, optional arguments can be provided to describe the number of elements \textquotedblleft dismissed\textquotedblright{} in each direction, and their respective location (left and/or right side). The available 1D transforms are DFT, alongside with 8 DTTs (for the present displacement-based algorithm, only $4$ DTTs are used), leading to $9^3 = 729$ combinations for the 3D transform. The DTTs available in the 2DECOMP\&FFT library correspond exactly to the transforms FFTW\_REDFT (\emph{i.e.,} cosine transforms) and FFTW\_RODFT (\emph{i.e.,} sine transforms) available in FFTW3 \citep{frigo_design_2005}.

The internal memory requirement for the parallel 3D transform in 2DECOMP\&FFT can reach five 3D temporary arrays. A reduction of this memory usage is possible, but it would increase the complexity of the 2DECOMP\&FFT code. To limit the impact of this increased memory usage, the library now provides a memory pool. When the caller (\emph{i.e.,} \emph{AMITEX$^\star$} in the present case) is not performing 3D transforms, he can temporarily read and write the memory blocks allocated by 2DECOMP\&FFT. Note that, the library being open source, all these developments are available to anyone, especially for applications in fluid/solid mechanics and multi-physics.

The implementation of the DTTs done in the parallel framework (MPI), have been validated by the following two-step methodology. First, considering a random 3D field (scalar $G$, vector $\ten{G}$ or tensor $\td{G}$) with various symmetries, the corresponding DTT computation of the field is undertaken to obtain a field in spectral space. Next, an inverse DTT calculation, denoted iDTT, is performed on the obtained spectral field. This sequence of operations (\emph{e.g.,} in the vector case iDTT(DTT)($\ten{G}$)) is expected to result in the recovery of the original random field (\emph{e.g.,} in the vector case $\ten{G}$). To ensure completeness, this validation is performed for any combination (\emph{i.e.,} symmetries and/or periodicity) of DTs (one per direction) varying also the grid sizes and the domain decomposition (number of processors).

\subsubsection{Discrete derivation \label{sec:discrete_derivation}}

\begin{itemize}
	\item \textbf{Derivation in real space}
\end{itemize}

The present displacement-based algorithm (see Alg. \ref{alg:algFFT}) differs from the classical strain-based algorithm used in the full periodic case \citep{Moulinec1998,amitex_fftp}. Indeed, the displacement-based algorithm has the advantage to reduce the memory footprint (by a factor of $2$ or $3$ whether small or finite transformation hypothesis are considered) of the array used to store the $4$ couples of solutions and residual fields used for the Anderson convergence acceleration. A further distinction between the displacement-based algorithm and the strain-based version implemented in \cite{amitex_fftp}, is that the former involves the evaluation of the divergence and the gradient in real space (see Alg. \ref{alg:algFFT}), whereas these derivation operations are not computed in the latter.

In the present distributed memory implementation, using pencil decomposition (see previously described Fig.~\ref{fig:Domain2Decomp}), computation of discrete derivation at points located at the frontier of the pencil domains can not be directly done. Indeed, the data of points located in the neighboring pencil are required for this derivation operation. Hence, such discrete derivation necessitates the use of halo-voxels: each pencil is then enlarged by one voxel (one is enough for HEX1 and TETRA2 schemes) and the corresponding data are transferred from one processor to another through a 2DECOMP\&FFT procedure relying on MPI non-blocking send/receive functions. In addition, the halo-voxels that are defined outside of the unit-cell must be evaluated according to the appropriate symmetry extension (associated with BCs). This last point is performed at \emph{AMITEX$^\star$} side.  

\begin{itemize}
	\item \textbf{Derivation in Fourier space}
\end{itemize}

The application of discrete derivation in Fourier space using DTs  (see Sec. \ref{sec:DerivationsOp}) requires the knowledge of modified frequency vectors defined in Eqs.~\eqref{eq:modified_freq_hex1},~\eqref{eq:freq_TETRA2} and Tab. \ref{Tab:BCsSymmetriesDTTDFT} (that exhibits the links between DTTs and DFT). These modified frequency vectors are Fourier fields which are evaluated and stored \textquotedblleft on the fly\textquotedblright{} as soon as required (for example when applying discrete divergence, gradient, or Green operators). Please note that, for the present algorithm, a modified frequency vector is stored per combination of BCs applied to the displacement field components. For full Dirichlet (or Neumann, or periodic) BC, the $3$ components of the displacement have the same BCs and a single vector is stored. In the worst case, the three components of the displacement have different BCs and three modified frequency vectors are stored.

\begin{itemize}
	\item \textbf{Cross validation}
\end{itemize}

As a result of the extension to non-periodic BCs, the implementation becomes more complex, prone to bugs, especially when considering the high number of possible BC combinations (\emph{i.e.,} the total number of BC combinations is $(5^3)^{n_{comp}}$, with $n_{comp}=3$ for a vector displacement field, hence approximately $2.10^6$ combinations). The validation of discrete derivation in real and Fourier spaces is essential. For that purpose, a cross validation strategy is implemented: a random field is initialized, then derived in real space on the one hand, and derived using DTs for the back and forth in Fourier space on the other hand. The relative squared error between the two derived fields is subsequently computed and verified to be lower than a tolerance of approximately $10^{-13}$. This strategy is repeated for all the possible combinations of BCs, and for, scalar, vector and tensor (symmetric or not) as inputs of the various discrete derivation operators (divergence, gradient, curl, Laplacian). We believe that this cross validation  is a major cornerstone in the development of the code: no need to go further if such cross validations are not performed.

\section{Numerical simulations \label{sec:numerical_simulations}}

A versatile parallel FFT-based solver, relying on the use of DTs, was introduced in the previous section to circumvent the limitations of periodic BCs. In the present section, an extensive investigation is conducted to highlight the extended capabilities and the robustness of this new solver. The objective is to demonstrate the versatility of the approach in addressing small and finite transformation problems with different loading scenarios of interest in material science and out of reach for standard (periodic) FFT-based solvers. The finite difference schemes introduced in this study are examined, and their outcomes are discussed, with a focus on the recently proposed TETRA2 scheme \citep{finel_tetrahedron-based_2025}, that demonstrates a better robustness than the HEX1 scheme \citep{willot_fourier-based_2015}. Leveraging on the parallel implementation, different types of unit-cell geometries and material behavior laws, including isotropic elasticity, isotropic perfect plasticity and crystal plasticity are explored. To ensure comprehensiveness, the behavior laws for each category of examples are briefly outlined in the corresponding section. When possible, the simulations are validated against analytical solutions. Unless otherwise precised, a tolerance $\epsilon = 10^{-5}$ is used for the equilibrium condition (see Eq.~\eqref{eq:equilibrium_condition}).

\subsection{Elastic cantilever beam under bending configurations}

With the aim of validating non-homogeneous loadings, the first category of loadings, consists of a bending loading (see \ref{sec:NeoHooken} for a homogeneous pure tension displacement-controlled analytical validation within finite transformation framework). Small and finite transformation frameworks are examined here using an elastic cantilever beam under two types of bending configurations. The small strain case for both the HEX1 and TETRA2 finite difference schemes is validated through the use of an analytical solution. The objective of the finite transformation case is to ascertain the most suitable candidate between the two presented finite difference schemes for such simulations. For simplicity, in the pure isotropic elastic cases, we consider standard steel elastic properties with the parameters: $\lambda = 120 \, \si{GPa}$ and $\mu = 80 \, \si{GPa}$. As the bending configurations to be treated are not compatible with \textquotedblleft normal-mixed\textquotedblright{} (see Eq.~\eqref{eq:SGO_BCs}), we then use the $\mathbb{B}_0$-discrete Green operator with $\lambda_0=\lambda$ and $\mu_0=\mu$.

\subsubsection{Cantilever beam under uniformly distributed applied stress \label{sec:NumEx1}}

\begin{figure}[t]
	\centering\subfloat[Bending setup for uniformly distributed applied stress.]{
		\includegraphics[width=0.48\textwidth,valign=c]{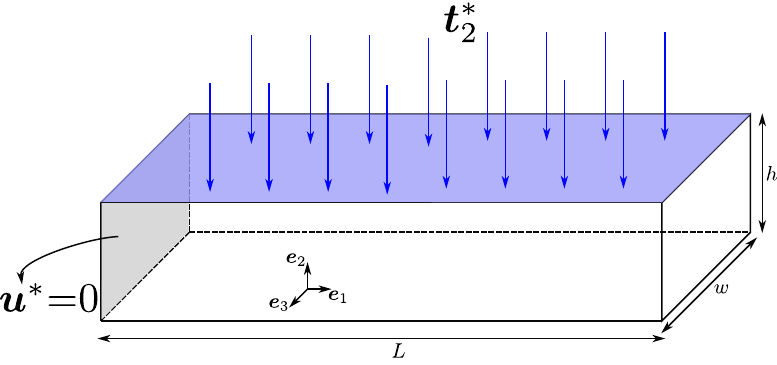}
		\label{fig:ElasticBendingSetup}
	}
	\hfill{} \subfloat[Comparison of analytical solution and numerical simulations within small strain framework.]{
		\includegraphics[width=0.48\textwidth,valign=c]{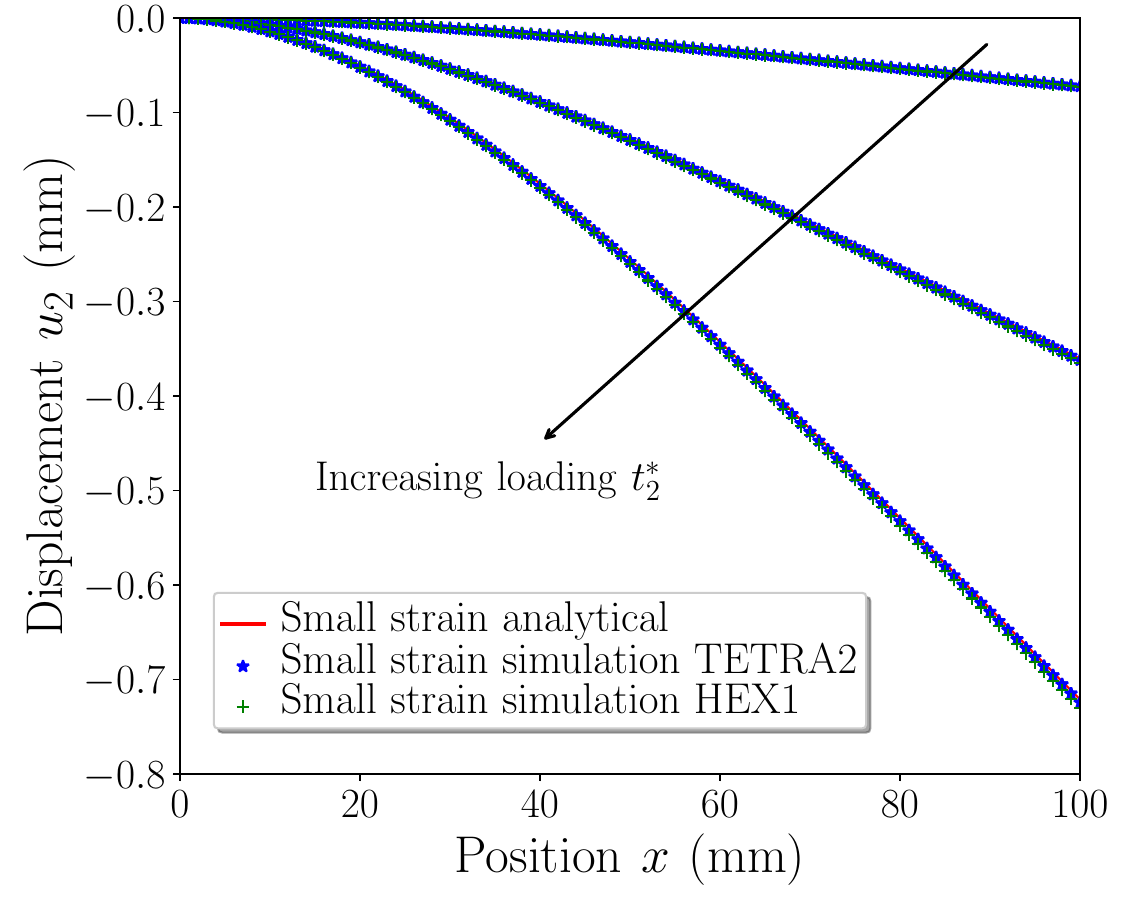}
		\label{fig:ElasticBendingAnalvsSimu}
	}\caption{Elastic bending configuration with a distributed uniform loading on the top face $S_{21}$. Dimensions: $L\times w \times h = 100 \times 10 \times 10 \,\mathrm{mm^3}$ with a discretization $n_1\times n_2 \times n_3 = 100 \times 10 \times 10$ voxels.}\label{fig:ElasticBending}
\end{figure}

Assuming small strain hypothesis and isotropic elasticity, the material behavior is given by 
\begin{equation}
	\td{\sigma} = \mathbb{C}:\td{\nabla^s u} = \mu \, \td{\nabla^s u} + \lambda \, \mathrm{tr}(\td{\nabla u}) \, \td{1} 
\end{equation} The well-known bending setup in Fig.~\ref{fig:ElasticBendingSetup} is used. The left side (face $S_{10}$) of the beam is fully clamped by enforcing Dirichlet BCs with a zero value for the applied displacement field, \emph{i.e.,} $\ten{u^*}=0$. On the top face $S_{21}$, a Neumann BC with an uniform applied stress vector in the direction $\ten{e}_2$ is imposed (\emph{i.e.,} $\ten{t^*} = -t_2^* \ten{e}_2$).

Considering that the length $L$ of the beam in the direction $\ten{e}_1$, is considerably greater than the other dimensions ($h$ and $w$ in the directions $\ten{e}_2$ and $\ten{e}_3$ respectively, thus allowing for the face load to be approximated by a linear load), the beam theory provides the analytical displacement solution, in the direction $\ten{e}_2$, of the neutral axis 
\begin{equation}
	u_2(x) = - \displaystyle \frac{t_2^*w}{EI} \left(\frac{L^2 x^2}{4} - \frac{Lx^3}{6} + \frac{x^4}{24}\right) 
	\label{eq:analytical_sol_bending_elastic_own_weight}
\end{equation} where $x$ denotes here the position along $\ten{e}_1$, $I = wh^3/12$ is the area moment of inertia and $E$ is the Young modulus. Fig.~\ref{fig:ElasticBendingAnalvsSimu} presents the comparison between the analytical solution (see Eq.~\eqref{eq:analytical_sol_bending_elastic_own_weight})and the FFT-based solver results for both finite difference schemes HEX1 and TETRA2. Three increasing loading steps are used to show that, within small strain framework, HEX1 and TETRA2 schemes give similar results in a very good agreement with the analytical solution.

\subsubsection{Cantilever beam under end load applied stress \label{sec:NumEx2}}

To check the robustness of the method, another kind of bending loading is investigated. More specifically, in addition, to small strain simulations, we also explore the capability of the solver for finite transformation simulations. The material behavior law is extended to a finite transformation formalism with the second Piola–Kirchhoff stress tensor given by the relationship   
\begin{equation}
	\td{\Pi} = \mathbb{C}:\td{E} = \mu \, \td{E} + \lambda \, \mathrm{tr}(\td{E}) \, \td{1} 
\end{equation} with the Green-Lagrange strain tensor $\td{E}$ defined as 
\begin{equation}
	\td{E} = \displaystyle \frac{1}{2} \left(\td{F}^{T}\td{F} - \td{1}\right)
\end{equation}

\begin{figure}[h!]
	\centering\subfloat[End load bending setup.]{
		\includegraphics[width=0.5\textwidth]{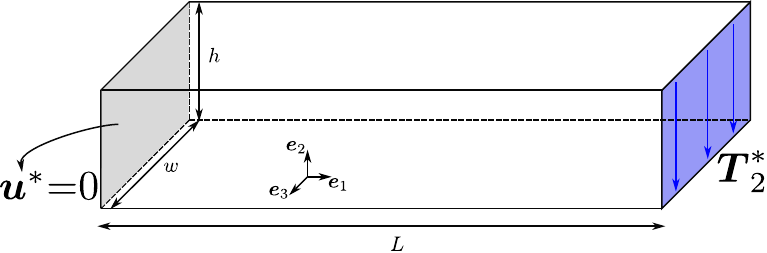}
		\label{fig:FiniteStrainLocalizedBeamsetupSetup}
	}
	\vfill{} \subfloat[Macroscopic responses in small strain versus finite transformation framework, the deformed shapes are shown without any amplification factor. These shapes are plotted at the level of stress where the HEX1-based simulation fails.]{
		\includegraphics[width=1.\textwidth]{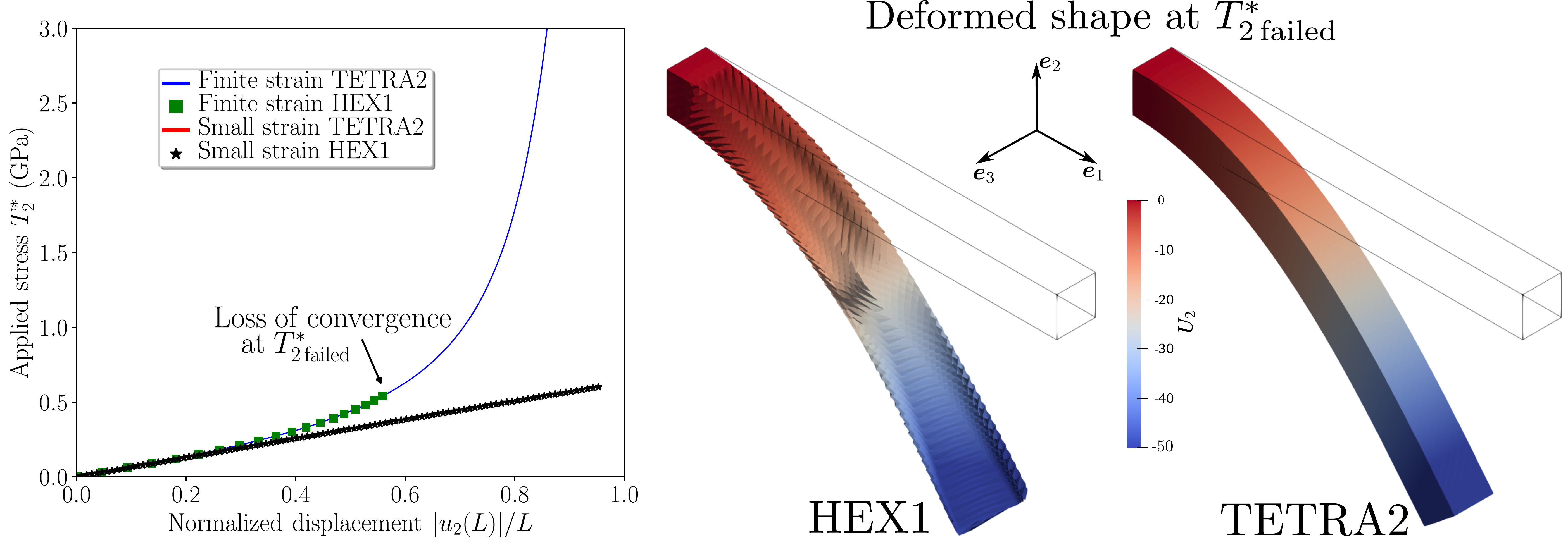}
		\label{fig:FiniteStrainLocalizedBeamsetupSimu}
	}\caption{Elastic bending: distributed loading on a surface. Dimensions: $L\times w \times h = 100 \times 10 \times 10 \,\mathrm{mm^3}$ with a discretization $n_1\times n_2 \times n_3 = 100 \times 10 \times 10$ voxels}\label{fig:FiniteStrainLocalizedBeamsetup}
\end{figure}

For the purpose of the present study, the bending setup with an end load applied stress as sketched in Fig.~\ref{fig:FiniteStrainLocalizedBeamsetupSetup} is used. This configuration prompts to a large displacement response as the loading increases. Fig.~\ref{fig:FiniteStrainLocalizedBeamsetupSimu} presents the macroscopic responses. In order to validate the implemented Lagrangian formulation for the finite transformation case, a comparison with the small strain framework simulations is also plotted. Fig.~\ref{fig:FiniteStrainLocalizedBeamsetupSimu} (left part) shows that both small and finite transformations lead to the same behavior at low applied stress. As expected, the choice of a small strain approximation leads to linear response (identical for TETRA2 and HEX1 schemes). Conversely, in the finite transformation framework, non-linear elastic responses are obtained. Both HEX1 and TETRA2 schemes yield similar non-linear behavior, allowing to reach more than $0.5$ normalized displacement. It is important to notice that the finite transformation simulation with HEX1 scheme fails to converge post a certain level of applied stress $T_{2\,\mathrm{failed}}^*$. Meanwhile, the TETRA2 scheme allows to converge well beyond the stress level $T_{2\,\mathrm{failed}}^*$ (up to a ratio $u_2(L)/L > 0.8 $). The total displacement profiles are plotted for both schemes at $T_{2\,\mathrm{failed}}^*$, \emph{c.f.,} Fig.~\ref{fig:FiniteStrainLocalizedBeamsetupSimu} (right part). It is visible that the HEX1 scheme shows spurious displacement oscillations, leading to the failed convergence while TETRA2 scheme yields smooth deformed configuration, enabling convergence at elevated strain levels. The TETRA2 scheme seems to be more robust than the HEX1 scheme in modeling such non-linearity.

\subsection{Perfect plasticity porous medium under a normal-mixed loading \label{sec:PorousPerfectPlasticity}}

\begin{figure}[t]
	\centering{}\includegraphics[width=0.3\textwidth]{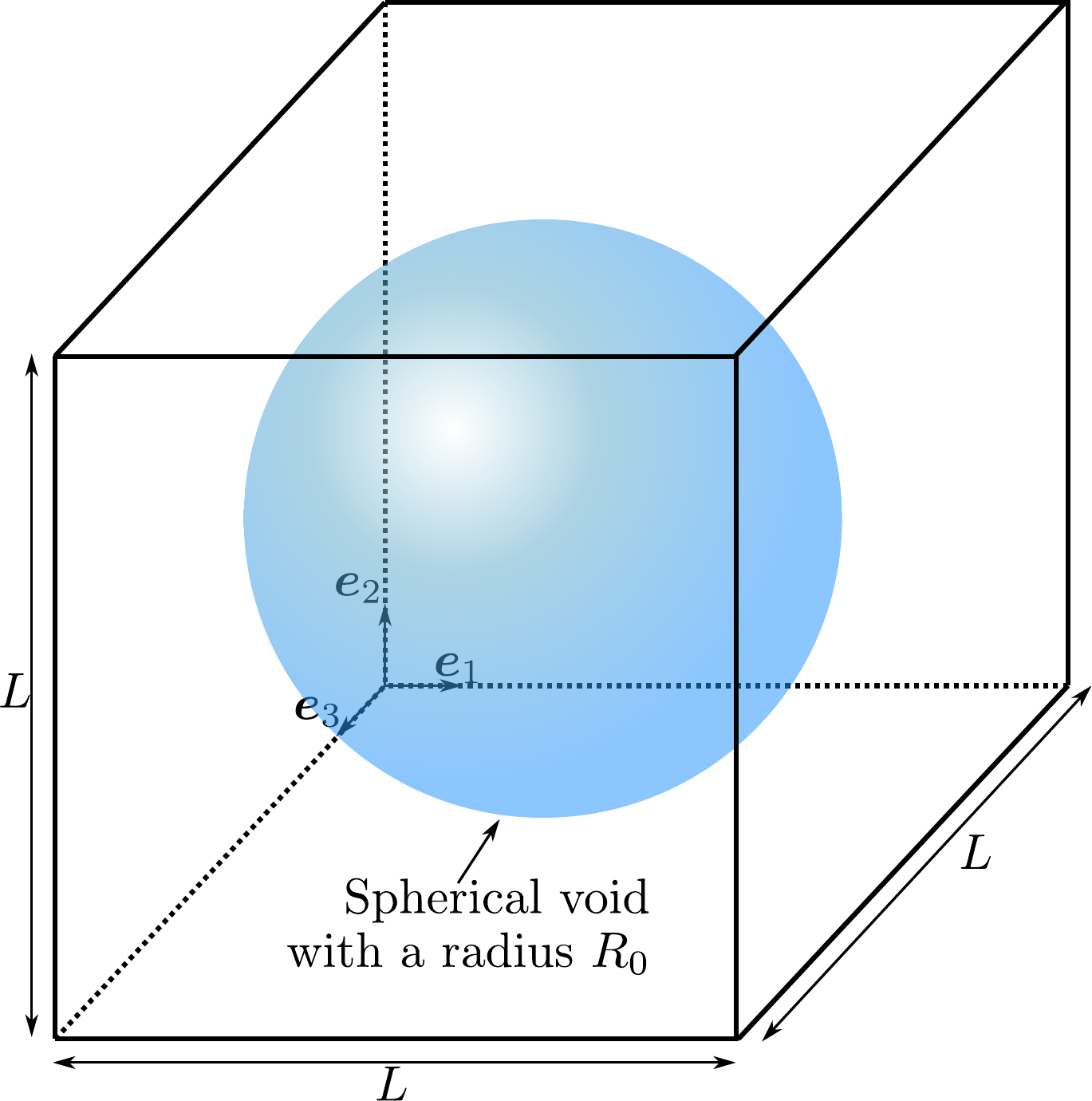}
	\caption{Illustration of a 3D cuboid unit-cell with a spherical void. The radius of the sphere used here is $R_0=0.4\,L$.}\label{fig:PorousBox}
\end{figure}

\begin{figure}[t]
	\centering{}
	\includegraphics[width=\textwidth]{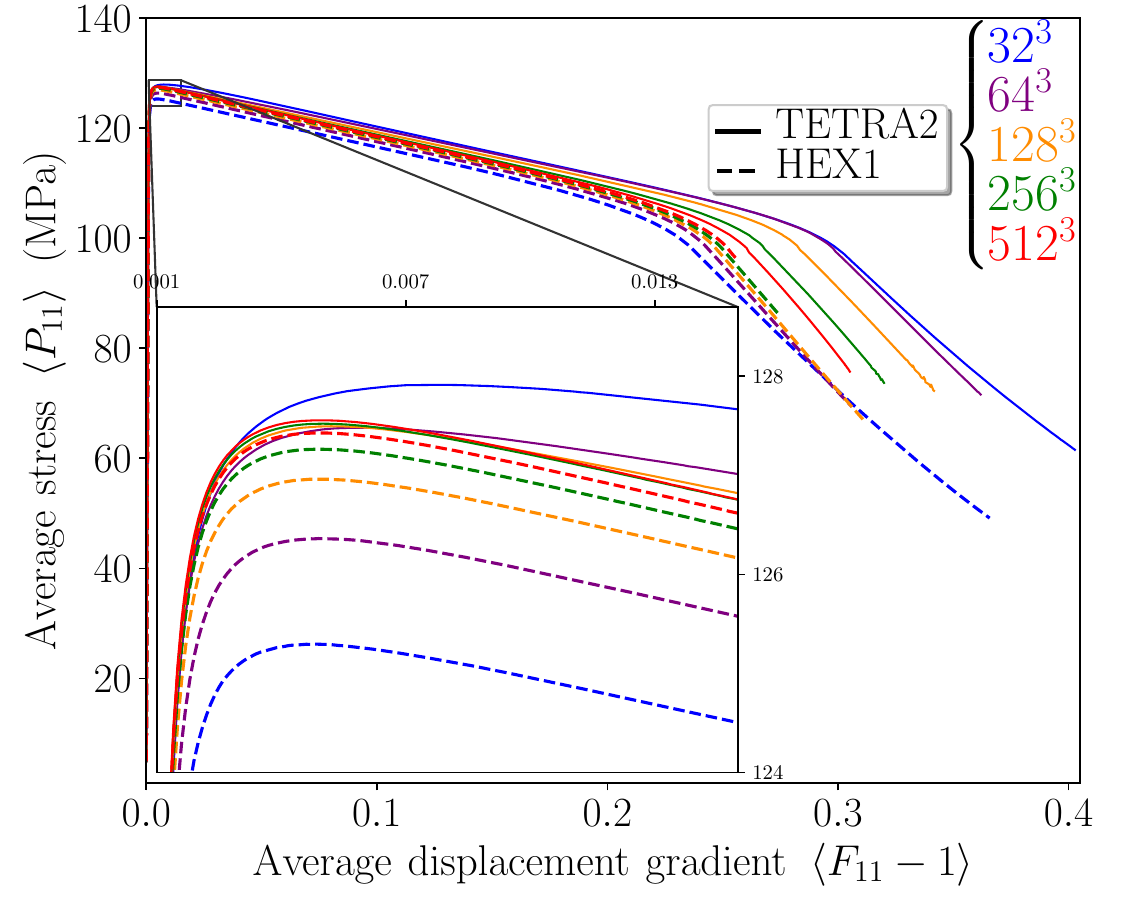}
	\caption{Macroscopic responses of a porous medium under the normal-mixed BCs of type 1 (NMBC1). Plain lines correspond to the TETRA2 scheme and dash lines to the HEX1 scheme. Increasing discretization is employed for both, TETRA2 and HEX1, finite difference schemes. A given color corresponds to a specific spatial discretization. Simulations are performed with the discrete Green operator based on the pre-conditioner $\mathbb{B}_0: \protect\td{\nabla u^f}$.  }
	\label{fig:Stress_Strain_Gop4_TETRA2_vs_HEX1_Discretization}
\end{figure}

Following the examples in elasticity, a perfect isotropic plasticity behavior in a finite transformation framework is considered in this section in order to study the deformation of a porous medium (see Fig.~\ref{fig:PorousBox}). FFT-based simulations of porous media in full periodic conditions have been reported in the literature  \citep{cruzado_effect_2026,hure_homogenized_2026} using \cite{amitex_fftp}, to reproduce cavity growth and coalescence. These kind of numerical simulations are quite challenging. We discuss here, in a non-periodic case, the effect of the finite difference scheme, of the discrete Green operator and of the spatial discretization. For the simulations in this section only, a tolerance of $\epsilon = 10^{-4}$ is employed for the equilibrium condition.

For the purpose, the small strain version of the classical von~Mises isotropic $J_2$ plasticity formulation is extended to finite transformation based on the logarithmic formalism \citep[Miehe–Apel–Lambrecht framework,][]{miehe_anisotropic_2002}. This requires the introduction of the Hencky strain tensor as
\begin{equation}
	\td{H} = \frac{1}{2} \log \left(\td{F}^{T}\td{F}\right)
\end{equation} Similarly to small strain framework an additive decomposition into an elastic part $\td{H}^e$ and a plastic contribution $\td{H}^p$ is considered for the Hencky strain tensor
\begin{equation}
	\td{H} = \td{H}^e + \td{H}^p
\end{equation} The Cauchy stress is written as
\begin{equation}
	\td{\sigma} = 2\mu \, \mathrm{dev}\left(\td{H}^e\right) + K \, \mathrm{tr}(\td{H}^e) \, \td{1}  
\end{equation} with $K = \lambda + 2 \mu /3 $. Perfect plasticity is accounted for through a yield surface defined as  
\begin{equation}
	f(\td{\sigma}) = \sigma_{eq} - \sigma_0 = 0
\end{equation} where $\sigma_{eq}$ is the equivalent stress written as
\begin{equation}
	\sigma_{eq} = \displaystyle \sqrt{\frac{3}{2} \, \td{s}:\td{s}} \qquad ; \qquad \td{s} = \mathrm{dev}\left(\td{\sigma}\right)
\end{equation} 
The normality flow rule is given by 
\begin{equation}
	\dot{\td{H}^p} = \dot{p} \pd{f}{\td{\sigma}} \qquad ; \qquad \dot{p} \geq 0
\end{equation} with $\dot{p}$ the rate of the cumulative plastic strain. The behavior law is implemented with MFront \citep{helfer_introducing_2015}. We use Young modulus $E=200 \, \mathrm{GPa}$, a Poisson ratio $\nu=0.3$ (\emph{i.e.,} $\lambda\approx 115 \, \mathrm{GPa}$, $\mu \approx 76 \, \mathrm{GPa}$) and the initial yield stress $\sigma_0 = 200\,\mathrm{MPa}$. The void is represented by an elastic material with null Lamé coefficients. The pre-conditioner parameters $\lambda_0$ and $\mu_0$ have to be chosen. To the best knowledge of the authors, no obvious optimized values for $\lambda_0$ and $\mu_0$ are demonstrated in the literature for non-periodic BCs in plasticity and finite transformation. Classical Moulinec and Suquet \citep{Moulinec1998} choice would result in $\beta_0 = 1/2 \left(\max (\beta) + \min (\beta)\right)$ with $\beta \in \{\lambda,\mu\}$. Nevertheless, for the present simulations, based on preliminary tests (not shown here), we choose the fixed value of $\beta_0=\max \left(\beta\right)$ (\emph{i.e.,} $\lambda_0 = \max (\lambda)$ and $\mu_0 = \max (\mu)$) for all three pre-conditioners.

In the present non-periodic case, a homogenized-like tension which satisfies the \textquotedblleft normal-mixed\textquotedblright{} condition in Eq.~\eqref{eq:SGO_BCs} is used. This gives the possibility to employ all three types of discrete Green operators. The BCs consist of normal-mixed BCs of type 1 (NMBC1) with Dirichlet conditions on the normal displacement components and Neumann conditions of the tangential stress, \emph{i.e.,}
	\begin{equation}
		\text{NMBC1} \,\,
		\left\{
		\begin{array}{l}
			\bullet \text{ Dirichlet BCs in directions } q=i,  \quad u^f_i = 0 \text{ and } \\[0.5em] 
			\bullet \text{ Null Neumann BCs in directions } q \neq i, \quad P_{iq} = 0
		\end{array}
		\right.
		\label{eq:NMBC1}
	\end{equation} The loading is prescribed by imposing a volume-average of the gradient of displacement $\td{\nabla u^*} = \td{G}$ with the component $G_{11}$ controlled and the other components $G_{rs}$, $\{r,s\} \neq \{1,1\}$ adjusted at each iteration to obtain $\left<P_{rs}\right> = 0 \text{ if } \{r,s\} \neq \{1,1\} $.

Considering both finite difference schemes (HEX1 and TETRA2) and also various spatial discretizations, Fig.~\ref{fig:Stress_Strain_Gop4_TETRA2_vs_HEX1_Discretization} gives the evolution of the average stress component, $\left<P_{11}\right>$, as a function of the average of the displacement gradient component, $\left<F_{11}-1\right>$. Overall, three regimes can be distinguished. The first regime consists of an elastic evolution up to an apparent yield stress $\sigma_y$. It is followed by a softening regime due to void growth with a significant irreversible deformation. A quasi-linear decrease of the stress field from $\sigma_y$ to a macroscopic \textquotedblleft coalescence-like\textquotedblright{} stress $\sigma_{c}$ is observed. The third regime characterizes an instability due to a strong plastic strain localization around the void, with a sudden drop of the macroscopic stress post-$\sigma_{c}$. Going forward with the HEX1 scheme, increasing the discretization (from $32^3$ to $512^3$ voxels) leads to a visible increase in the apparent yield stress $\sigma_y$ (see the zoom part in Fig.~\ref{fig:Stress_Strain_Gop4_TETRA2_vs_HEX1_Discretization}). On the contrary with the TETRA2 scheme, the apparent yield stress $\sigma_y$ varies less with the discretization. Continuing with Fig.~\ref{fig:Stress_Strain_Gop4_TETRA2_vs_HEX1_Discretization}, we can see that both TETRA2 and HEX1 schemes, predict identical level of \textquotedblleft coalescence-like\textquotedblright{} stress $\sigma_{c}$ regardless of the discretization. It is possible to conclude that, with a reasonable discretization (\emph{e.g.,} $64^3$ voxels), acceptable approximations of $\sigma_y$ and $\sigma_{c}$ are obtained with the TETRA2 scheme. Furthermore, examining the deformation level corresponding to the beginning of the \textquotedblleft coalescence-like\textquotedblright{} transition, the TETRA2 scheme shows a higher sensitivity than the HEX1 scheme \emph{w.r.t.} the discretization. However, with an increasing discretization, both the TETRA2 and HEX1 schemes seem to converge towards a macroscopic response. The TETRA2 scheme has the advantage of allowing convergence up to larger deformations in the \textquotedblleft coalescence-like\textquotedblright{} regime in comparison with the HEX1 scheme.

\begin{figure}[t]
	\centering\subfloat[HEX1 case.]{
		\includegraphics[width=0.48\textwidth]{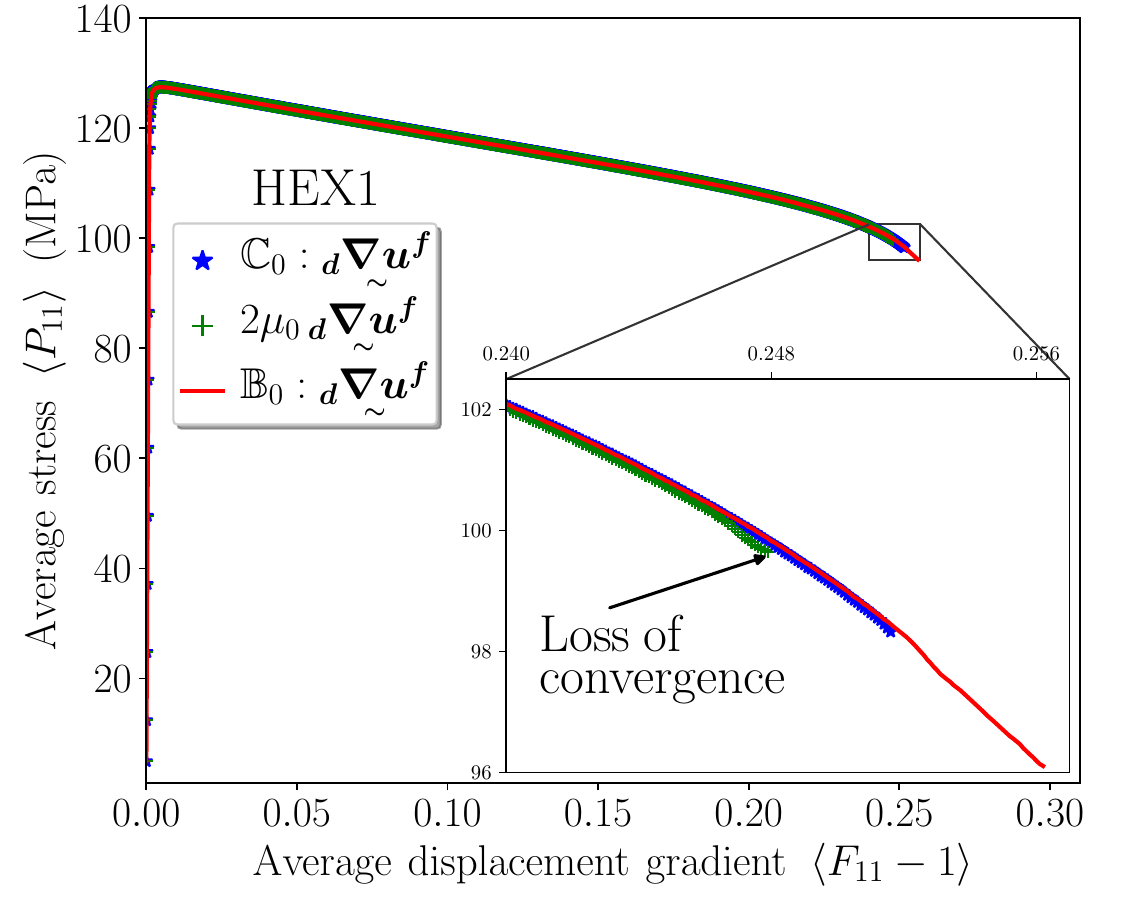}
		\label{fig:Stress_Strain_HEX1_Gop2_vs_Gop3_vs_Gop4}
	}
	\hfill{} \subfloat[TETRA2 case.]{
		\includegraphics[width=0.48\textwidth]{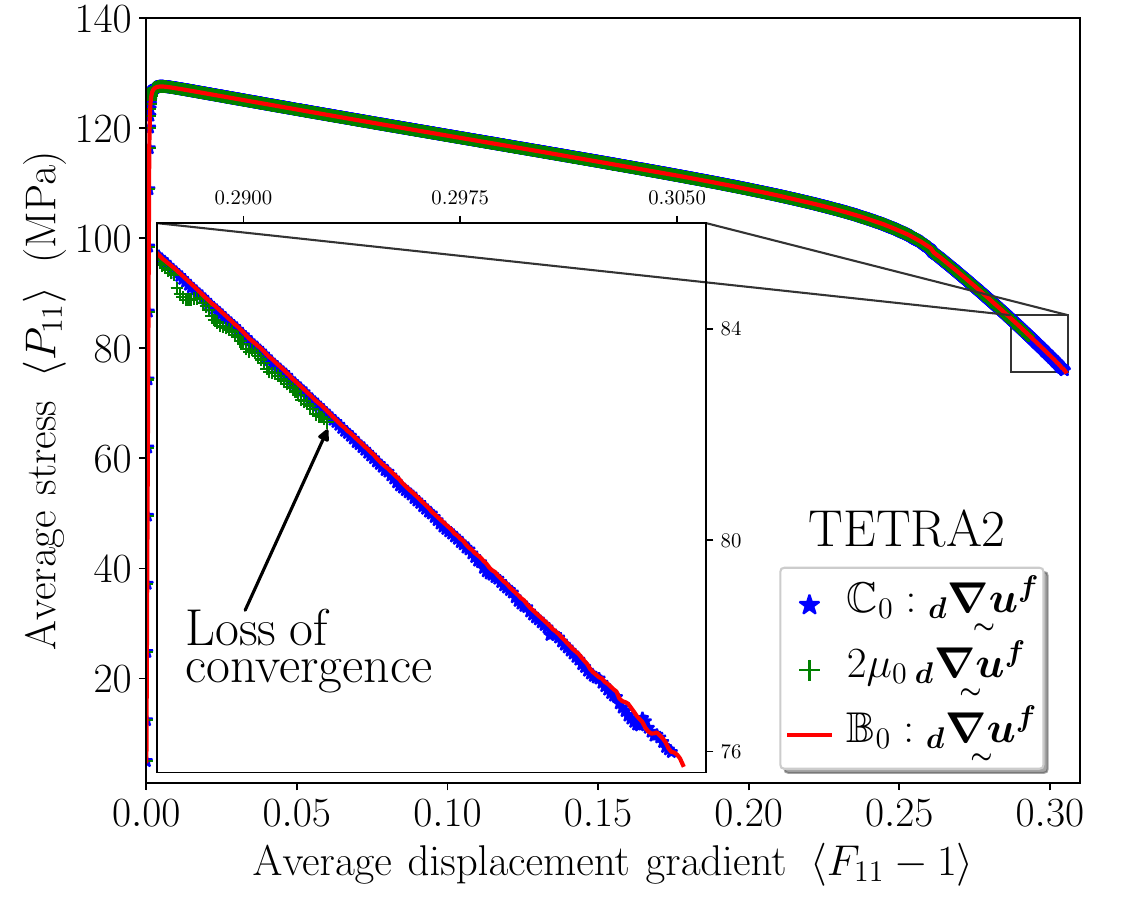}
		\label{fig:Stress_Strain_TETRA2_Gop2_vs_Gop3_vs_Gop4}
	}\caption{Macroscopic response curves depending on the choice of the pre-conditioner and the finite difference scheme. For the simulations, a discretization of $n_x\times n_y \times n_z = 512^3$ voxels is used.}\label{fig:Stress_Strain_HEX1_TETRA2_Gop2_vs_Gop3_vs_Gop4}
	
\end{figure}
\begin{table}[t]
	\centering{}
	\caption{Evolution of the total convergence iteration as function of the finite difference schemes and the choice of the discrete Green operators at the deformation value $F_{11}-1=0.2479$. For the simulations, a discretization of $n_x\times n_y \times n_z = 512^3$ voxels is used.}
	\begin{tabularx}{\linewidth}{|Y|Y|Y|}
		\hline
		Pre-conditioner to build the discrete Green operator  & Finite difference scheme &  Total of iterations \\
		\hline
		\multirow{2}{*}{$ \mathbb{C}_0 :\td{{}_d\nabla u^f} $} & HEX1 & $8809$ \\[0.2em]
		\cline{2-3} 
		& TETRA2 &  $5894$ \\[0.2em]
		\hline
		\multirow{2}{*}{$2 \mu_0\,\td{{}_d\nabla u^f} $} & HEX1 &  $12643$ \\[0.2em]
		\cline{2-3} 
		& TETRA2 &  $8701$ \\[0.2em]
		\hline
		\multirow{2}{*}{$\mathbb{B}_0 :\td{{}_d\nabla u^f} $} & HEX1 & $8879$ \\[0.2em]
		\cline{2-3} 
		& TETRA2 &  $6468$ \\[0.2em]
		\hline
	\end{tabularx}
	\label{Tab:FDS_Gop_It}
\end{table}

Fig.~\ref{fig:Stress_Strain_HEX1_TETRA2_Gop2_vs_Gop3_vs_Gop4} presents for a given discretization, \emph{i.e.,} $n_x\times n_y \times n_z = 512^3$ voxels, the comparison of the three presented pre-conditioners (leading to the choice of discrete Green operator) considering the HEX1 and TETRA2 finite difference schemes. Globally, the invariance of the macroscopic responses is demonstrated with the selection of the pre-conditioners. More precisely, the simulations with the pre-conditioner $2 \mu_0\td{\nabla u^f}$ failed to converge first, then follows the ones with the pre-conditioner $\mathbb{C}_0: \td{\nabla u^f}$ and finally the simulations with the pre-conditioner $\mathbb{B}_0: \td{\nabla u^f}$. The latter pre-conditioner helps to reach higher level of deformations. Tab. \ref{Tab:FDS_Gop_It} summarizes the consequence of the choice of a discrete Green operator based on the finite difference scheme. Focusing on a deformation range where all three discrete Green operators converge, we can say that, for a given pre-conditioner, in order to reach a desired deformation, the TETRA2 scheme converges faster than the HEX1 scheme (approximatively $1.4$ factor in terms of the iteration numbers). Based on a given finite difference scheme, the pre-conditioners $\mathbb{C}_0: \td{\nabla u^f}$, $\mathbb{B}_0: \td{\nabla u^f}$ and $2 \mu_0\td{\nabla u^f}$ converge respectively one faster than the other. We recall that the pre-conditioner $\mathbb{C}_0: \td{\nabla u^f}$ can only be used for BCs that are compatible with \textquotedblleft normal-mixed\textquotedblright{} BCs in Eq.~\eqref{eq:SGO_BCs}. 

To sum up, these results demonstrate the robustness of the novel FFT-based solver in handling complex simulations (non-periodic, infinite contrast, perfect plasticity, finite transformation). The TETRA2 scheme seems to demonstrate greater robustness than HEX1, in predicting the apparent yield stress (at small strain). However, the TETRA scheme exhibits higher discretization sensitivity in the \textquotedblleft coalescence-like\textquotedblright{} strain level than the HEX1 scheme. A deeper analysis of the performance of the TETRA2 scheme and an optimization of the parameters $\lambda_0$ and $\mu_0$ are of interests but are beyond the scope of the present paper. It would also be interesting in a future work to enhance this study by including multiple voids with different shapes and apply the composite voxels technique \citep{gelebart_filtering_2015,kabel_use_2015} specially in TETRA2 case.

\subsection{Single crystal L-beam under torsion-bending loading \label{sec:Singlecrystal}}

As it is thoroughly documented in the literature, the classical FFT-based solver has proven to be a valuable modeling technique for periodic single crystal and polycrystal materials \citep{zecevic_new_2022,cocke_implementation_2023}. It is worthwhile to investigate the behavior of such anisotropic materials within the novel non-periodic framework. We aim to discuss observations within continuum single crystal plasticity, building on experimental studies such as shear-compression \citep{mu_micro-pillar_2014}, torsion-bending \citep{zhang_toward_2023}, non-proportional orthogonal bending \citep{zhang_non-proportional_2026}. As part of the present work, these complex experimental-like simulations have been successfully tested. For conciseness, only the L-beam torsion-bending case inspired from the work of \cite{zhang_toward_2023}, is reported in this paper. Note that, we do not attempt here a direct point-to-point comparison with the experimental macroscopic curves. Such a comparison would require the utilization of higher-order type models (see \emph{e.g.,} a strain gradient crystal plasticity \citep[][and references therein]{amouzou-adoun_advanced_2024,amouzou-adoun_enhanced_2025,mukherjee_quantitative_2026} or a Cosserat continuum framework \citep{kucharski_size_2024}). These types of gradient-based plasticity theories are out of the scope of this paper and will be the subject of future work. A local crystal viscoplasticity model, which is based on the evolution of dislocation densities \citep{roters_overview_2010,scherer_deformation_2026}, is employed for the purpose of the present analysis. 

The finite transformation crystal plasticity model used in \cite{scherer_deformation_2026}, is briefly recalled here. The deformation gradient, $\td{F} = \td{1} + \td{\nabla u}$, is decomposed in a product of an elastic and a plastic parts \citep{mandel_equations_1973}, respectively noted $\td{F}^e$ and $\td{F}^p$
\begin{equation}
	\td{F} = \td{F}^{e}\,\td{F}^{p} 
\end{equation} so that the Green-Lagrange elastic strain measure $\td{E}^e$ is written as  
\begin{equation}
\td{E}^e = \displaystyle \frac{1}{2} \left(\td{F}^{eT}\td{F}^e - \td{1}\right)
\end{equation} In order to account for plasticity on each slip system, the total velocity gradient $\td{L}=\dot{\td{F}} \td{F}^{-1}$ is additively decomposed as 
\begin{equation}
	\td{L} = \td{L}^e + \td{F}^e \td{L}^p {\td{F}^e}^{-1}
\end{equation}
with 
\begin{equation}
	 \td{L}^{e} = \dot{\td{F}^{e}} {\td{F}^{e}}^{-1} \qquad;\qquad \td{L}^{p} = \dot{\td{F}^{p}}  {\td{F}^{p}}^{-1} = \displaystyle \sum_{s=1}^{M_{ss}} \dot{\gamma}^s \ten{m}^s \otimes \ten{n}^s
\end{equation} where $M_{ss}$ is the total number of slip systems, $\dot{\gamma}^s$ denotes the plastic slip rate on the $s$th slip system, $\ten{m}^s$ and $\ten{n}^s$ are respectively the slip direction and slip plane normal of the  $s$th slip system. Face-centered cubic (FCC) copper material is considered so that $M_{ss}=12$ with $\langle 110 \rangle\{111\}$ slip systems. The constitutive law is expressed here by the second Piola–Kirchoff stress tensor
\begin{equation}
	\td{\Pi} = \mathbb{C}:\td{E}^{e} 
\end{equation} It follows that the Mandel stress tensor is given by
\begin{equation}
	\td{M} = {\td{F}^{e}}^{T} \, \td{F}^{e} \, \td{\Pi}
\end{equation}
For each slip system, plasticity is activated through a yield function 
\begin{equation}
	f^s(\td{M}) = \rvert \tau^s\lvert - \tau^s_c
\end{equation} where $\tau^s$, is the resolved shear stress on the $s$th slip system and is related to the Mandel stress based on the relation  
\begin{equation}
	\tau^s = \td{M} : \left(\ten{m}^s \otimes \ten{n}^s\right)
\end{equation} and $\tau_c^s$ denotes the critical resolved shear stress (CRSS) for the $s$th slip system, accounting for the yielding stress and a hardening. $\tau_c^s$ evolves following a dislocation density-based hardening 
\begin{equation}
	 \tau^s_c = \tau^s_0 + \mu_h \, b \sqrt{ \displaystyle  \sum_{u=1}^{M_{ss}} a^{su}\rho^u}
\end{equation} with $\tau^s_0$ the the thermal component of critical resolved shear stress of the $s$th slip system, $\mu_h$ a hardening parameter, $b$ the Burgers vector magnitude, $a^{su}$ the interaction coefficient between dislocations on slip systems $s$ and $u$ and $\rho^u$ the dislocation density on the slip system $u$. In general, the same value of $\tau^s_0$ is used for all the slip systems, \emph{i.e.,} $\tau^s_0 = \tau_0$. The rate of the dislocation density on a given slip system $s$ accounts for the multiplication and the annihilation through the formula 
\begin{equation}
	\dot{\rho}^s = \displaystyle \frac{\lvert \dot{\gamma}^s\rvert}{b} \left( \displaystyle \frac{\sqrt{\displaystyle \sum_{u\neq s} \rho^u}}{\kappa} - yb \rho^s \right)
\end{equation} with $\kappa$ and $y$ dimensionless coefficients which respectively characterize the dislocation mean free path and the dynamic recovery. A rate-dependent flow rule evolution is used
\begin{equation}
	\dot{\gamma}^s = \mathrm{sign}(\tau^s) \left(\max \left(\displaystyle \frac{f^s}{K_v},0\right)\right)^{n_{vp}}
\end{equation} where the parameters $K_v$ and $n_{vp}$ govern viscosity. These constitutive equations have been implemented in the MFront code generator \citep{helfer_introducing_2015}. In this example, we choose the following values: for the self hardening $a^{ss} = 0.1$, for the latent hardening $a^{su} = 0.12, s\neq u$, for the viscosity parameters $n_{vp}=30$, $K_v=0.1\,\mathrm{MPa.s}^{1/n_{vp}}$ (note that the choice of the viscosity parameters leads a quasi-rate-independent behavior), for the critical resolved shear stress $\tau_0=30\,\mathrm{MPa}$, for the hardening parameter $\mu_h=9400\,\mathrm{MPa}$, for the dislocation density evolution, $\kappa = 20$, $b=2.54 \,\si{\angstrom}$, $y=10$ and $\rho_0^{tot} = 10^{12}\,\mathrm{m^{-2}}$. The elastic moduli are $C_{11}=168\,\mathrm{GPa}$, $C_{12}=121\,\mathrm{GPa}$ and $C_{44}=75\,\mathrm{GPa}$.

\begin{figure}[t]
	\centering{}
	\subfloat[Example of copper L-beam in the initial configuration: experimental setup from \cite{zhang_toward_2023}.]{
		\includegraphics[width=0.4\textwidth,valign=c]{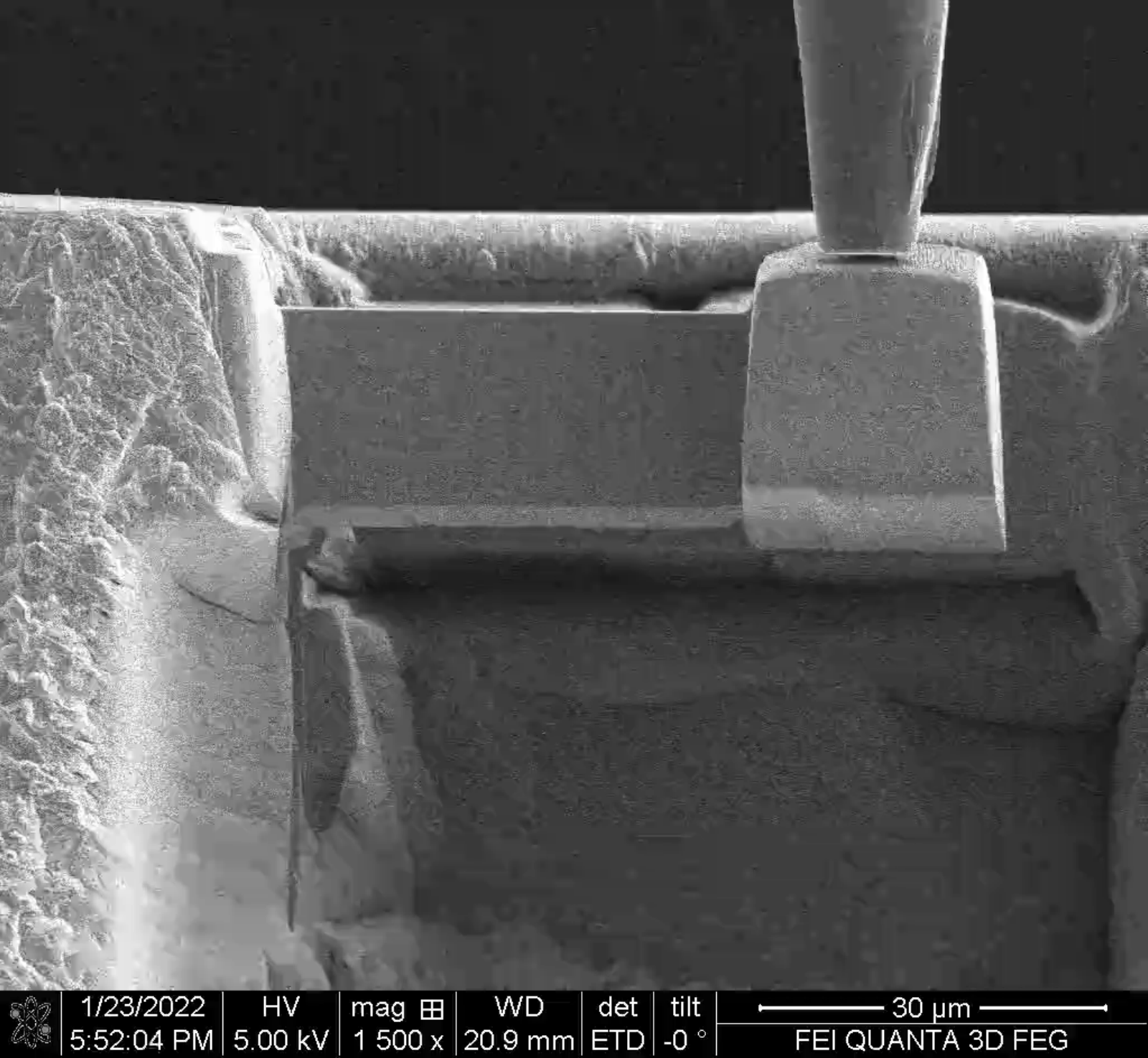}
		\label{fig:TorsionBendingExp}
	}
	\hfill{} \subfloat[Numerical setup of L-beam, the region in green is a void (represented by an elastic material with a null elastic moduli). A discretization with $n_1\times n_2 \times n_3 = 150\times50\times175$ cuboid voxels is used.]{
		\centering
		\includegraphics[width=0.5\textwidth,valign=c]{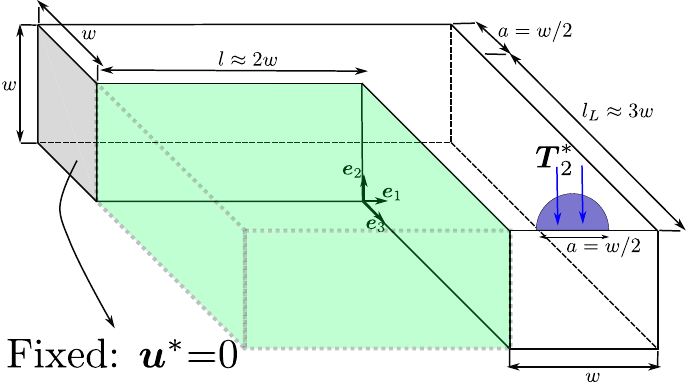}
		\label{fig:TorsionBendingSimu}
	}
	\caption{Torsion-bending setup (inspired from \cite{zhang_toward_2023}). Dimensions: $w = 20\,\mathrm{\mu m}$, $l \approx 2w$, $ l_L \approx 3w$.}
	\label{fig:TorsionBending}
\end{figure}
\begin{figure}[h!]
	\centering\subfloat[Macroscopic response curves depending on the finite difference scheme.]{
		\includegraphics[width=0.48\textwidth]{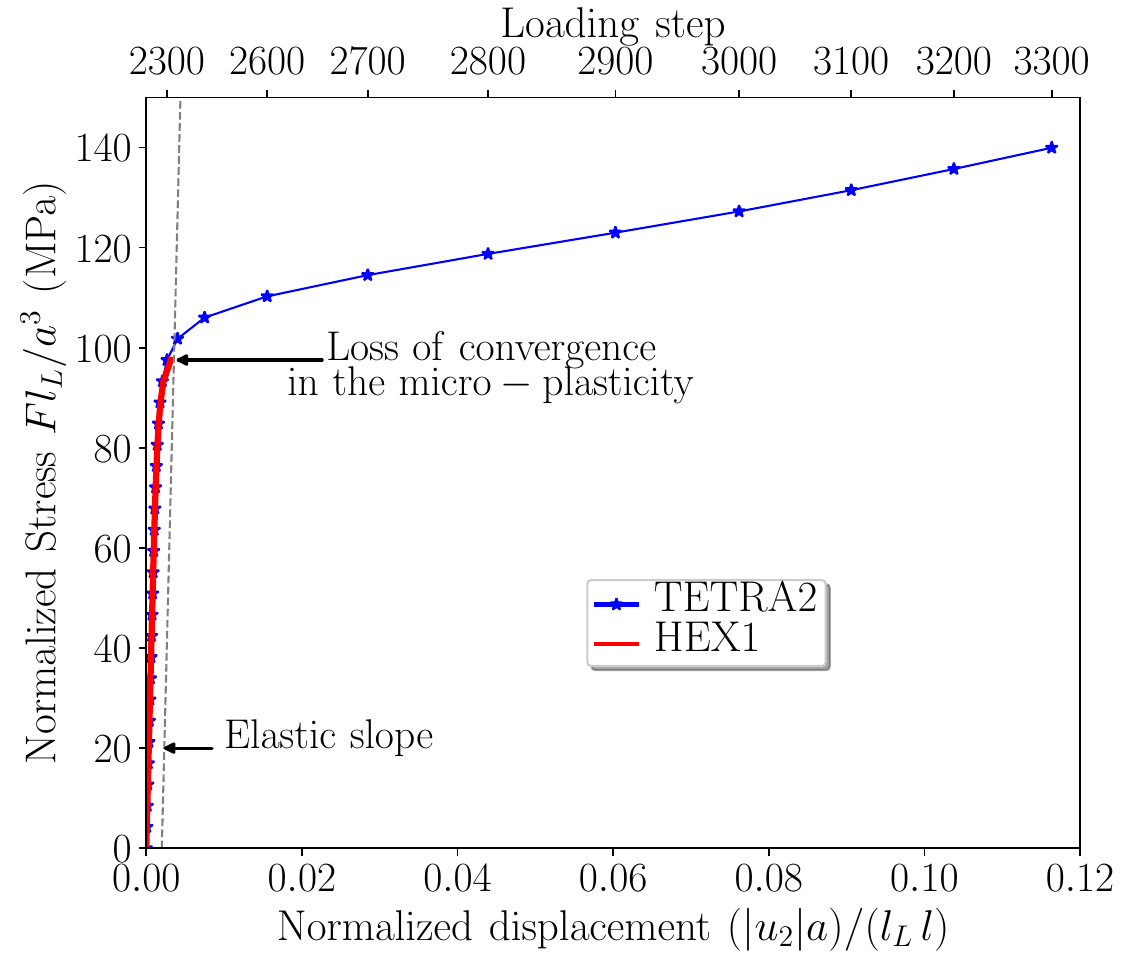}
		\label{fig:TorsionBendingresponses1}
	}
	\hfill{} \subfloat[Convergence during the loading depending on the finite difference scheme.]{
		\includegraphics[width=0.48\textwidth]{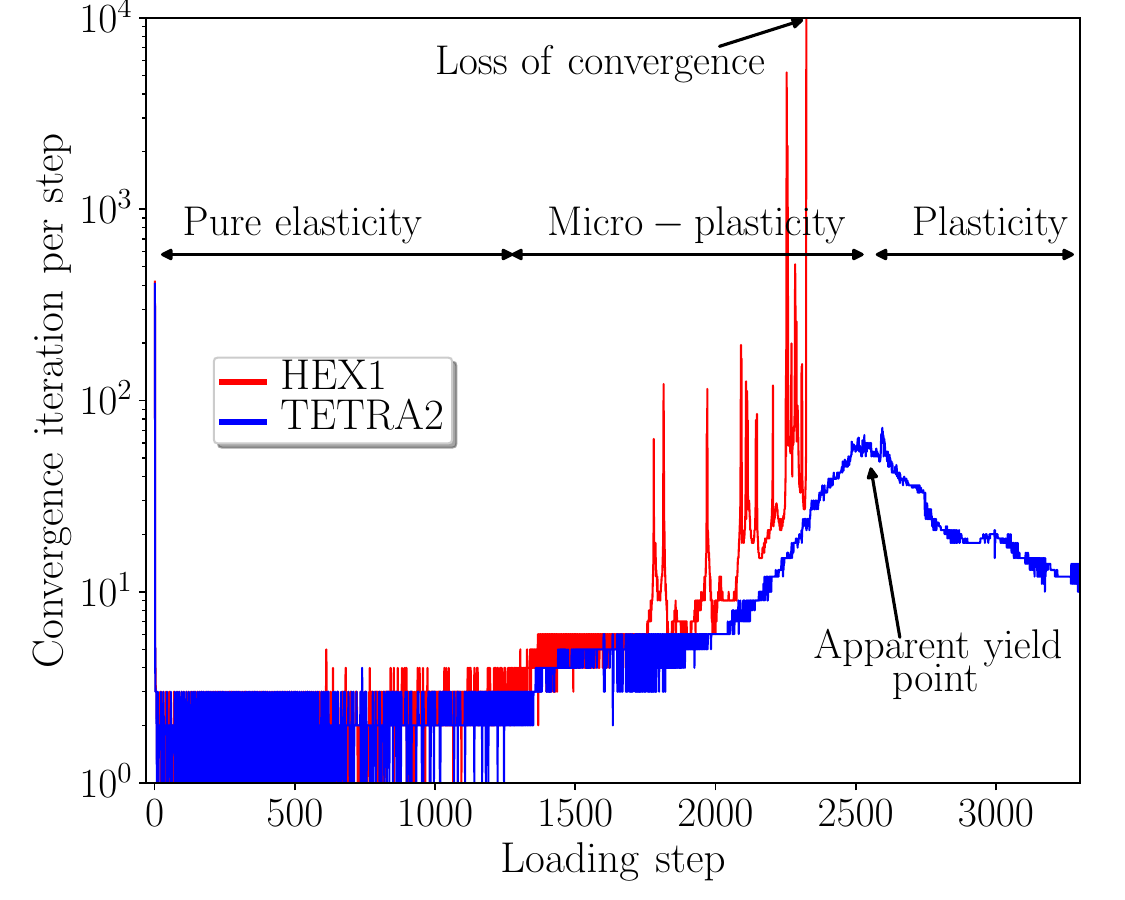}
		\label{fig:TorsionBendingresponses2}
	}\caption{Torsion-bending simulations.}\label{fig:TorsionBendingresponses}
	
\end{figure}

\begin{figure}[t]
	\centering
	\includegraphics[width=\textwidth]{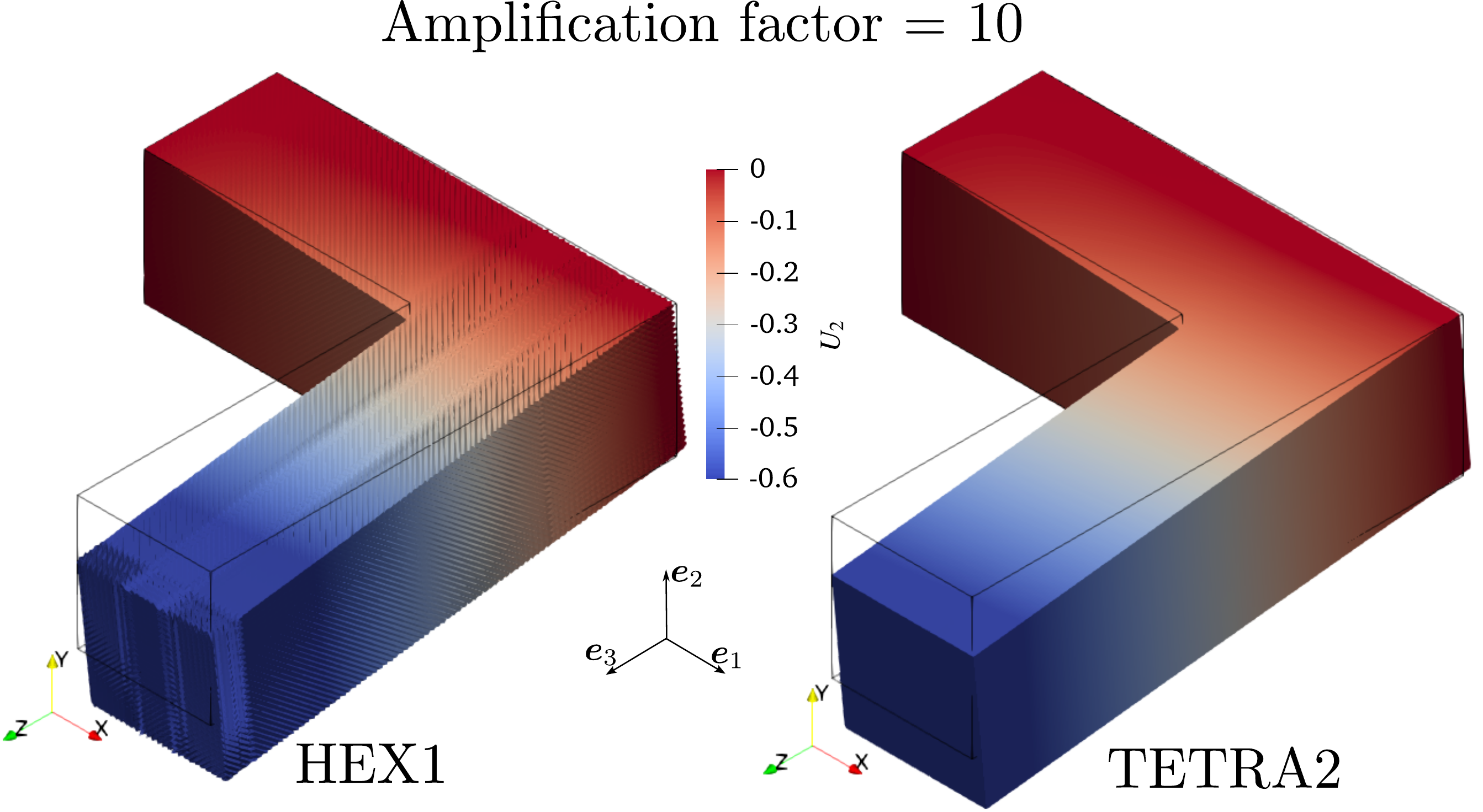}
	\caption{Comparison of the finite difference schemes: displacement $u_2$ around the apparent yield stress (loading step = 2300). The deformed shape are amplified here by a factor 10. A discretization of $n_1\times n_2 \times n_3 = 150\times50\times175$ voxels is used.}
	\label{fig:TorsionBendingdeformedshape}
\end{figure}
\begin{figure}[h!]
	\centering{}
	\subfloat[Numerical displacement $u_2$ field at the end load (loading step = 3300) of copper L-beam: the deformed shape is shown without any amplification. A discretization of $n_1\times n_2 \times n_3 = 150\times50\times175$ voxels is used.]{
		\includegraphics[width=0.49\textwidth,valign=c]{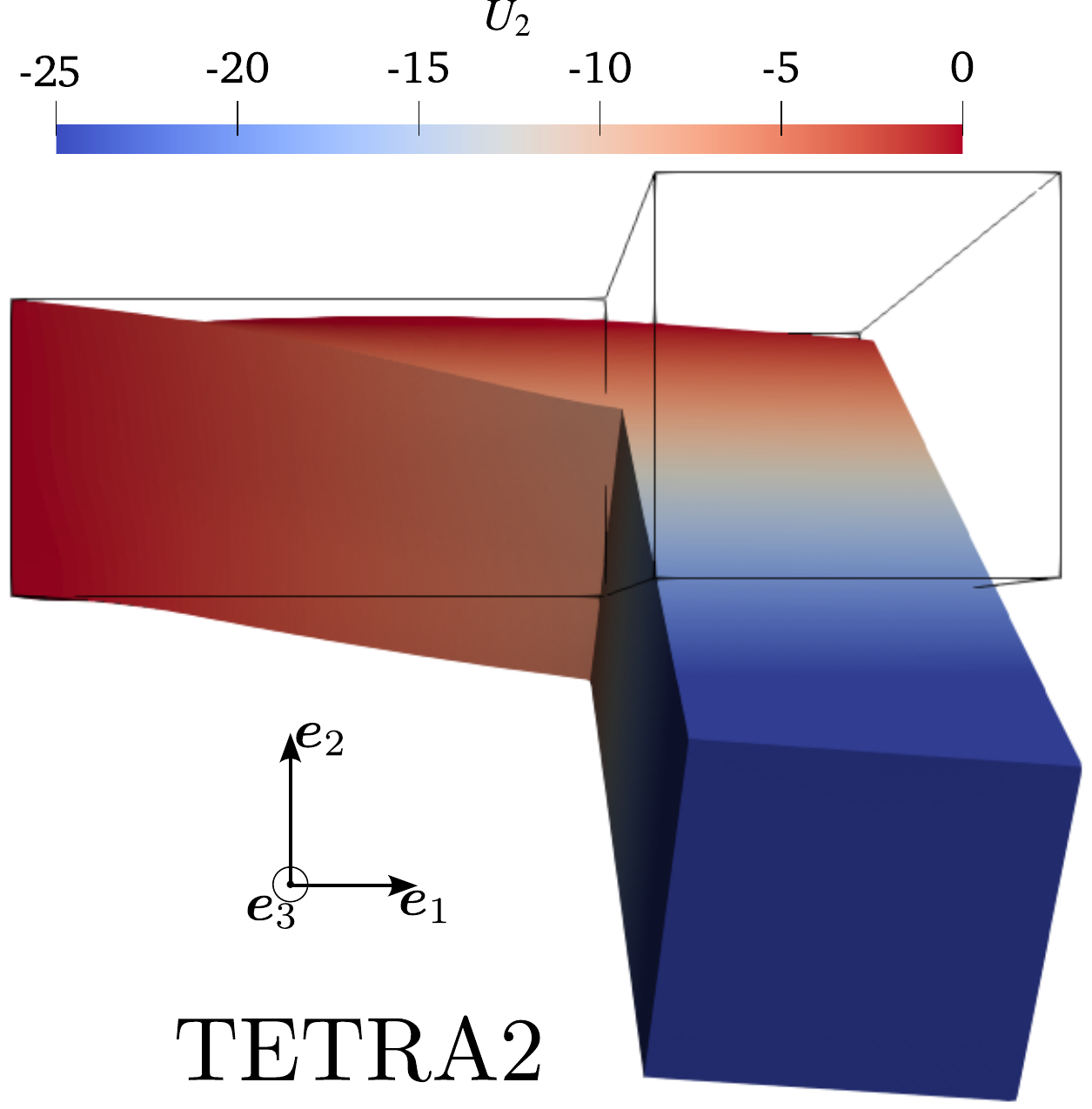}
		\label{fig:TorsionBendingSimu2}
	}
	\subfloat[Example of copper L-beam: experimental setup from \cite{zhang_toward_2023} showing the deformed shape as consequence of the actuator.]{
		\includegraphics[width=0.49\textwidth,valign=c]{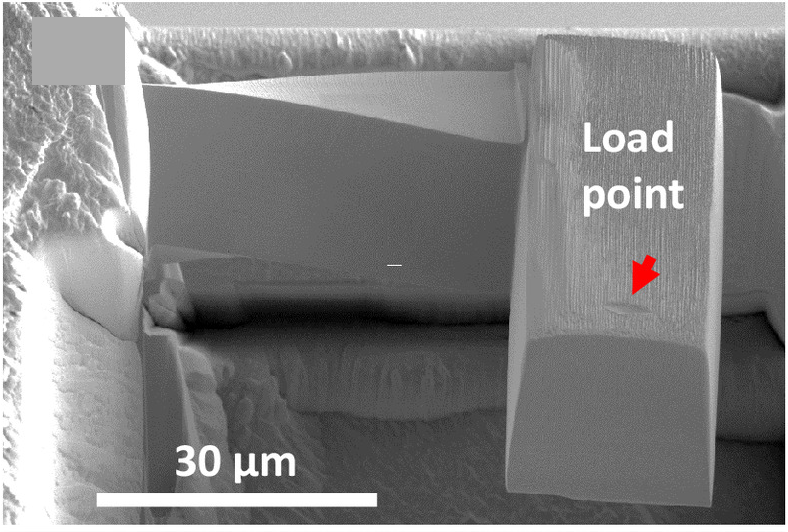}
		\label{fig:TorsionBendingExp2}
	}
	\caption{Deformed shape under the torsion-bending loading.}
	\label{fig:TorsionBending2}
\end{figure}

\cite{zhang_toward_2023} designed a torsion-bending experiment to study size effects on a $\left< 111\right>$ oriented single crystal copper L-beam. Fig.~\ref{fig:TorsionBendingExp} presents an experimental setup example of a L-beam under a force-controlled actuator (applied force $F$). For the numerical simulations, the setup in Fig.~\ref{fig:TorsionBendingSimu} is employed. Null Dirichlet BCs are applied on the left face $S_{10}$ (normal to the axis $\ten{e}_1$) to mimic embedding. A Neumann BC is applied on the top face $S_{21}$ (normal to the axis $\ten{e}_2$) with a non-null stress vector $\ten{T^*}=-T^*_2 \ten{e}_2$, in a restricted half-circle region (\emph{i.e.,} an applied force $F = T^*_2 \,A_{area}$ with the actuator area $A_{area}=\pi a^2/8$, see Fig.~\ref{fig:TorsionBendingSimu}). We applied a maximum level of stress of $T^*_{2\,max}=60\,\mathrm{MPa}$ in $3300$ loading steps. Furthermore, in order to have a unit-cell in a parallelepiped shape, the region outside the interested L-beam arm is filled with a void (\emph{i.e.,} an elastic material with null elastic moduli, see green part in Fig.~\ref{fig:TorsionBendingSimu}). 

Unlike in the previous section, $\mathbb{C}_0$-discrete Green operator can not be used for the present torsion-bending loading. In the context of such complex simulations, the optimal choice of pre-conditioner parameter(s), $\mu_0$ (and $\lambda_0$ for the $\mathbb{B}_0$-discrete Green operator) remains an open question. A first option is the classical Moulinec and Suquet \citep{Moulinec1998} choice, \emph{i.e.,} $\beta_0 = 1/2 \left(\max (\beta) + \min (\beta)\right)$ with $\beta \in \{\lambda,\mu\}$. With this choice, the $2\mu_0$-discrete Green operator is outperformed by the $\mathbb{B}_0$-discrete Green operator. The same observation is made with regard to the choice $\beta_0 = \max (\beta) $ with $\beta \in \{\lambda,\mu\}$. In addition, the latter option results in a total of approximately $37500$ iterations for the entire loading, as opposed to a total of $45000$ iterations for the classical Moulinec and Suquet option, while employing the $\mathbb{B}_0$-discrete Green operator. For the purpose, we decide to continue all analysis with the  $\beta_0 = \max \left(\beta\right)$ and with the $\mathbb{B}_0$-discrete Green operator.

Fig.~\ref{fig:TorsionBendingresponses1} presents the macroscopic responses under the torsion-bending loading. Both finite difference schemes TETRA2 and HEX1 are used for comparison purpose. It turns out that TETRA2 scheme is once again more robust than HEX1 scheme. Actually the TETRA2 scheme allows to reach much larger plastic deformation level. This elastic-plastic response is qualitatively consistent with the experimental observations. Meanwhile, the HEX1 scheme fails to converge around the apparent yield stress. 

Fig.~\ref{fig:TorsionBendingresponses2} displays the corresponding number of iterations required at each time step to satisfy the convergence criterion. After the very first loading step and for the most part of the elastic regime, the convergence iteration decreases with the loading so that a convergence on the equilibrium is reached with less than $5$ iterations. The transition from elastic to plastic regime (micro-plasticity) leads to an increase in the required number of iterations for the convergence. It is shown that with HEX1 scheme a total of $10^4$ iterations has not been enough to reach the convergence around the apparent yield stress.With the TETRA2, the convergence degrades around the yield stress but is reached with less than $70$ iterations. In comparison with the HEX1 scheme, the TETRA2 scheme is demonstrated to produce a more stable evolution of the iterations from one loading step to another. Focusing on the results obtained with the TETRA2 scheme, within the plastic regime, a decrease of the convergence iteration is then observed (see Fig. Fig.~\ref{fig:TorsionBendingresponses2}).

The non-convergence of the HEX1 scheme at the apparent yield stress can be understood by visualizing the displacement field as shown in Fig.~\ref{fig:TorsionBendingdeformedshape} (please note that the deformed shape is amplified for this particular Fig.~\ref{fig:TorsionBendingdeformedshape} by a factor 10 only for visualization purpose). HEX1 leads to strong oscillations of the displacement field, preventing in the meantime a convergence as the loading increases. TETRA2 contributes to a smoother displacement field. Fig.~\ref{fig:TorsionBending2} presents the deformed shape (this time without any amplification) of the L-beam in the present simulation (see Fig.~\ref{fig:TorsionBendingSimu2}) and in the experiment (see Fig.~\ref{fig:TorsionBendingExp2}). In both cases, based on the geometrical design of the studied crystal, one can see that the arm in contact with the actuator is bent while the embedded arm is twisted, allowing to generate a strong strain gradient. A qualitatively good agreement with the experiment is observed. 

\begin{figure}[t]
	\centering\subfloat[Total simulation time as function of the number of processors.]{
		\includegraphics[width=0.48\textwidth,valign=c]{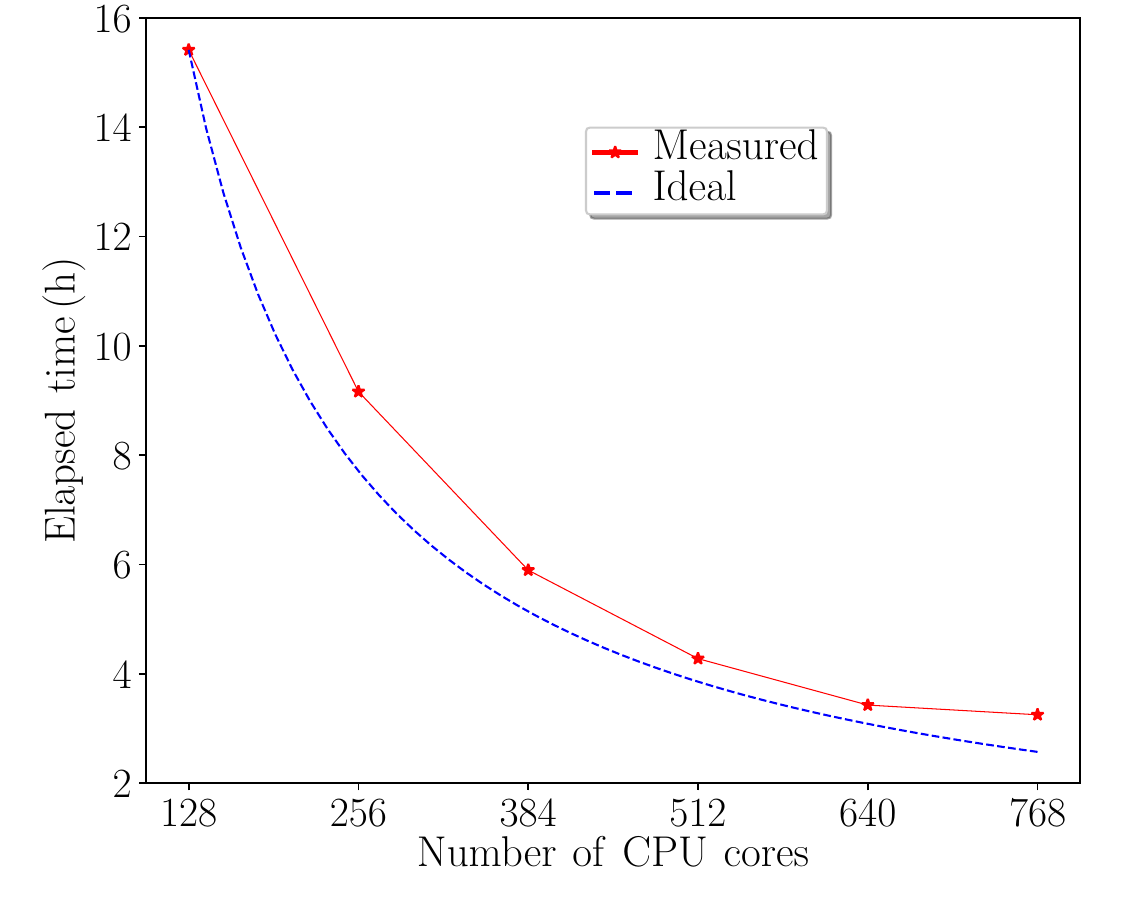}
		\label{fig:Raw_Perf_Time_vs_Nproc_L_beam}
	}
	\hfill{} \subfloat[Efficiency as function of the number of processors.]{
		\includegraphics[width=0.48\textwidth,valign=c]{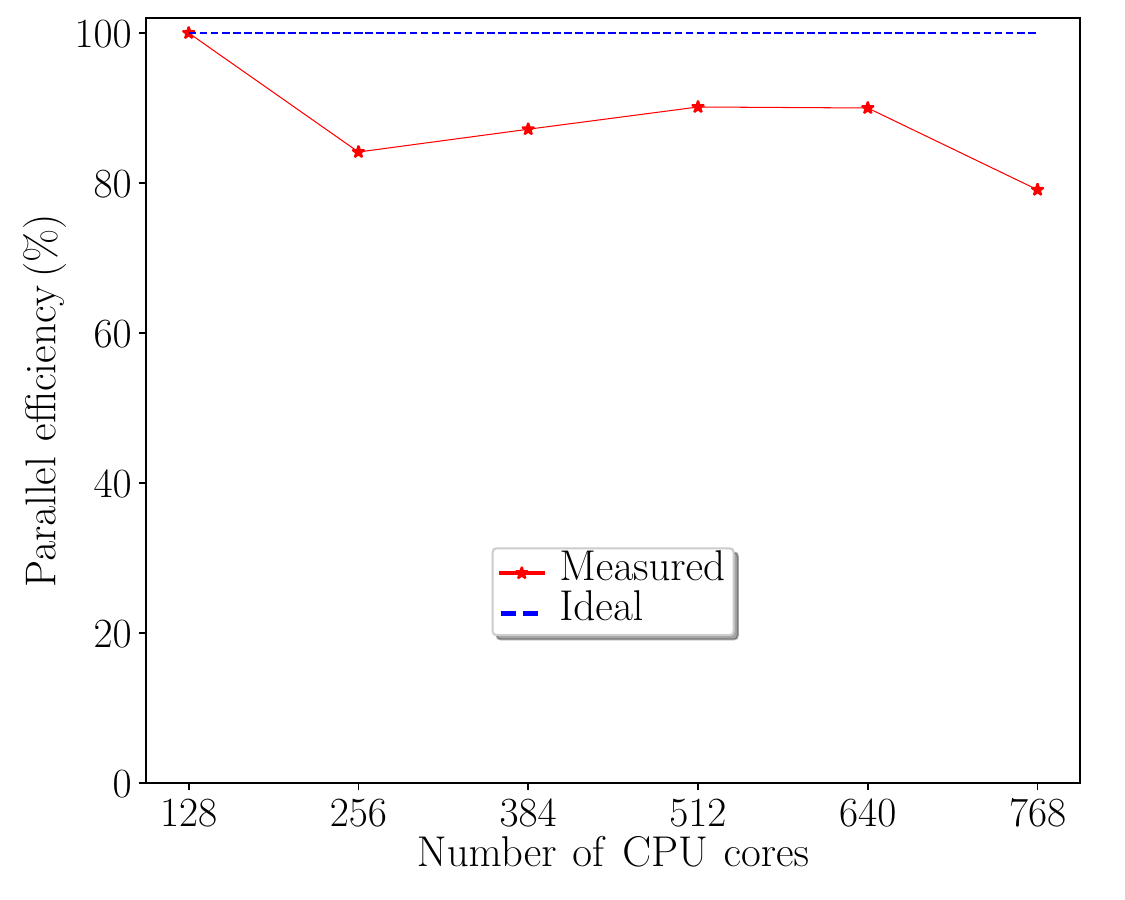}
		\label{fig:Efficiency_L_beam}
	}\caption{Parallel implementation outcomes (strong scaling) for the single crystal L-beam under torsion-bending loading with a discretization of $n_1\times n_2 \times n_3 = 150\times50\times175$ voxels. The result with the $128$ processors (available on a single cluster node) is used as reference for the expected ideal performances.}\label{fig:Parallel_L_beam}
\end{figure}

In addition to the macroscopic behavior, it is worthwhile to examine the performance of the present MPI-based solver, for such simulations (finite transformation crystal plasticity with quasi-rate-independent). It is well known that these types of simulations are notably time-consuming (CPU time). Strong scalability tests are performed for the given discretization of $n_1\times n_2 \times n_3 = 150\times50\times175$ voxels. The simulations are conducted on an in-house cluster that possesses $10$ compute nodes with each node having the following characteristics: $512\,\mathrm{Go}$ DDR5-RAM and $2$ sockets with $64$ cores per socket (AMD-EPYC9554) running at up to $3.1\,\mathrm{GHz}$ frequency (\emph{i.e.,} a total of $128$ cores per node). Fig.~\ref{fig:Raw_Perf_Time_vs_Nproc_L_beam} shows the raw computation time $T_{\mathrm{simu}}$ as function of the number of CPU cores $N_{\mathrm{CPUs}}$ (in practice, we increase the number of compute nodes from $1$ to $6$). Considering the result with $N_{\mathrm{CPUs}}^{\mathrm{ref}}=128$ CPU cores (\emph{i.e.,} a single node) as reference, the measured simulation time is compared to the ideal expected computation time, $\dfrac{N_{\mathrm{CPUs}}^{\mathrm{ref}}\times T_{\mathrm{simu}}(N_{\mathrm{CPUs}}^{\mathrm{ref}})}{N_{\mathrm{CPUs}}}$. The measured time follows closely the ideal time. A significant decrease of the simulation time is observed. From $128$ to $640$ CPU cores, the simulation time drops from approximately from $16$ hours to $3.4$ hours while from $640$ processors to $768$ processors, the simulation time goes from $3.4$ hours to $3.2$ hours. Fig.~\ref{fig:Efficiency_L_beam} presents these results in terms of the evaluation of the parallel efficiency, $100\dfrac{N_{\mathrm{CPUs}}^{\mathrm{ref}}\times T_{\mathrm{simu}}(N_{\mathrm{CPUs}}^{\mathrm{ref}})}{N_{\mathrm{CPUs}} \times T_{\mathrm{simu}}(N_{\mathrm{CPUs}})}$. We obtain an excellent strong scaling maintaining approximately $90\,\%$ efficiency up to $640$ CPU cores. A drop in efficiency to $79\,\%$ is observed at $768$ processors. This could be the consequence of load imbalance inherent to the heterogeneous plastic strain distribution within the unit-cell. For the moment, the 2DECOMP\&FFT library and \emph{AMITEX$^\star$} consequently, only accounts for 2D pencils of fixed size during the simulation and load balance is not yet available. This point could be a future prospect for both 2DECOMP\&FFT library and \emph{AMITEX$^\star$} code.

\section{Conclusion and recommendations \label{sec:conclusion}}

In this work, the FFT-based solver classically used to solve periodic mechanical problems \citep{Moulinec1994,Moulinec1998}, is naturally extended to account for general (non-periodic and/or periodic) BCs, non-linear behaviors, small and finite transformations, within a parallel programming environment. The new developments are implemented within the \emph{AMITEX$^\star$} code, which is a new version (in progress) of the open-source \emph{AMITEX\_FTTP} code \citep{amitex_fftp}. These are based on new upgrades of the 2DECOMP\&FFT library \citep{li_2decompfft_nodate,rolfo_2decompfft_2023}, to account for discrete trigonometric transforms (DTTs). 

The mechanical problem (small or finite transformation frameworks for Cauchy medium local theories) is reformulated with the fluctuation vector displacement as unknown. Leveraging between the link in the non-periodic BCs and the symmetry extensions of the fluctuation displacement, DTTs are introduced in a displacement-based accelerated fixed-point algorithm. The symmetry extensions of the concerned fields allow to build \textquotedblleft virtual\textquotedblright{} extended periodic fields which discrete Fourier transform (DFT) can be related to the DTTs of non-extended fields. This helps to re-adapt the general aspects of the fixed-point algorithm with a minimal cost. Furthermore, finite difference schemes are used here to build the discrete Green operator. For this purpose, the recently introduced double tetrahedron \citep[TETRA2,][]{finel_tetrahedron-based_2025,gelebart_accurate_2025} and the classical hexahedral \citep[HEX1,][]{willot_fourier-based_2015,schneider_fftbased_2017} schemes are employed in the context of non-periodic BCs. Depending on the loading type and the small or finite transformation frameworks, three discrete Green operators, relying on different pre-conditioners (\emph{i.e.,} reference material), are discussed.

The present analysis draws upon the parallel programming concept to explore a range of numerical simulations inspired by realistic experimental non-trivial loading configurations. These simulations are employed to investigate the consequences of selecting a finite difference scheme and a discrete Green operator. We validated the presented approach against analytical solutions for elastic unit-cells loaded in bending (small strain) and in tension (finite transformation). For these examples, both the TETRA2 and HEX1 finite difference scheme results are in a very good agreement with the analytical solutions. Subsequent to this, isotropic plasticity and crystal plasticity material behaviors are employed within heterogeneous unit-cells. Based on the discussed cases (porous media, single-crystals), we show that for strong non-homogeneous loading in finite transformation framework, the TETRA2 scheme appears more robust than the HEX1 scheme. Concerning the discrete Green operator, its $\mathbb{C}_0$-based version converges faster than the two others, $2\mu_0$-based and $\mathbb{B}_0$-based versions, when the loading consists of \textquotedblleft normal-mixed \textquotedblright{} BCs (compatible with the usage of $\mathbb{C}_0$-discrete Green operator). In other cases, the $\mathbb{B}_0$-discrete Green operator, is found to perform better. However, the optimization of the pre-conditioner parameters requires a deeper analysis, specially in the context of finite transformation. To conclude, the different applications demonstrated the versatility, robustness and parallel capabilities of the proposed FFT-based solver.

Going forward with the presented solver, it will be interesting to impose desired displacement and/or force values on interior nodes allowing to simulate for instance a compact tension specimen for which the loading is applied inside the unit-cell. Building on the various treated examples, it would be valuable, in a future work, to integrate the non-uniform grid procedures \citep{bellis_numerical_2024,bignonnet_fourier-based_2026,zecevic_achieving_2026} and to enrich the finite difference schemes (\emph{e.g.,} with C3D20 FE type scheme) within the current generic FFT-based solver. Moreover, non-local models \citep{amouzou-adoun_advanced_2024,mukherjee_quantitative_2026,rys_selecting_2026} could benefit from the versatile presented framework to allow for finite transformation scenarios and three-dimensional calculations, thereby enabling comparison with experimental high-resolution results. For multi-physic applications, it could also be interesting to discuss Robin-type BCs with the present framework.

\section*{Data availability}

\emph{AMITEX$^\star$} code used for the simulations, is available on request. It will be soon made open-source similarly to \emph{AMITEX\_FTTP} \citep{amitex_fftp}.
\section*{Acknowledgements}

This research project was funded by the PTC-SN program (Programmes Transversaux de Competences-Simulation Numérique) of the French Alternative Energies and Atomic Energy Commission (CEA), France. Authors are grateful to Jean-Michel Scherer for providing the Mfront implementation of the crystal plasticity behavior law in finite transformation framework.



\appendix

\section{Symmetric gradient-conjugate $\mathbb{C}_0$-discrete Green operator \label{sec:SS_DGO1}}

Following the detailed analysis for the expression of the discrete Green operator ${}_d\widehat{\mathrm{DGO}}(\widehat{\ten{p}})$ in Sec. \ref{sec:DG01} within finite transformation framework, we outline in the present appendix the changes within small strain framework. With small strain hypothesis, the pre-conditioner is considered as 
\begin{equation}
	\mathbb{R}_0:{}_d\td{\nabla^s u^f} = \mathbb{C}_0:{}_d\td{\nabla^s u^f} = 2\mu_0 \, {}_d\td{\nabla^s u^f} + \lambda_0  \, \mathrm{tr}\left({}_d\td{\nabla^s u^f}\right) \ten{1} 
\end{equation} It implies that 
\begin{equation}
	\widehat{\ten{p}} = \mu_0 \, \widehat{\ten{{}_d\nabla}\cdot\td{{}_d\nabla u^f}} +  (\mu_0 + \lambda_0) \, \widehat{\ten{{}_d\nabla}\cdot\td{{}_d\nabla^T u^f}}
\end{equation} For a given finite difference scheme and for \textquotedblleft normal-mixed\textquotedblright{} BCs, we show that the ${}_d\widehat{\mathrm{DGO}}(\widehat{\ten{p}})$ is given by
\begin{itemize}
	\item \textbf{HEX1 scheme}:
	\begin{equation}
		\widehat{\ten{u^{f}}} = - \displaystyle \frac{1}{\mu_0 \lVert {}_{H}\ten{\xi} \rVert^2} \left(\widehat{\ten{p}} - \frac{ \mu_0 + \lambda_0}{2\mu_0 + \lambda_0} \frac{\widehat{\ten{p}} \cdot {}_{H}\ten{\xi}}{ \lVert  {}_{H}\ten{\xi} \rVert^2} {}_{H}\ten{\xi} \right) = {}_H\widehat{\mathrm{DGO}}(\widehat{\ten{p}})
	\end{equation}
	\item \textbf{TETRA2 scheme}:
	\begin{equation}
		\widehat{\ten{u^{f}}} = \displaystyle - \frac{1}{\mu_0\,\lVert {}_{T_2}\ten{\xi} \rVert^2} \left[\widehat{\ten{p}} + \displaystyle \frac{ \mu_0 +  \lambda_0}{2}\left(p_{\mathrm{int}1}\,\overline{{}_{T_2}\ten{\xi}} + p_{\mathrm{int}2}\,{}_{T_2}\ten{\xi} \right)\right] = {}_{T_1T_2}\widehat{\mathrm{DGO}}(\widehat{\ten{p}})
	\end{equation} with $p_{\mathrm{int}1}$ and $p_{\mathrm{int}2}$ intermediate fields defined as
	\begin{equation}
		\left\{
		\begin{array}{l}
			
			p_{\mathrm{int}1}  = \displaystyle \frac{1}{D_s} \left[\displaystyle -\frac{3\mu_0+\lambda_0}{2}\,\lVert {}_{T_2}\ten{\xi} \rVert^2\,\left( \widehat{\ten{p}} \cdot {}_{T_2}\ten{\xi}\right)  + \displaystyle \frac{\mu_0+\lambda_0}{2}\, \left({}_{T_2}\ten{\xi} \cdot {}_{T_2}\ten{\xi}\right) \left( \widehat{\ten{p}} \cdot \overline{{}_{T_2}\ten{\xi}}\right)\right]  \\[1em]
			
			p_{\mathrm{int}2} = \displaystyle \frac{1}{D_s} \left[ \displaystyle \frac{\mu_0+\lambda_0}{2}\,\left(\overline{{}_{T_2}\ten{\xi}} \cdot \overline{{}_{T_2}\ten{\xi}}\right)\,\left( \widehat{\ten{p}} \cdot {}_{T_2}\ten{\xi}\right) -  \displaystyle\frac{3\mu_0+\lambda_0}{2} \, \lVert {}_{T_2}\ten{\xi} \rVert^2 \left( \widehat{\ten{p}} \cdot \overline{{}_{T_2}\ten{\xi}}\right)\right] \\[1em]
			\qquad \text{where } D_s = \displaystyle \left(\frac{3\mu_0+\lambda_0}{2}\,\lVert {}_{T_2}\ten{\xi} \rVert^2\right)^2 - \left(\frac{\mu_0+\lambda_0}{2}\,\lvert {}_{T_2}\ten{\xi} \cdot {}_{T_2}\ten{\xi} \rvert\right)^2
		\end{array}
		\right.
	\end{equation}
\end{itemize} We recall that the modified frequencies ${}_{H}\ten{\xi}$ (real values), ${}_{T_1}\ten{\xi}$ and ${}_{T_2}\ten{\xi}$ (complex values) depend on the BCs.
	
\section{Neo-Hooken beam under pure tension with Dirichlet BCs: numerical simulation versus (semi-)analytical finite transformation solution \label{sec:NeoHooken}}

\begin{figure}[t]
	\centering\subfloat[Pure tension setup: applied displacement. Dimensions: $l\times w \times h = 100 \times 10 \times 10 \,\mathrm{mm^3}$ with a discretization $n_1\times n_2 \times n_3 = 100 \times 10 \times 10$ voxels.]{
		\includegraphics[width=0.45\textwidth,valign=c]{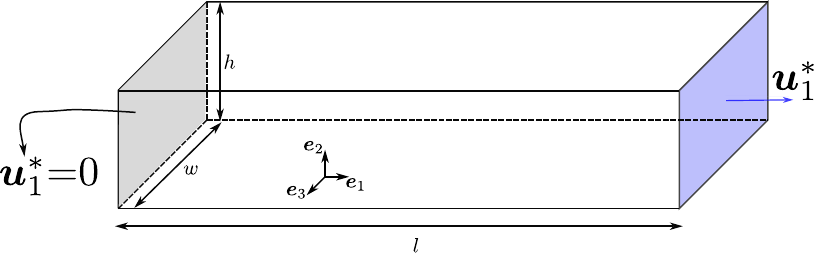}
		\label{fig:ElasticFSTension}
	}
	\hfill{} \subfloat[Analytical solution versus numerical simulations.]{
		\includegraphics[width=0.5\textwidth,valign=c]{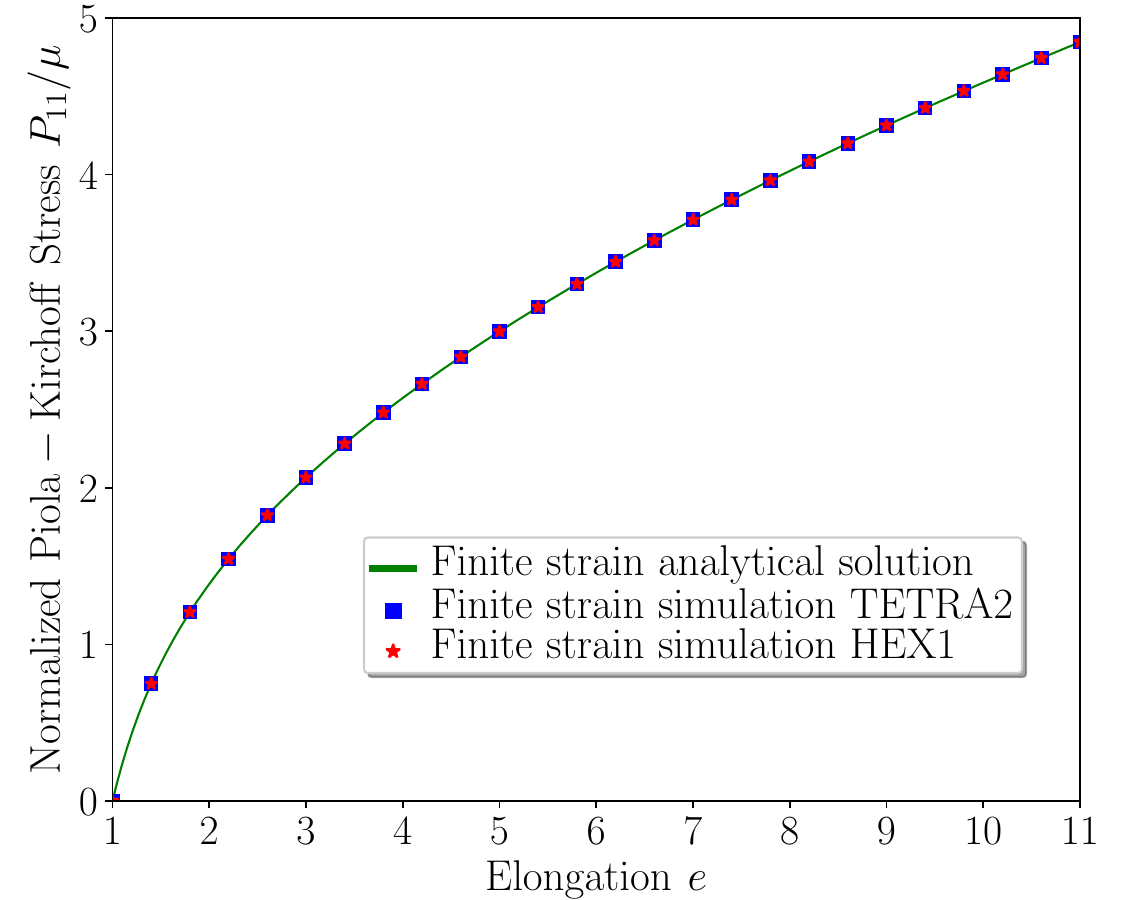}
		\label{fig:FSNeoHookenPureTensionAnalvsSimu}
	}\caption{Finite transformation pure tension of an elastic Neo-Hooken beam.}\label{fig:ElasticFSTensionNumvsAna}
\end{figure}

This section proposes a validation of a displacement-controlled loading numerical simulation based on the presented FFT-based solver against an analytical solution. We consider an elasticity finite transformation framework to study a beam under a pure tension as sketched in Fig.~\ref{fig:ElasticFSTension}. In order to diversify the behavior law, a Neo-Hooken compressible elastic evolution is used in this example.

The strain energy density function $\psi$ is given as
\begin{equation}
	\psi = \displaystyle \frac{\mu}{2} \left(\overline{I_1}-3\right) + \frac{K}{2}\left(J-1\right)^2
\end{equation} where $\mu$ and $K$ are materials parameters respectively the shear and bulk moduli, $\overline{I_1}$ and $J$ are principal invariants defined as
\begin{equation}
	\left\{
	\begin{array}{c}
		I_1 =\mathrm{tr}\left(\td{B}\right) =  \mathrm{tr}\left(\td{F}\td{F}^T\right) \\[0.2em]
		I_2 = \displaystyle \frac{1}{2} \left(\mathrm{tr}\left(\td{B}\right)^2 - \mathrm{tr}\left(\td{B}^2\right)\right) \\[0.2em]
		I_3 = \mathrm{det}\left(\td{B}\right) = J^2
	\end{array}
	\right.  \Longrightarrow 
	\left\{
	\begin{array}{c}
		\overline{I_1} = \displaystyle J^{-2/3} I_1 \\[0.2em]
		\overline{I_2} = \displaystyle J^{-4/3} I_2 
	\end{array}
	\right. 
\end{equation} The Piola–Kirchhoff stress tensors are given by
\begin{equation}
	\td{\Pi} = \displaystyle \pd{\psi}{\td{E}} \quad; \quad \td{P} = \td{F} \,\td{\Pi}
\end{equation} This behavior law is also implemented with MFront \citep{helfer_introducing_2015}.

We demonstrate that the pure tension configuration leads to the analytical solution on the components $P_{11}$ and $P_{22}$ of the first Piola–Kirchhoff stress tensor in the form
\begin{equation}
	\left\{
	\begin{array}{c}
		P_{11} = \displaystyle \frac{2}{3} \mu e_T^{2} \left( e\, e_T^{2} \right)^{-5/3}
		\left( e^{2} - e_T^{2} \right)
		+ K \left( e\, e_T^{4} - e_T^{2} \right) \\[0.5em]
		P_{22} = \displaystyle \frac{1}{3} \mu \,e\, e_T
		\left( e\, e_T^{2} \right)^{-5/3}
		\left( e_T^{2} - e^{2} \right)
		+ K e\, e_T \left( e_T^{2} - 1 \right)
	\end{array}
	\right.
\end{equation} with $e = 1 + \displaystyle \frac{u^*_1}{l}$ the longitudinal elongation and $e_T$ the transversal elongation. The pure tension configuration results in the condition $P_{22}=0$. Solving the non-trival equation $P_{22} = 0$, helps to deduce the transversal elongation $e_T$ from the elongation $e$. In order to facilitate the calculation of $e_T$, a simple Newton-Raphson method is employed in Python. Subsequently, a comparison can be drawn between the (semi-)analytical solution and the numerical simulations, depending on the finite difference scheme. Fig.~\ref{fig:FSNeoHookenPureTensionAnalvsSimu} presents the macroscopic responses. One can observe that the numerical simulations are in perfect accordance with the (semi-)analytical solution. Both schemes converge to the exact same non-linear responses.

\footnotesize{}
\bibliographystyle{elsarticle-harv}
\bibliography{Bibliography}

@article{Moulinec1998,
   author = {H Moulinec and P Suquet},
   isbn = {457825(97)00218},
   journal = {Computer Methods in Applied Mechanics and Engineering},
   pages = {69-94},
   title = {A numerical method for computing the overall response of nonlinear composites with complex microstructure},
   volume = {157},
   year = {1998}
}

@article{zhang_toward_2023,
	title = {Toward the development of plasticity theories for application to small-scale metal structures},
	volume = {120},
	issn = {0027-8424, 1091-6490},
	url = {https://pnas.org/doi/10.1073/pnas.2312538120},
	doi = {10.1073/pnas.2312538120},
	language = {en},
	number = {44},
	urldate = {2025-09-30},
	journal = {Proceedings of the National Academy of Sciences},
	author = {Zhang, Bin and Nielsen, K. L. and Hutchinson, J. W. and Meng, W. J.},
	month = oct,
	year = {2023},
	pages = {e2312538120},
}

@article{zhang_non-proportional_2026,
	title = {Non-proportional plastic deformation at the micron scale: {Single} crystal {Cu} cantilever beams subjected to orthogonal bending},
	volume = {206},
	issn = {00225096},
	shorttitle = {Non-proportional plastic deformation at the micron scale},
	url = {https://linkinghub.elsevier.com/retrieve/pii/S0022509625003497},
	doi = {10.1016/j.jmps.2025.106375},
	language = {en},
	urldate = {2025-10-16},
	journal = {Journal of the Mechanics and Physics of Solids},
	author = {Zhang, Bin and Dahlberg, Carl F. O. and Fischer, Tim and Hutchinson, J. W. and Meng, W. J.},
	month = jan,
	year = {2026},
	pages = {106375},
}

@article{scherer_deformation_2026,
	title = {Deformation band patterns and dislocation structures in finite strain crystal viscoplasticity},
	volume = {303},
	issn = {13596454},
	url = {https://linkinghub.elsevier.com/retrieve/pii/S1359645425009334},
	doi = {10.1016/j.actamat.2025.121647},
	language = {en},
	urldate = {2025-12-09},
	journal = {Acta Materialia},
	author = {Scherer, Jean-Michel},
	month = jan,
	year = {2026},
	pages = {121647},
}

@article{mu_micro-pillar_2014,
	title = {Micro-pillar measurements of plasticity in confined {Cu} thin films},
	volume = {1},
	issn = {23524316},
	url = {https://linkinghub.elsevier.com/retrieve/pii/S2352431614000121},
	doi = {10.1016/j.eml.2014.12.001},
	language = {en},
	urldate = {2025-12-09},
	journal = {Extreme Mechanics Letters},
	author = {Mu, Yang and Hutchinson, J.W. and Meng, W.J.},
	month = dec,
	year = {2014},
	pages = {62--69},
}

@article{bellis_numerical_2024,
	title = {Numerical homogenization by an adaptive {Fourier} spectral method on non-uniform grids using optimal transport},
	volume = {419},
	issn = {00457825},
	url = {https://linkinghub.elsevier.com/retrieve/pii/S0045782523007818},
	doi = {10.1016/j.cma.2023.116658},
	language = {en},
	urldate = {2025-12-11},
	journal = {Computer Methods in Applied Mechanics and Engineering},
	author = {Bellis, Cédric and Ferrier, Renaud},
	month = feb,
	year = {2024},
	pages = {116658},
}

@article{zecevic_new_2022,
	title = {New large-strain {FFT}-based formulation and its application to model strain localization in nano-metallic laminates and other strongly anisotropic crystalline materials},
	volume = {166},
	issn = {01676636},
	url = {https://linkinghub.elsevier.com/retrieve/pii/S0167663621004154},
	doi = {10.1016/j.mechmat.2021.104208},
	language = {en},
	urldate = {2025-12-11},
	journal = {Mechanics of Materials},
	author = {Zecevic, Miroslav and Lebensohn, Ricardo A. and Capolungo, Laurent},
	month = mar,
	year = {2022},
	pages = {104208},
}

@article{paux_discrete_2025,
	title = {A discrete sine–cosine based method for the elasticity of heterogeneous materials with arbitrary boundary conditions},
	volume = {433},
	issn = {00457825},
	url = {https://linkinghub.elsevier.com/retrieve/pii/S0045782524007424},
	doi = {10.1016/j.cma.2024.117488},
	language = {en},
	urldate = {2025-10-23},
	journal = {Computer Methods in Applied Mechanics and Engineering},
	author = {Paux, Joseph and Morin, Léo and Gélébart, Lionel and Amadou Sanoko, Abdoul Magid},
	month = jan,
	year = {2025},
	pages = {117488},
}

@article{gelebart_fft-based_2024,
	title = {{FFT}-based simulations of heterogeneous conducting materials with combined non-uniform {Neumann}, periodic and {Dirichlet} boundary conditions},
	volume = {105},
	issn = {09977538},
	url = {https://linkinghub.elsevier.com/retrieve/pii/S0997753824000287},
	doi = {10.1016/j.euromechsol.2024.105248},
	language = {en},
	urldate = {2025-10-27},
	journal = {European Journal of Mechanics - A/Solids},
	author = {Gélébart, Lionel},
	month = may,
	year = {2024},
	pages = {105248},
}

@article{nkoumbou_kaptchouang_multiscale_2022,
	title = {Multiscale coupling of {FFT}-based simulations with the {LDC} approach},
	volume = {394},
	issn = {00457825},
	url = {https://linkinghub.elsevier.com/retrieve/pii/S0045782522001967},
	doi = {10.1016/j.cma.2022.114921},
	language = {en},
	urldate = {2025-12-15},
	journal = {Computer Methods in Applied Mechanics and Engineering},
	author = {Nkoumbou Kaptchouang, Noé Brice and Gélébart, Lionel},
	month = may,
	year = {2022},
	pages = {114921},
}

@article{finel_tetrahedron-based_2025,
	title = {A tetrahedron-based discretization for {FFT}-based computational homogenization with smooth solution fields},
	volume = {436},
	issn = {00457825},
	url = {https://linkinghub.elsevier.com/retrieve/pii/S0045782524009575},
	doi = {10.1016/j.cma.2024.117703},
	language = {en},
	urldate = {2025-10-23},
	journal = {Computer Methods in Applied Mechanics and Engineering},
	author = {Finel, A.},
	month = mar,
	year = {2025},
	pages = {117703},
}

@article{willot_fourier-based_2015,
	title = {Fourier-based schemes for computing the mechanical response of composites with accurate local fields},
	volume = {343},
	issn = {1631-0721, 1873-7234},
	url = {https://comptes-rendus.academie-sciences.fr/mecanique/articles/10.1016/j.crme.2014.12.005/},
	doi = {10.1016/j.crme.2014.12.005},
	language = {en},
	number = {3},
	urldate = {2025-10-28},
	journal = {Comptes Rendus Mécanique},
	author = {Willot, François},
	month = jan,
	year = {2015},
	pages = {232--245},
}

@article{risthaus_fftbased_2024,
	title = {{FFT}‐based computational micromechanics with {Dirichlet} boundary conditions on the rotated staggered grid},
	issn = {0029-5981, 1097-0207},
	url = {https://onlinelibrary.wiley.com/doi/10.1002/nme.7569},
	doi = {10.1002/nme.7569},
	language = {en},
	urldate = {2025-10-23},
	journal = {International Journal for Numerical Methods in Engineering},
	author = {Risthaus, Lennart and Schneider, Matti},
	month = jul,
	volume = {125},
	number = {21},
	year = {2024},
	pages = {e7569},
}

@article{lucarini_fft_2022,
	title = {{FFT} based approaches in micromechanics: fundamentals, methods and applications},
	volume = {30},
	issn = {0965-0393, 1361-651X},
	shorttitle = {{FFT} based approaches in micromechanics},
	url = {https://iopscience.iop.org/article/10.1088/1361-651X/ac34e1},
	doi = {10.1088/1361-651X/ac34e1},
	language = {en},
	number = {2},
	urldate = {2025-12-15},
	journal = {Modelling and Simulation in Materials Science and Engineering},
	author = {Lucarini, S and Upadhyay, M V and Segurado, J},
	month = mar,
	year = {2022},
	pages = {023002},
}

@article{schneider_review_2021,
	title = {A review of nonlinear {FFT}-based computational homogenization methods},
	volume = {232},
	issn = {0001-5970, 1619-6937},
	url = {https://link.springer.com/10.1007/s00707-021-02962-1},
	doi = {10.1007/s00707-021-02962-1},
	language = {en},
	number = {6},
	urldate = {2025-12-15},
	journal = {Acta Mechanica},
	author = {Schneider, Matti},
	month = jun,
	year = {2021},
	pages = {2051--2100},
}

@article{zecevic_extended_2025,
	title = {Extended {FFT}-based micromechanical formulation to consider general non-periodic boundary conditions},
	volume = {311},
	issn = {00207683},
	url = {https://linkinghub.elsevier.com/retrieve/pii/S0020768325000113},
	doi = {10.1016/j.ijsolstr.2025.113225},
	language = {en},
	urldate = {2025-12-15},
	journal = {International Journal of Solids and Structures},
	author = {Zecevic, Miroslav and Lebensohn, Ricardo A.},
	month = apr,
	year = {2025},
	pages = {113225},
}

@article{lebensohn_orientation_2008,
	title = {Orientation image-based micromechanical modelling of subgrain texture evolution in polycrystalline copper},
	volume = {56},
	issn = {13596454},
	url = {https://linkinghub.elsevier.com/retrieve/pii/S1359645408002887},
	doi = {10.1016/j.actamat.2008.04.016},
	language = {en},
	number = {15},
	urldate = {2025-12-15},
	journal = {Acta Materialia},
	author = {Lebensohn, Ricardo A. and Brenner, Renald and Castelnau, Olivier and Rollett, Anthony D.},
	month = sep,
	year = {2008},
	pages = {3914--3926},
}

@article{gelebart_modified_2020,
	title = {A modified {FFT}-based solver for the mechanical simulation of heterogeneous materials with {Dirichlet} boundary conditions},
	volume = {348},
	issn = {1631-0721, 1873-7234},
	url = {https://comptes-rendus.academie-sciences.fr/mecanique/articles/10.5802/crmeca.54/},
	doi = {10.5802/crmeca.54},
	language = {en},
	number = {8-9},
	urldate = {2025-12-15},
	journal = {Comptes Rendus Mécanique},
	author = {Gélébart, Lionel},
	month = dec,
	year = {2020},
	pages = {693--704},
}

@article{grimmstrele_fftbased_2021,
	title = {{FFT}‐based homogenization with mixed uniform boundary conditions},
	volume = {122},
	issn = {0029-5981, 1097-0207},
	url = {https://onlinelibrary.wiley.com/doi/10.1002/nme.6830},
	doi = {10.1002/nme.6830},
	language = {en},
	number = {23},
	urldate = {2025-12-15},
	journal = {International Journal for Numerical Methods in Engineering},
	author = {Grimm‐Strele, Hannes and Kabel, Matthias},
	month = dec,
	year = {2021},
	pages = {7241--7265},
}

@techreport{wiegmann_fast_1999,
	title = {Fast {Poisson}, {Fast} {Helmholtz} and fast linear elastostatic solvers on rectangular parallelepipeds},
	url = {http://www.osti.gov/servlets/purl/982430-4tTDDX/},
	language = {en},
	number = {LBNL-43565, 982430},
	urldate = {2025-12-15},
	author = {Wiegmann, A.},
	month = jun,
	year = {1999},
	doi = {10.2172/982430},
	pages = {LBNL--43565, 982430},
}

@article{risthaus_robust_2025,
	title = {Robust and {Efficient} {FFT}‐{Based} {Solvers} for {Unit}‐{Cell} {Problems} {With} {Voids} and {Pores} {Under} {Displacement} {Boundary} {Conditions}},
	volume = {126},
	issn = {0029-5981, 1097-0207},
	url = {https://onlinelibrary.wiley.com/doi/10.1002/nme.70124},
	doi = {10.1002/nme.70124},
	language = {en},
	number = {18},
	urldate = {2025-12-15},
	journal = {International Journal for Numerical Methods in Engineering},
	author = {Risthaus, Lennart and Schneider, Matti},
	month = sep,
	year = {2025},
	pages = {e70124},
}

@article{chen_analysis_2019,
	title = {Analysis of the damage initiation in a {SiC}/{SiC} composite tube from a direct comparison between large-scale numerical simulation and synchrotron {X}-ray micro-computed tomography},
	volume = {161},
	issn = {00207683},
	url = {https://linkinghub.elsevier.com/retrieve/pii/S0020768318304554},
	doi = {10.1016/j.ijsolstr.2018.11.009},
	language = {en},
	urldate = {2025-10-29},
	journal = {International Journal of Solids and Structures},
	author = {Chen, Yang and Gélébart, Lionel and Chateau, Camille and Bornert, Michel and Sauder, Cédric and King, Andrew},
	month = apr,
	year = {2019},
	pages = {111--126},
}

@article{ramiere_iterative_2015,
	title = {Iterative residual-based vector methods to accelerate fixed point iterations},
	volume = {70},
	issn = {08981221},
	url = {https://linkinghub.elsevier.com/retrieve/pii/S0898122115004046},
	doi = {10.1016/j.camwa.2015.08.025},
	language = {en},
	number = {9},
	urldate = {2026-01-15},
	journal = {Computers \& Mathematics with Applications},
	author = {Ramière, Isabelle and Helfer, Thomas},
	month = nov,
	year = {2015},
	pages = {2210--2226},
}

@article{gelebart_accurate_2025,
	title = {An accurate and robust {FFT}-based solver for transient diffusion in heterogeneous materials},
	volume = {353},
	copyright = {https://creativecommons.org/licenses/by/4.0/},
	issn = {1631-0721, 1873-7234},
	url = {https://comptes-rendus.academie-sciences.fr/mecanique/articles/10.5802/crmeca.281/},
	doi = {10.5802/crmeca.281},
	language = {en},
	number = {G1},
	urldate = {2025-10-23},
	journal = {Comptes Rendus Mécanique},
	author = {Gélébart, Lionel},
	month = jan,
	year = {2025},
	pages = {113--125},
}

@article{gehrig_elementbased_2025,
	title = {Element‐{Based} {Internal} {Variable} {Formulations} for {Finite} {Element} {Discretizations} in {FFT}‐{Based} {Homogenization} {Methods}},
	volume = {126},
	issn = {0029-5981, 1097-0207},
	url = {https://onlinelibrary.wiley.com/doi/10.1002/nme.70170},
	doi = {10.1002/nme.70170},
	language = {en},
	number = {21},
	urldate = {2026-01-16},
	journal = {International Journal for Numerical Methods in Engineering},
	author = {Gehrig, Flavia and Schneider, Matti},
	month = nov,
	year = {2025},
	pages = {e70170},
}

@article{cocke_implementation_2023,
	title = {Implementation and experimental validation of nonlocal damage in a large-strain elasto-viscoplastic {FFT}-based framework for predicting ductile fracture in {3D} polycrystalline materials},
	volume = {162},
	issn = {07496419},
	url = {https://linkinghub.elsevier.com/retrieve/pii/S0749641922002856},
	doi = {10.1016/j.ijplas.2022.103508},
	language = {en},
	urldate = {2026-01-30},
	journal = {International Journal of Plasticity},
	author = {Cocke, C.K. and Mirmohammad, H. and Zecevic, M. and Phung, B.R. and Lebensohn, R.A. and Kingstedt, O.T. and Spear, A.D.},
	month = mar,
	year = {2023},
	pages = {103508},
}

@article{amouzou-adoun_enhanced_2025,
	title = {Enhanced plastic distortion based micromorphic model for flexible control of scaling effects},
	volume = {322},
	issn = {00207683},
	url = {https://linkinghub.elsevier.com/retrieve/pii/S0020768325003543},
	doi = {10.1016/j.ijsolstr.2025.113568},
	language = {en},
	urldate = {2026-01-30},
	journal = {International Journal of Solids and Structures},
	author = {Amouzou-Adoun, Yaovi Armand and Abatour, Mohamed and Jebahi, Mohamed and Forest, Samuel and Fivel, Marc},
	month = nov,
	year = {2025},
	pages = {113568},
}

@article{kucharski_size_2024,
	title = {Size effects in spherical indentation of single crystal copper},
	volume = {272},
	issn = {00207403},
	url = {https://linkinghub.elsevier.com/retrieve/pii/S0020740324001814},
	doi = {10.1016/j.ijmecsci.2024.109138},
	language = {en},
	urldate = {2026-01-30},
	journal = {International Journal of Mechanical Sciences},
	author = {Kucharski, S. and Maj, M. and Ryś, M. and Petryk, H.},
	month = jun,
	year = {2024},
	pages = {109138},
}

@article{helfer_introducing_2015,
	title = {Introducing the open-source mfront code generator: {Application} to mechanical behaviours and material knowledge management within the {PLEIADES} fuel element modelling platform},
	volume = {70},
	issn = {08981221},
	shorttitle = {Introducing the open-source mfront code generator},
	url = {https://linkinghub.elsevier.com/retrieve/pii/S0898122115003132},
	doi = {10.1016/j.camwa.2015.06.027},
	language = {en},
	number = {5},
	urldate = {2026-01-30},
	journal = {Computers \& Mathematics with Applications},
	author = {Helfer, Thomas and Michel, Bruno and Proix, Jean-Michel and Salvo, Maxime and Sercombe, Jérôme and Casella, Michel},
	month = sep,
	year = {2015},
	pages = {994--1023},
}

@article{mandel_equations_1973,
	title = {Equations constitutives et directeurs dans les milieux plastiques et viscoplastiques},
	volume = {9},
	copyright = {https://www.elsevier.com/tdm/userlicense/1.0/},
	issn = {00207683},
	url = {https://linkinghub.elsevier.com/retrieve/pii/0020768373901200},
	doi = {10.1016/0020-7683(73)90120-0},
	language = {fr},
	number = {6},
	urldate = {2026-01-30},
	journal = {International Journal of Solids and Structures},
	author = {Mandel, J.},
	month = jun,
	year = {1973},
	pages = {725--740},
}

@article{amouzou-adoun_advanced_2024,
	title = {Advanced modeling of higher-order kinematic hardening in strain gradient crystal plasticity based on discrete dislocation dynamics},
	volume = {193},
	issn = {00225096},
	url = {https://linkinghub.elsevier.com/retrieve/pii/S0022509624003417},
	doi = {10.1016/j.jmps.2024.105875},
	language = {en},
	urldate = {2026-01-30},
	journal = {Journal of the Mechanics and Physics of Solids},
	author = {Amouzou-Adoun, Yaovi Armand and Jebahi, Mohamed and Forest, Samuel and Fivel, Marc},
	month = dec,
	year = {2024},
	pages = {105875},
}

@article{miehe_anisotropic_2002,
	title = {Anisotropic additive plasticity in the logarithmic strain space: modular kinematic formulation and implementation based on incremental minimization principles for standard materials},
	volume = {191},
	issn = {0045-7825},
	url = {https://www.sciencedirect.com/science/article/pii/S0045782502004383},
	doi = {10.1016/S0045-7825(02)00438-3},
	number = {47},
	journal = {Computer Methods in Applied Mechanics and Engineering},
	author = {Miehe, C. and Apel, N. and Lambrecht, M.},
	month = nov,
	year = {2002},
	pages = {5383--5425},
}

@article{frigo_design_2005,
	title = {The {Design} and {Implementation} of {FFTW3}},
	volume = {93},
	copyright = {https://ieeexplore.ieee.org/Xplorehelp/downloads/license-information/IEEE.html},
	issn = {0018-9219},
	url = {http://ieeexplore.ieee.org/document/1386650/},
	doi = {10.1109/JPROC.2004.840301},
	language = {en},
	number = {2},
	urldate = {2026-03-18},
	journal = {Proceedings of the IEEE},
	author = {Frigo, M. and Johnson, S.G.},
	month = feb,
	year = {2005},
	pages = {216--231},
}

@article{mukherjee_quantitative_2026,
	title = {Quantitative prediction of size effects using an advanced micromorphic approach},
	volume = {218},
	issn = {01676636},
	url = {https://linkinghub.elsevier.com/retrieve/pii/S0167663626000839},
	doi = {10.1016/j.mechmat.2026.105679},
	language = {en},
	urldate = {2026-04-17},
	journal = {Mechanics of Materials},
	author = {Mukherjee, Anjan and Jebahi, Mohamed and Abatour, Mohamed and Forest, Samuel},
	month = jul,
	year = {2026},
	pages = {105679},
}

@article{kabel_use_2015,
	title = {Use of composite voxels in {FFT}-based homogenization},
	volume = {294},
	issn = {00457825},
	url = {https://linkinghub.elsevier.com/retrieve/pii/S0045782515001954},
	doi = {10.1016/j.cma.2015.06.003},
	language = {en},
	urldate = {2026-04-18},
	journal = {Computer Methods in Applied Mechanics and Engineering},
	author = {Kabel, Matthias and Merkert, Dennis and Schneider, Matti},
	month = sep,
	year = {2015},
	pages = {168--188},
}

@article{gelebart_filtering_2015,
	title = {Filtering material properties to improve {FFT}-based methods for numerical homogenization},
	volume = {294},
	issn = {00219991},
	url = {https://linkinghub.elsevier.com/retrieve/pii/S002199911500203X},
	doi = {10.1016/j.jcp.2015.03.048},
	language = {en},
	urldate = {2026-04-18},
	journal = {Journal of Computational Physics},
	author = {Gélébart, Lionel and Ouaki, Franck},
	month = aug,
	year = {2015},
	pages = {90--95},
}

@article{rys_selecting_2026,
	title = {Selecting deformation patterns in {Cosserat} crystal plasticity by incremental energy minimization},
	volume = {212},
	issn = {00225096},
	url = {https://linkinghub.elsevier.com/retrieve/pii/S0022509626000694},
	doi = {10.1016/j.jmps.2026.106569},
	language = {en},
	urldate = {2026-04-20},
	journal = {Journal of the Mechanics and Physics of Solids},
	author = {Ryś, M. and Kursa, M. and Petryk, H.},
	month = jun,
	year = {2026},
	pages = {106569},
}

@article{schneider_fftbased_2017,
	title = {{FFT}‐based homogenization for microstructures discretized by linear hexahedral elements},
	volume = {109},
	copyright = {http://onlinelibrary.wiley.com/termsAndConditions\#vor},
	issn = {0029-5981, 1097-0207},
	url = {https://onlinelibrary.wiley.com/doi/10.1002/nme.5336},
	doi = {10.1002/nme.5336},
	language = {en},
	number = {10},
	urldate = {2026-04-20},
	journal = {International Journal for Numerical Methods in Engineering}, 
	author = {Schneider, Matti and Merkert, Dennis and Kabel, Matthias},
	month = mar,
	year = {2017},
	pages = {1461--1489},
}

@article{anderson_iterative_1965,
	title = {Iterative {Procedures} for {Nonlinear} {Integral} {Equations}},
	volume = {12},
	issn = {0004-5411, 1557-735X},
	url = {https://dl.acm.org/doi/10.1145/321296.321305},
	doi = {10.1145/321296.321305},
	language = {en},
	number = {4},
	urldate = {2026-04-24},
	journal = {Journal of the Association for Computing Machinery},
	author = {Anderson, Donald G.},
	month = oct,
	year = {1965},
	pages = {547--560},
}

@article{bignonnet_fourier-based_2026,
	title = {Fourier-based computational micromechanics on boundary-conforming transformed grids},
	volume = {354},
	issn = {1631-0721, 1873-7234},
	url = {https://comptes-rendus.academie-sciences.fr/mecanique/articles/10.5802/crmeca.354/},
	doi = {10.5802/crmeca.354},
	language = {en},
	number = {G1},
	urldate = {2026-04-24},
	journal = {Comptes Rendus. Mécanique},
	author = {Bignonnet, François and Ladecký, Martin and Pultarová, Ivana and Zeman, Jan},
	month = mar,
	year = {2026},
	pages = {227--256},
}

@article{rolfo_2decompfft_2023,
	title = {The {2DECOMP}\&{FFT} library: an update with new {CPU}/{GPUcapabilities}},
	volume = {8},
	copyright = {http://creativecommons.org/licenses/by/4.0/},
	issn = {2475-9066},
	shorttitle = {The {2DECOMP}\&{FFT} library},
	url = {https://joss.theoj.org/papers/10.21105/joss.05813},
	doi = {10.21105/joss.05813},
	language = {en},
	number = {91},
	urldate = {2026-04-27},
	journal = {Journal of Open Source Software},
	author = {Rolfo, Stefano and Flageul, Cédric and Bartholomew, Paul and Spiga, Filippo and Laizet, Sylvain},
	month = nov,
	year = {2023},
	pages = {5813},
}

@article{li_2decompfft_nodate,
	title = {{2DECOMP}\&{FFT} - {A} {Highly} {Scalable} {2D} {Decomposition} {Library} and {FFT} {Interface}},
	language = {en},
	journal = {Cray User Group 2010 conference},
	year = {2010},
	author = {Li, Ning and Laizet, Sylvain},
}

@online{amitex_fftp,
  author   = {{AMITEX\_FFTP}},
  title    = { A {FFT}-based solver for non-linear mechanical problems},
  year     = {2020},
  url      = {https://amitexfftp.github.io/AMITEX/index.html},
  urldate  = {2026-04-27}
}

@article{Moulinec1994,
author  = {Moulinec, Hervé and Suquet, Pierre},
title   = {A Fast Numerical Method for Computing the Linear and Nonlinear Mechanical Properties of Composites},
journal = {Comptes Rendus de l'Académie des Sciences. Série II},
volume   = {318},
number   = {11},
pages    = {1417--1423},
year     = {1994}
}

@article{zecevic_achieving_2026,
	title = {Achieving geometric accuracy in {FFT}-based micromechanical models using conformal grid},
	volume = {212},
	issn = {01676636},
	url = {https://linkinghub.elsevier.com/retrieve/pii/S0167663625002741},
	doi = {10.1016/j.mechmat.2025.105512},
	language = {en},
	urldate = {2026-06-09},
	journal = {Mechanics of Materials},
	author = {Zecevic, Miroslav and Lebensohn, Ricardo A. and Capolungo, Laurent},
	month = jan,
	year = {2026},
	pages = {105512},
}

@article{gierden_review_2022,
	title = {A {Review} of {FE}-{FFT}-{Based} {Two}-{Scale} {Methods} for {Computational} {Modeling} of {Microstructure} {Evolution} and {Macroscopic} {Material} {Behavior}},
	volume = {29},
	issn = {1134-3060, 1886-1784},
	url = {https://link.springer.com/10.1007/s11831-022-09735-6},
	doi = {10.1007/s11831-022-09735-6},
	language = {en},
	number = {6},
	urldate = {2026-06-10},
	journal = {Archives of Computational Methods in Engineering},
	author = {Gierden, Christian and Kochmann, Julian and Waimann, Johanna and Svendsen, Bob and Reese, Stefanie},
	month = oct,
	year = {2022},
	pages = {4115--4135},
}

@article{risthaus_imposing_2024,
	title = {Imposing {Dirichlet} boundary conditions directly for {FFT}-based computational micromechanics},
	volume = {74},
	issn = {0178-7675, 1432-0924},
	url = {https://link.springer.com/10.1007/s00466-024-02469-1},
	doi = {10.1007/s00466-024-02469-1},
	language = {en},
	number = {5},
	urldate = {2026-06-11},
	journal = {Comput Mech},
	author = {Risthaus, Lennart and Schneider, Matti},
	month = nov,
	year = {2024},
	pages = {1089--1113},
}

@article{monchiet_fft_2024,
	title = {{FFT} based iterative schemes for composite conductors with uniform boundary conditions},
	volume = {103},
	issn = {09977538},
	url = {https://linkinghub.elsevier.com/retrieve/pii/S0997753823002383},
	doi = {10.1016/j.euromechsol.2023.105146},
	language = {en},
	urldate = {2026-06-11},
	journal = {European Journal of Mechanics - A/Solids},
	author = {Monchiet, V. and Bonnet, G.},
	month = jan,
	year = {2024},
	pages = {105146},
}

@article{cao_modeling_2026,
	title = {Modeling pseudo-ductile failure in hybrid fiber composites using an {FFT}-based solver with non-periodic boundary conditions},
	volume = {203},
	issn = {1359835X},
	url = {https://linkinghub.elsevier.com/retrieve/pii/S1359835X26000461},
	doi = {10.1016/j.compositesa.2026.109599},
	language = {en},
	urldate = {2026-06-11},
	journal = {Composites Part A: Applied Science and Manufacturing},
	author = {Cao, Rui and Wang, Xiaoqiang and Yu, Zhongliang and Cong, Chaonan and Wei, Xiaoding and Ren, Huiqi},
	month = apr,
	year = {2026},
	pages = {109599},
}

@article{brisard_combining_2012,
	title = {Combining {Galerkin} approximation techniques with the principle of {Hashin} and {Shtrikman} to derive a new {FFT}-based numerical method for the homogenization of composites},
	volume = {217-220},
	issn = {00457825},
	url = {https://linkinghub.elsevier.com/retrieve/pii/S0045782512000059},
	doi = {10.1016/j.cma.2012.01.003},
	language = {en},
	urldate = {2026-06-12},
	journal = {Computer Methods in Applied Mechanics and Engineering},
	author = {Brisard, S. and Dormieux, L.},
	month = apr,
	year = {2012},
	pages = {197--212},
}

@article{vondrejc_fft-based_2014,
	title = {An {FFT}-based {Galerkin} method for homogenization of periodic media},
	volume = {68},
	issn = {08981221},
	url = {https://linkinghub.elsevier.com/retrieve/pii/S0898122114002077},
	doi = {10.1016/j.camwa.2014.05.014},
	language = {en},
	number = {3},
	urldate = {2026-06-12},
	journal = {Computers \& Mathematics with Applications},
	author = {Vondřejc, Jaroslav and Zeman, Jan and Marek, Ivo},
	month = aug,
	year = {2014},
	pages = {156--173},
}

@online{castem,
  author   = {{Cast3M}},
  year     = {2026},
  url      = {https://www-cast3m.cea.fr/},
  urldate  = {2026-06-12}
}

@article{hure_homogenized_2026,
	title = {A homogenized model for porous materials with an inhomogeneous matrix: {Application} to the modelling of strain hardening},
	volume = {206},
	issn = {00225096},
	shorttitle = {A homogenized model for porous materials with an inhomogeneous matrix},
	url = {https://linkinghub.elsevier.com/retrieve/pii/S0022509625003746},
	doi = {10.1016/j.jmps.2025.106400},
	language = {en},
	urldate = {2026-06-24},
	journal = {Journal of the Mechanics and Physics of Solids},
	author = {Hure, J.},
	month = jan,
	year = {2026},
	pages = {106400},
}

@article{cruzado_effect_2026,
	title = {Effect of nonuniform porosity on the plastic flow and ductility of metals},
	volume = {316},
	issn = {00207403},
	url = {https://linkinghub.elsevier.com/retrieve/pii/S0020740326003103},
	doi = {10.1016/j.ijmecsci.2026.111454},
	language = {en},
	urldate = {2026-07-01},
	journal = {International Journal of Mechanical Sciences},
	author = {Cruzado, A. and Gélébart, L. and Benzerga, A.A.},
	month = apr,
	year = {2026},
	pages = {111454},
}

@article{to_fourier_2025,
	title = {Fourier {Transform} {Approach} to {Boundary} {Domain} {Integral} {Equations} for {Elastic} {Composites} {With} {Domain} {Decomposition} and {Multi} {Reference} {Parameters}},
	volume = {126},
	issn = {0029-5981, 1097-0207},
	url = {https://onlinelibrary.wiley.com/doi/10.1002/nme.7601},
	doi = {10.1002/nme.7601},
	language = {en},
	number = {1},
	urldate = {2026-07-01},
	journal = {International Journal for Numerical Methods in Engineering},
	author = {To, Quy‐Dong and Bonnet, Guy},
	month = jan,
	year = {2025},
	pages = {e7601},
}

@article{roters_overview_2010,
	title = {Overview of constitutive laws, kinematics, homogenization and multiscale methods in crystal plasticity finite-element modeling: {Theory}, experiments, applications},
	volume = {58},
	copyright = {https://www.elsevier.com/tdm/userlicense/1.0/},
	issn = {13596454},
	shorttitle = {Overview of constitutive laws, kinematics, homogenization and multiscale methods in crystal plasticity finite-element modeling},
	url = {https://linkinghub.elsevier.com/retrieve/pii/S1359645409007617},
	doi = {10.1016/j.actamat.2009.10.058},
	language = {en},
	number = {4},
	urldate = {2026-07-02},
	journal = {Acta Materialia},
	author = {Roters, F. and Eisenlohr, P. and Hantcherli, L. and Tjahjanto, D.D. and Bieler, T.R. and Raabe, D.},
	month = feb,
	year = {2010},
	pages = {1152--1211},
}

\end{document}